\documentclass[english]{JHEP3}
\pdfoutput=1
\usepackage{url}
\usepackage{epsfig}
\usepackage{float}
\usepackage{amsmath}
\usepackage{amsfonts}
\usepackage{multirow}

\def\spaa#1.#2.#3{\langle\mskip-1mu{#1}|#2|{#3}\mskip-1mu\rangle}
\def\spbb#1.#2.#3{[\mskip-1mu{#1}|#2|{#3}\mskip-1mu]}
\def\spa#1.#2{\left\langle#1\,#2\right\rangle}
\def\spb#1.#2{\left[#1\,#2\right]}
\def\spab#1.#2.#3{\left\langle#1|#2|#3\right]}
\def\spba#1.#2.#3{\left[#1|#2|#3\right\rangle}

\def\P3b{\bar{P}_3}

\def\g0{\gamma_0}

\newcommand{\beq}{\begin{equation}}
\newcommand{\eeq}{\end{equation}}
\newcommand{\beqn}{\begin{eqnarray}}
\newcommand{\eeqn}{\end{eqnarray}}

\newcommand{\pt}{p_t}

\def\vec#1{{\mbox{\boldmath$#1$}}}

\newcommand\POWHEGBOX{{\tt POWHEG BOX}}
\newcommand\POWHEG{{\tt POWHEG}}
\newcommand\PYTHIASIX{{\tt PYTHIA 6}}
\newcommand\PYTHIA{{\tt PYTHIA}}
\newcommand\PYTHIAEIGHT{{\tt PYTHIA 8}}
\newcommand\Herwig{{\tt HERWIG}}
\newcommand\Herwigpp{{\tt HERWIG++}}
\newcommand\MINLO{MiNLO}
\newcommand\diff{\mathrm{d}}

\newcommand\ptl{p_t^{(l)}}
\newcommand\etal{\eta_l}
\newcommand\ptnu{p_t^{(\nu)}}

\newcommand\et{E_t}
\newcommand\kt{k_t}

\newcommand\ptmin{p_t^{\rm min}}
\newcommand\VJJ{{\tt W/ZJJ}}
\newcommand\WJJ{{\tt WJJ}}
\newcommand\HJJ{{\tt HJJ}}
\newcommand\WJ{{\tt WJ}}
\newcommand\ZJ{{\tt ZJ}}
\newcommand\HJ{{\tt HJ}}
\newcommand\W{{\tt W}}
\newcommand\ZJJ{{\tt ZJJ}}
\newcommand\Vjj{{$W/Zjj$}}
\newcommand\Vj{{$W/Zj$}}
\newcommand\V{{$W/Z$}}
\newcommand\Wjj{{$Wjj$}}
\newcommand\Wjjj{{$Wjjj$}}
\newcommand\Zjj{{$Zjj$}}
\newcommand\HTot{$\hat{H}_{\scriptscriptstyle\rm T}/2$}
\newcommand\HT{$\hat{H}_{\scriptscriptstyle\rm T}$}
\newcommand\MPI{MPI}
\newcommand\MCFM{{\tt MCFM}}
\newcommand\MadLoop{{\tt MadLoop}}

\makeatletter

\preprint{FERMILAB-PUB-13-075-T,OUTP-13-08P}
\title{W and Z bosons in association with two jets using the POWHEG method}

\author{
John M. Campbell and R. Keith Ellis\\
    Fermilab, Batavia, IL 60510, USA\\
    E-mail: 
    \email{johnmc@fnal.gov}, 
    \email{ellis@fnal.gov}\\
}
\author{Paolo Nason\\
  INFN, Sezione di Milano Bicocca,  Piazza della Scienza 3, 20126 Milan, Italy\\
  E-mail: \email{paolo.nason@mib.infn.it}\\
}
\author{Giulia Zanderighi\\
  Rudolf Peierls Centre for Theoretical Physics, 1 Keble Road, University of Oxford, UK\\
  E-mail: \email{g.zanderighi1@physics.ox.ac.uk\\}
}

\abstract{In this work we present the implementation of generators for
  $W$ and $Z$ bosons in association with two jets interfaced to parton
  showers using the \POWHEGBOX{} method.  We incorporate matrix
  elements from the parton-level Monte Carlo program MCFM in the
  POWHEG-BOX, allowing for a considerable improvement in speed
  compared to previous implementations.  We address certain problems
  that arise when processes that are singular at the Born level are
  implemented in a shower framework using either a generation cut or a
  Born suppression factor to yield weighted events. In such a case,
  events with very large weights can be generated after the shower
  through a number of mechanisms. Events with very small transverse
  momentum at the Born level
  can develop large transverse momentum either after the hardest emission,
  after the shower, or after the inclusion of multi-parton
  interactions. We present a solution to this problem that can be
  easily implemented in the \POWHEGBOX{}. We also show that a full
  solution to this problem can only be achieved if the generator
  maintains physical validity also when the transverse momentum of the
  emitted partons becomes unresolved. One such scheme is the recently-proposed
  \MINLO{} method for the choice of scale and the exponentiation of
  Sudakov form factors in NLO computations.  We present a validation
  study of our generators, by comparing their output to available LHC
  data. Furthermore, we suggest an observable that is very sensitive
  to the modeling of multi-parton interactions, that may be studied in
  both $W$ and $Z$ production in association with two jets.}

\keywords{QCD, Phenomenological Models, Hadronic Colliders, LHC}

\begin{document}
\maketitle

\section{Introduction} 
The production of a vector boson at large transverse momentum,
i.e.\ in association with jet activity, played an important role as a
testing ground for perturbative QCD calculations and event
generators. Nowadays, these processes continue to be used for model
testing. However, their primary importance derives from their role as
a background to rarer physics processes.  Thus the production of a
Higgs boson decaying to $W^+W^-$, where one of the $W$ bosons decays
hadronically, has as its principal background the $W$ + 2 jet process.
The $W$ + 2 jet process is also a background to $WW$ and $WZ$
production where one $W$ or the $Z$ boson decays hadronically.  With
higher jet multiplicities the $W$+jets process is the primary
background to top pair production.  In addition the $Z$ + 2-jet
process with the $Z$-boson decaying to neutrinos gives a signature of
two jets and missing energy that constitutes a background to searches
for weakly interacting particles, such as those that occur in
supersymmetry~\cite{Mangano:2008ha} or in the search for dark
matter~\cite{Fox:2012ee}.

Ten years ago, the next-to-leading order (NLO) corrections to
$W$+2-jet and $Z$+2-jet production (from now on \Wjj{} and \Zjj){} at
hadron colliders were
published~\cite{Campbell:2002tg,Campbell:2003hd}.
These calculations were performed at the parton level, so the
predicted final states are jets of partons, rather than the hadrons
observed in the experiment.  NLO results for vector bosons in
association with 3 and 4 jets have also been
obtained~\cite{KeithEllis:2009bu,Berger:2009zg,Berger:2010vm,Berger:2010zx,Ita:2011wn}.

Although parton level predictions have been
successfully compared with
experiment~\cite{Aaltonen:2007ip,Abazov:2011rf,Aad:2011qv,Aad:2012en}
and continue to provide a useful benchmark,
NLO calculations are most useful when matched with parton showers such
as \PYTHIA{}~\cite{Sjostrand:2006za,Sjostrand:2007gs} or
\Herwig{}~\cite{Corcella:2002jc,Bahr:2008pv}, including an implementation
of hadronization models, that describe the transition from partons to
the observed hadrons.  Viable methods for matching NLO calculations
with parton shower generators have been given in
refs.~\cite{Frixione:2002ik,Nason:2004rx,Frixione:2007vw}.

Both the \Wjj{} and \Zjj{} processes at NLO matched with parton
showers have been previously considered in the literature.  
Using the aMC@NLO method, 
ref.~\cite{Frederix:2011ig} gives results for the \Wjj{}
process at the Tevatron, in order to more reliably assess the
principal Standard Model background to the Tevatron
anomaly~\cite{Aaltonen:2011mk}.\footnote{A recent CDF analysis of a larger data set~\cite{CDFnote10973}, accounting
for a number of subtle systematic effects -- most notably improving the treatment of jet energy scale
corrections, finds no evidence for the earlier anomaly.}  More recently, results for \Wjj{} and
\Wjjj{} processes have been presented using a variant of the MC@NLO
method~\cite{Frixione:2002ik} in the SHERPA framework~\cite{Hoeche:2012ft}. Meanwhile, the
\Zjj{} process has been considered previously in the context of the
\POWHEGBOX{} by Re in ref.~\cite{Re:2012zi}.
The electroweak production of \Wjj{~\cite{Schissler:2013nga} and \Zjj{}~\cite{Schissler:2013nga,Jager:2012xk} matched with parton showers has also been considered recently. 

In this paper we present new \POWHEG{} generators, based upon the
\POWHEGBOX{}~\cite{Alioli:2010xd} framework, for the \Wjj{} and \Zjj{}
processes. We will refer to these henceforth as the \WJJ{} and \ZJJ{}
generators.  These generators present a considerable advantage in
speed (more than a factor of three improvement)
as compared to the previous ones, thanks to the use of the matrix
elements taken from the \MCFM{} program. These rely upon the calculation of
ref.~\cite{Bern:1997sc} for the virtual amplitudes, and of
ref.~\cite{Nagy:1998bb} for the real contributions.

Our generators, being based upon \POWHEG{}, offer the possibility of
investigating the effects of matching with all parton shower models
that comply with the Les Houches requirements for interfacing showers
to user processes. In this work we have mainly used \PYTHIASIX{},
but we have also checked that no substantial differences are seen
with \PYTHIAEIGHT{}. The programs can be interfaced as easily to
\Herwig{} and \Herwigpp{}. In this case one should also supply
truncated showers~\cite{Nason:2004rx} in order for coherence in soft
emissions to be correctly treated. The full impact of truncated showers,
in particular for complex processes, remains to be properly assessed.
However, in the \POWHEG{} simulations where
they have been implemented and studied, their effect was found to be basically
negligible~\cite{Hamilton:2010mb,Hamilton:2009za,Hamilton:2008pd,LatundeDada:2006gx,Hamilton:2009ne}.
It was also shown that truncated shower effects can be mimicked well simply by a
clever choice of shower starting scales \cite{Schofield:2011zi}.

In this work we also address some physics issues, that are 
relevant to the matching of NLO and parton showers 
for all processes with associated jets present 
at the Born level.\footnote{In the present
work, by processes with associated jets we mean processes obtained by adding
jets to a more basic reaction. Typically, these processes become singular
when the associated jets become collinear or soft.
Thus we explicitly exclude, for example,
vector boson fusion, where jets are present at the Born level
and cannot be eliminated.} As is well known, in the
framework of shower models, a generation cut is needed in these cases
at the Born level, in order to build a finite sample. In fact, the
Born level cross section diverges when the associated jet has small
transverse momentum. One then imposes that, at the Born level, the
transverse momenta of the jets must be above a generation cut. Thereafter 
it is necessary to ensure
that events that do not pass the generation cut, 
would not, after showering and hadronization, contribute appreciably
to the final output after the analysis cuts are imposed. In other words,
the generation cut should be low enough so that events lying just 
below the $\pt$ cut at the Born level,
would only rarely be promoted to high $\pt$ events after showering
and hadronization.  In the \POWHEG{} approach, this ``promotion'' of
low $\pt$ to high $\pt$ events can take place at three levels: at the
generation of the hardest radiation, at the shower level, and at the
hadronization level, especially after accounting for multi-parton interactions.
This last point is particularly insidious, since (for
example) a $W$ production event with very little jet activity after
the hardest emission and parton shower, may acquire hard jets because
of multi-parton interactions. In this paper we examine in detail these
problems and propose a suitable solution that makes use of the
recently developed \MINLO{}
method~\cite{Hamilton:2012np,Hamilton:2012rf}.

In our current implementation of the \ZJJ{} and \WJJ{} generators, we
have incorporated several extensions and improvements to the
\POWHEGBOX{} framework, that will become eventually a ``Version 2'' of
the \POWHEGBOX{}. These extensions are described in the following section.

\section{Description of the implementation}
\label{implementation}
The routines for the generation of the Born phase space have been
taken, with minor modifications, from ref.~\cite{Campbell:2012am}.
The remaining ingredients needed for the \POWHEGBOX{} implementation
are as follows: the lowest order (Born approximation) matrix element
squared; the
real matrix element squared for the emission of one extra parton; the
virtual contributions coming from the interference of the one-loop
matrix element with the lowest order matrix element; the color
correlated Born and spin correlated Born matrix elements squared;
the Born amplitudes in the large $N_c$ limit (where $N_c$ is the
number of colours) needed for assigning a colour structure to the
generated Les Houches events.

Previous implementations of \Vjj{} processes have used automatically
generated Madgraph-style calculations for all or some of these
ingredients. The Madgraph-\POWHEGBOX{} interface presented in
ref.~\cite{Campbell:2012am} can be used to generate automatically all
the above ingredients, except for the phase space and the virtual
corrections.

Our implementation uses the simpler analytic expressions for all
matrix elements  that are used in the \MCFM{} program,
leading to an improvement in computing time
performance. These rely upon the
calculation of ref.~\cite{Nagy:1998bb} for the real, and of
ref.~\cite{Bern:1997sc} for the virtual contributions. The new
analytic routines have been checked by comparing with the routines
generated automatically using the Madgraph-\POWHEGBOX{} interface.  We
have also compared the virtual matrix elements of the \ZJJ{} generator
with those derived from the BlackHat results and found agreement.  The
\MCFM{} matrix elements have also been checked with the \MadLoop{}
results~\cite{Hirschi:2011pa}.

\subsection{Improved aspects of the \POWHEGBOX{}}
Our implementation makes use of improved aspects of the \POWHEGBOX{},
that at some point will lead to a new version of the package. These
aspects are:
\begin{itemize}
\item The possibility of generating the importance sampling grids by
  parallel runs.  This was not possible with the standard \POWHEGBOX{}
  version. Without this feature, the generation of the importance
  sampling grid for the integration becomes a bottleneck for complex
  processes, where a single processor run requires too much time.
\item The possibility of performing scale variation or PDF variation
  studies using reweighting.  Again, for complex processes this is the
  only practical way to study uncertainties.
\item The possibility to adopt the
  \MINLO{}~\cite{Hamilton:2012np,Hamilton:2012rf} procedure for
  processes with associated jets.
\item An improved structure of the generation of the underlying Born
  kinematics, first introduced in ref.~\cite{Melia:2011gk}, and now
  made available for all processes, which largely increases the
  generation efficiency.
\end{itemize}

\subsection{Generation cuts and Born suppression factor}
The Born matrix elements for the \Zjj{} and \Wjj{} processes contain
both soft and collinear divergences.  These require a special
treatment, in order to avoid the predominant generation of soft or
collinear events that will be removed by phase space cuts on the jets.
This problem is typically handled by introducing generation cuts that
avoid the soft and collinear region in the underlying Born.
Alternatively one can introduce a suppression
factor~\cite{Alioli:2010qp} that depends upon the kinematics of the
underlying Born configuration.  Here we use this second alternative.
We use a modified (Born suppressed) $\bar{B}$ function of the form
\beq
\label{Bornsupp}
\bar{B}_{\mbox{supp}}(\Phi_n) = \bar{B}(\Phi_n) F(\Phi_n)\,.
\eeq
Events are then generated as usual, except that a variable weight
equal to $1/F(\Phi_n)$ is applied, leading to a weighted event sample.

We chose as the reweighting factor the form
\beq
F(\Phi_n)= \left[1+ \frac{\Lambda^2}{p_1^2}+\frac{\Lambda^2}{p_2^2} +\frac{\Lambda^2}{p_{12}^2} \right]^{-1}\,,
\eeq
where $p_1^2$, $p_2^2$ are the transverse momentum squared of the two
recoiling partons $1$ and $2$ and
\beq
p_{12}^2= \frac{2 p_1 \cdot p_2 \; E_1 E_2}{(E_1^2+E_2^2)}\,.
\eeq
The cutoff parameter $\Lambda$ is set of the order of the
typical jet transverse momentum cutoff that we need in the analysis.
\subsection{Scale choice}
We consider in the present work two schemes for the factorization and
renormalization scale. The first one uses \HTot{} as a central scale,
with
\begin{equation}
\hat{H}_T=\sqrt{M_V^2+p_{t,V}^2}+\sum_{j=1}^2 \pt^{(j)},
\end{equation}
where $M_V$ is the mass of the $W$ or $Z$ boson, $p_{t,V}$ its
transverse momentum and $\pt^{(j)}$ denotes the transverse momentum of
parton $j$.  As customary in \POWHEG{}, the \HT{} value is computed
using the underlying Born configuration of the event. We have checked
that there is little difference when using real momenta in the \HT{}
definition.
The second scheme relies on the \MINLO{} method, illustrated in detail
in ref.~\cite{Hamilton:2012np}. We remark here that, when using the
\MINLO{} method, it is no longer necessary to include a suppression
factor.  The Sudakov form factors introduced in the \MINLO{} procedure
suppress the region of soft and collinear emission, leading to finite
predictions. Furthermore, as shown in refs.~\cite{Hamilton:2012np,
  Hamilton:2012rf}, when \MINLO{} is applied, the predictions for
\Vjj{} merge smoothly into the prediction for \Vj{} and even for the
inclusive \V{} production when one or two partons become unresolved.

\section{Problems with singular underlying Born configurations}\label{sec:problems}
The Born suppression method works if, after imposing the analysis
cuts, the events with larger weight appear with a much smaller
frequency.  We note that the method of Born suppression is in
fact more general than the commonly adopted method of using generation
cuts. In fact, one may argue that these are equivalent to a Born
suppression factor that is infinitesimally small in the regions below
the cut. This leads to the possibility of events with infinitely large weights,
that however will never appear in any finite number of events. Thus,
when generation cuts are used, one should study the dependence of the
result upon the cuts to make sure that events right below the
generation cut do not contribute appreciable effects to relevant
kinematic observables. On the other hand, if a suppression factor is
used, it is guaranteed that for a sufficiently long run one obtains
the right result.  Thus, the use of a Born suppression factor has the
advantage that one does not need to check the dependence of the result
upon the generation cut.

However if the Born suppression factor is very severe then,
in a run with a limited number of events, it is possible that no events
with very large weights are generated. In this
case these events will only appear for a sufficiently large sample and,
because of their very large weight, they will lead to large spikes in
the computed distributions.  This is a problem that has occurred in practice
while studying our \VJJ{} generators with high statistics
runs.

\subsection{Statement of the problems}
In the present work we have extensively studied this ``spikes''
problem, and have found essentially three problematic areas that have
to be dealt with in order to solve it. These are:
\begin{enumerate}
\item Given an underlying Born configuration near a singularity, the
  \POWHEG{} hardest radiation may generate a phase space point that is
  away from any singularity. It may thus happen that a point passing
  the analysis cuts at the Les Houches level, may in fact carry a
  large weight, because the corresponding underlying Born
  configuration is soft or collinear.
\item An event that originates from an underlying Born configuration
  near a singularity, and that after the generation of the hardest
  radiation is still near the singularity, may lead to an event that
  passes the analysis cuts after the parton shower. In other words the
  shower may lead, in rare cases, to radiation that is harder than the
  one in the Les Houches event.
\item An event that originates from an underlying Born configuration
  near a singularity, and that reaches the shower stage still
  remaining near the singular region, may become a hard event, passing
  the analysis cuts, after hadronization and the inclusion of the
  underlying event.  This possibility is especially due to multi-parton
  interactions, that can generate pairs of hard jets accompanying the
  basic hard interaction.
\end{enumerate}

\subsection{Problem 1}
We illustrate the first problem using the simpler example of vector
boson production in association with one jet. In \POWHEG{}, one begins
by generating an underlying Born configuration, consisting of the
vector boson $V$ in association with one parton. In the partonic CM
system, this configuration is singular when the emitted parton has
small transverse momentum. Let us now consider the configuration when
the parton has appreciable energy, but nearly zero emission
angle. This event is passed to the mechanism for the generation of the
hardest radiation in \POWHEG{}. It may happen that the hardest
radiation is from the initial state. In this case, the full event
kinematics is constructed by boosting the underlying Born event in the
transverse direction, in order to balance the transverse momentum of
the radiation~\cite{Frixione:2007vw}.  Following this, the real
emission configuration would be away from all singular regions, since
the transverse boost will give a transverse momentum also to the parton
in the underlying Born configuration.  In the case of final state
radiation, a similar circumstance may arise, with the real radiation
originating from the large angle splitting of the Born parton
(collinear to the beam) into two partons, balanced in transverse
momentum.  Also in this case the real emission configuration is
non-singular.

The \POWHEG{} formula for the hardest radiation is schematically given
by
\begin{equation}\label{eq:powheg}
d\sigma = \bar{B}\, \diff\Phi_{\mathrm{B}} \Delta(k_T) \frac{R}{B}\,  \diff\Phi_{\mathrm{rad}},
\end{equation}
where we have used the standard notation of the \POWHEG{} literature~\cite{Nason:2004rx,Frixione:2007vw}.
The Sudakov form factor $\Delta(k_T)$ is such that
\begin{equation}
\int \Delta(k_T) \frac{R}{B} \, \diff\Phi_{\mathrm{rad}} = 1\,,
\end{equation}
up to corrections that are suppressed by typical hadronic scales.  Let
us assume first that no generation cuts or Born suppression factors
are used.  The \POWHEG{} generator works in a two-stage process. First
the underlying Born kinematics is generated with a probability
proportional to $\bar{B}$. Then the radiation kinematics is generated
with a probability given by eq.~\eqref{eq:powheg}. So, if there is a
singular region for the underlying Born, that region will have a
largely enhanced generation. However radiations that are not as singular
will be suppressed by the $R/B$ ratio, leading to a finite, regular
contribution.

We now observe that, although formally correct, the above approach may
lead to practical problems in the implementation. Let us suppose first
that we use a generation cut, avoiding the regions of underlying Born
kinematics where the parton in the Born configuration has small
transverse momentum. By doing this, we are also cutting off the
corresponding region of real radiation that has a singular underlying
Born.

A very clear example is given again in the \Vj{} example, when in the
real emission configuration the vector boson has nearly zero
transverse momentum, while the two emitted partons have a relatively
large, balanced transverse momenta. This real event admits an
underlying Born structure where the vector boson has nearly zero
transverse momentum, and the two partons are merged into a single
parton with small transverse momentum. Clearly, when using a
generation cut, one removes this type of configuration. The more one
lowers the cut, the better these configurations are covered, at the
cost of increasing the rate of production of soft events that will
never pass the final cuts. If instead we are using a Born suppression
factor then these configurations will always be produced, but they will
arise rarely and with a very large weight, leading to undesirable
spikes with large errors in the final distributions.

A simple way to avoid this problem is to modify the association of the
real contributions with an underlying Born, in such a way that the regular
real contributions associated with singular Born ones are suppressed.
In \POWHEG{} the separation of regions in the real cross section is
performed as follows
\begin{equation}\label{eq:moddalpha}
R=\sum_\alpha R_{\alpha}\,,\quad\quad  
R_{\alpha}=R\times\frac{\frac{1}{d_{\alpha}}}{\sum_\alpha \frac{1}{d_{\alpha}}}\,,
\end{equation}
where $\alpha$ labels each possible collinear singular region of the
real amplitude and $d_\alpha$ are functions of the event kinematics
that vanish in the singular regions. They are typically given by (some
power of) the transverse momentum of the parton for the ISR collinear
regions, and by the relative transverse momentum of the partons in the
FSR ones.

We thus replace the functions $d_{\alpha}$ with new functions
$\tilde{d}_{\alpha}$,
given by the following prescription:
\begin{equation}\label{eq:newdij}
\tilde{d}_{\alpha} =
 d_{\alpha}\times \sum_\beta\frac{1}{d^{\rm{ub}}_{\alpha,\beta}}\,,
\end{equation}
where now with $\beta$ we label the singular regions of the underlying
Born configuration associated with the singular region $\alpha$ in the
real contribution, and $d^{\mathrm ub}_{\alpha,\beta}$ are functions
of the underlying Born configuration kinematics that vanish when such
a configuration approaches the singular region $\beta$. In practice, we
choose them of the same form as the $d_\alpha$. This
modification has the effect of slowing the vanishing of the $d_\alpha$
coefficient near the singular region, if the corresponding underlying
Born is also singular.

It is instructive to see the effect of the modification
eq.~(\ref{eq:newdij}) in our previous example of a vector boson
produced at very small transverse momentum, associated with two
partons with relatively large transverse momenta that balance each
other.  Labelling the two partons with 1 and 2, we have three regions:
region 1, with parton 1 collinear to the beam, region 2, with 2
collinear to the beam, and region 12 with 1 collinear to 2, with
their associated $d_\alpha$ functions $d_1$, $d_2$ and $d_{12}$. With
the old $d_\alpha$ definition the $d_{12}$ term is of the same order
as the others.  With the new definition, instead, $\tilde{d}_{12}$ is
much bigger, because the underlying Born configuration of the $12$
region is singular, and, as dictated by eq.~(\ref{eq:newdij}),
$\tilde{d}_{12}$ includes a large extra factor $1/d^{\rm{ub}}_{12,0}$,
where 0 labels the (only) singular region of the underlying Born
corresponding to the Born parton becoming collinear to the beam.  On
the other hand, the underlying Born corresponding to regions 1 and 2
is non-singular. In fact, if for example we consider region 1, the
underlying Born in the \POWHEG{} formalism is obtained by boosting the
$V$-2 system (i.e. the system consisting of the vector boson plus
parton 2) in the transverse direction, in such a way that after the
boost the $V$-2 system is balanced, and the vector boson has non
vanishing transverse momentum. Thus, in the partition of $R$ according
to the singular regions, the 12 region is largely suppressed.

In the following we show the effect of the implementation of the
prescription in eq.~\eqref{eq:newdij}.  We consider $W^+\to \mu^+\nu$
production at the 7 TeV LHC, with a standard set of cuts on the
lepton, missing energy and jets, requiring two jets with transverse
momentum larger than 20~GeV.  The cuts are given later in
table~\ref{tab:wcuts}, but their precise form is irrelevant for the
present considerations.  In order to speed up the computation, we have
not used in our run the ``folding''~\cite{Alioli:2010xd,Nason:2007vt}
option of the \POWHEGBOX{}. By doing so, we have a relatively large
fraction of negative weighted events.

In fig.~\ref{fig:newdij-olddij} we plot the cross section as a
function of the event weight, as obtained with the old definition
of the $d_\alpha$ functions and the new one
(i.e. the $\tilde{d}_\alpha$ of eq.~(\ref{eq:newdij})).
\begin{figure}[htb]
\begin{center}
\epsfig{file=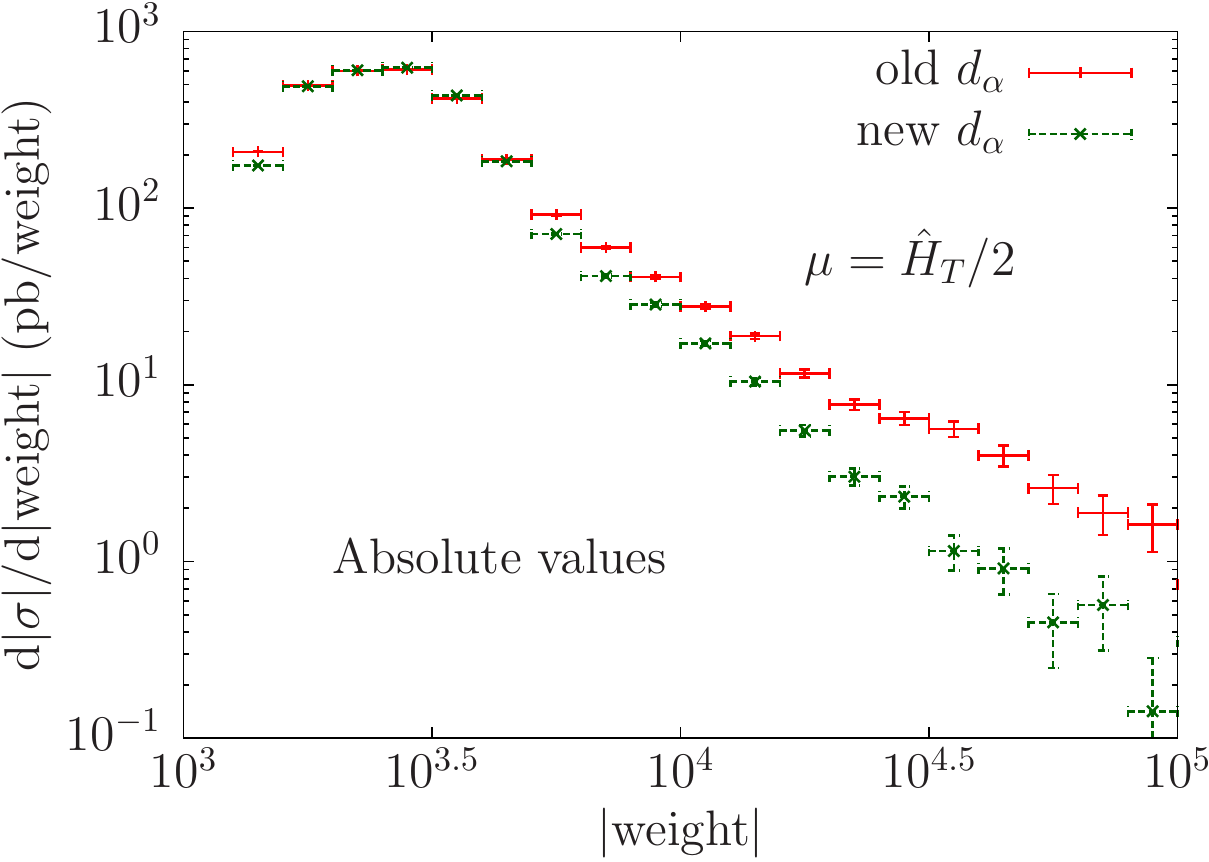,width=0.48\textwidth}
\epsfig{file=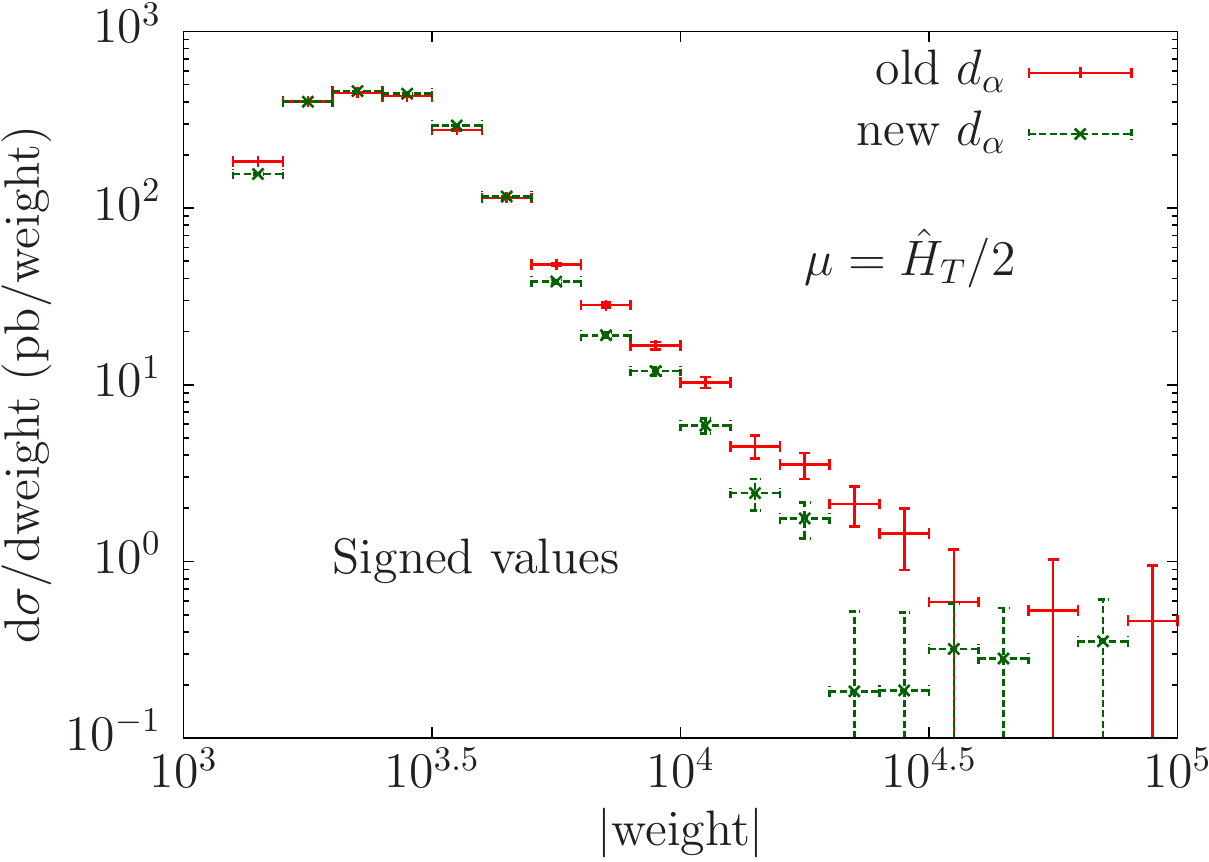,width=0.48\textwidth}
\end{center}
\caption{Cross section as a function of the absolute value of the
  event weight for $W^+$ production in association with two jets 
  (where cuts and jet definition are specified in
  table~\protect\ref{tab:wcuts}) at the 7 GeV LHC.  The \POWHEG{}
  generation uses the \HTot{} scale choice, and the analysis is
  performed at the Les Houches Event level, that is to say, no further
  shower is applied to the event.  The effect of using the old or the
  new $d_\alpha$ definition is shown. In the left plot, the absolute
  value of the cross section is histogrammed, while on the right plot
  the signed value is used.}
\label{fig:newdij-olddij}
\end{figure}
In the figure we plot both the absolute value of the cross section
(left) and the signed cross section (right). We see that there is a
tail of events with large weight, their contribution becoming smaller
as the weight increases. In order to have convergence, the
contribution to the cross section should decrease faster than the
inverse of the weight.  We see that this is not the case for the old
$d_\alpha$ definition as far as the absolute value of the cross
section is concerned.  The large tail of weights causes the ``spike''
problem when building histograms using the old $d_\alpha$ definition.

It is likely that, by using the folding feature to reduce the impact
of negative weights, the ``spike'' problem would also be reduced with
the old $d_\alpha$ definition. However it is also clear from fig.~\ref{fig:newdij-olddij}
that the new definition, besides being more correct from a formal point of view (as
we have shown previously), also performs better in all cases.

\subsection{Problem 2}
The second kind of problem may arise if the \POWHEG{} generator passes 
to the shower program 
a transverse momentum upper limit for subsequent
radiation that is too high. 
(In practice the transverse momentum upper limit 
is the value for the {\tt scalup} 
variable in the Les Houches common block).
This problem was seen to
affect especially the \POWHEG{} dijet production
generator~\cite{Alioli:2010xa}, and has been definitely solved with
the introduction of the {\tt doublefsr} option~\cite{doublefsr}.
In the case of vector boson
production in association with two jets, this option does not seem to
have visible consequences. However, we have set this feature as
default in our \WJJ{} and \ZJJ{} generators.

\subsection{Problem 3}
We now examine how an event that is soft at the parton level may
become hard after the inclusion of hadronization effects.  In our
studies, we have considered a large set of physical distributions at
three levels: at the parton level after shower, after the
hadronization but without multi-parton interactions (MPI from now on),
and after hadronization including MPI.  Our observables typically
require jets with a transverse momentum above $20$~GeV. With this
requirement we see only small effects when comparing the first two cases.
However we do see important effects at the third stage, when MPI are
turned on. This is not surprising, since an event of inclusive vector
boson production with relatively modest radiation activity may be
accompanied by a pair of relatively hard balanced jets due to a
secondary partonic collision. The cross section for such events is
given by the product of the inclusive $W$ cross section and the two jet cross
section for dijets with transverse momentum
above $20$~GeV, divided by the effective cross section for multi-parton
interactions, $\sigma_{\rm eff}$.  The probability for MPI events in our sample
is enhanced due to the fact that the
inclusive $W$ cross section is about a factor of 30 larger than the
cross section for $W$ production in association with at least two
20~GeV jets.  A back-of-the-envelope
estimate yields a probability of the order of
10\%{}.  This genuine physical effect should be accounted for correctly in a
generator, and thus the NLO generator should also provide a fair
description of the inclusive production of the vector boson.

In the
present work we will rely upon the \MINLO{} method, introduced in
ref.~\cite{Hamilton:2012np}.
  With this method the inclusive cross
section for the generation of the vector boson comes out remarkably
close to the NLO inclusive cross section. This property is discussed
in detail in the relevant references~\cite{Hamilton:2012np,Hamilton:2012rf}.
Here we only give a brief reminder of how this is possible, and why it is the
case. In the \MINLO{} method, generated parton level kinematic configurations
are clustered (using, for example, the $\kt$-clustering algorithm)
in such a way that a sort of branching history is constructed. Coupling
constants and Sudakov form factors are assigned to the branching vertices,
according to the CKKW procedure~\cite{Catani:2001cc},
taking care to subtract from the NLO corrections all contributions
that are overcounted. Once Sudakov
form factors are included, if we integrate out an unresolved emission, the combination
of the Sudakov form factor, with the soft or collinear factorization of the amplitude
yields an approximate identity, often referred to as the ``unitarity'' of
the splitting process. Thus the amplitude approaches an amplitude with one
less emission. This property works recursively, so that the generator
for the production of a heavy system plus two jets, when integrating over the
jets, yields a good approximation to the distributions for the inclusive
production of the heavy system alone.

In fig.~\ref{fig:mpi-nompi-ht}
\begin{figure}[htb]
\begin{center}
\epsfig{file=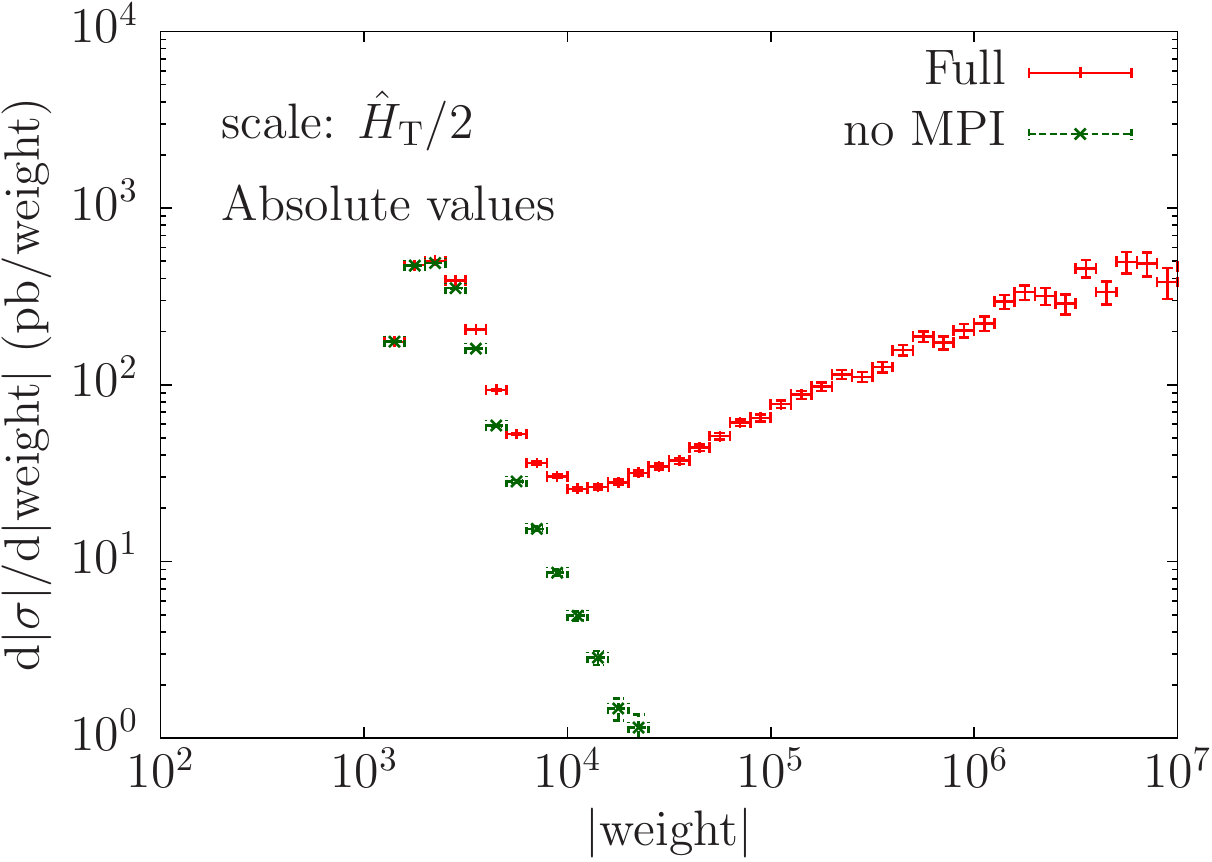,width=0.48\textwidth}
\epsfig{file=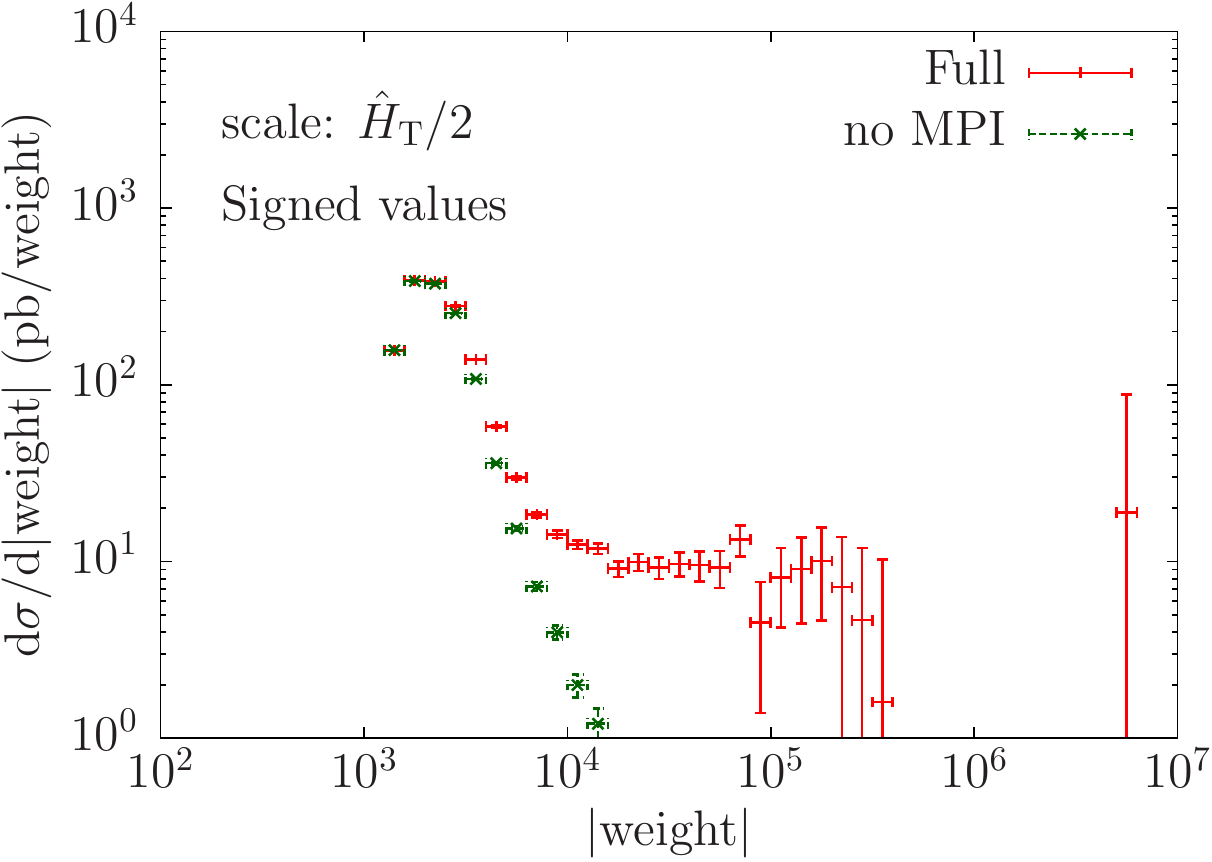,width=0.48\textwidth}
\end{center}
\caption{Fraction of events as a function of the absolute value of the
  event weight for $W^+$ production, with the standard cuts of
  table~\protect\ref{tab:wcuts}, in association with two jets with
  transverse momentum larger than 20~GeV, at the 7 GeV LHC.  The
  \POWHEG{} generation uses the \HTot{} scale choice, and the events are
  fully showered with \PYTHIASIX{}. The effect of switching on and off
  the multi-parton interaction is shown. In the left plot, the
  absolute value of the cross section is histogrammed, while on the
  right plot the signed value is used.
\label{fig:mpi-nompi-ht} }
\end{figure}
we plot the cross section for events that have at least two jets at
the hadron level, as a function of the event weight, comparing the results
obtained by switching on and off the multi-parton interaction
mechanism in \PYTHIA{}. The results are obtained with the
prescription of eq.~\eqref{eq:newdij}, which from now on will be our
default, and with the \HTot{} scale choice. From the previous discussion, it is clear that we
don't expect to give a reasonable description of events where
relatively soft QCD radiation is accompanied by relatively hard
multiple scattering events. It is however interesting to see what goes
wrong.  From the absolute value plot, it is clear that the inclusion
of MPI leads to the presence of a large number of events with large
weights. When histogramming standard observables, these large weights
lead to very poor results. From the plot of the signed value of the
cross section we see that these large weights largely
cancel. Nevertheless, the statistics required in order to produce
reasonable plots of standard physics observables becomes prohibitively
large.

The large, rising tail in the left plot of
Figure~\ref{fig:mpi-nompi-ht} can only arise from events that are soft
at the parton level, and that acquire hard jets because of secondary
collisions. By comparing this tail with the one on the right plot, we
deduce that there are a large number of negative weighted events that
cancel a large fraction of positive weighted ones, all being at low
$\pt$. This is as expected, since the calculation performed with the
\HTot{} scale becomes unreliable in the low $\pt$ region, where
negative virtual corrections can overwhelm the Born cross section and
lead to negative results. In this case, even the folding feature may
not be sufficient to get positive events.

Fig.~\ref{fig:mpi-nompi-minlo}
\begin{figure}[htb]
\begin{center}
\epsfig{file=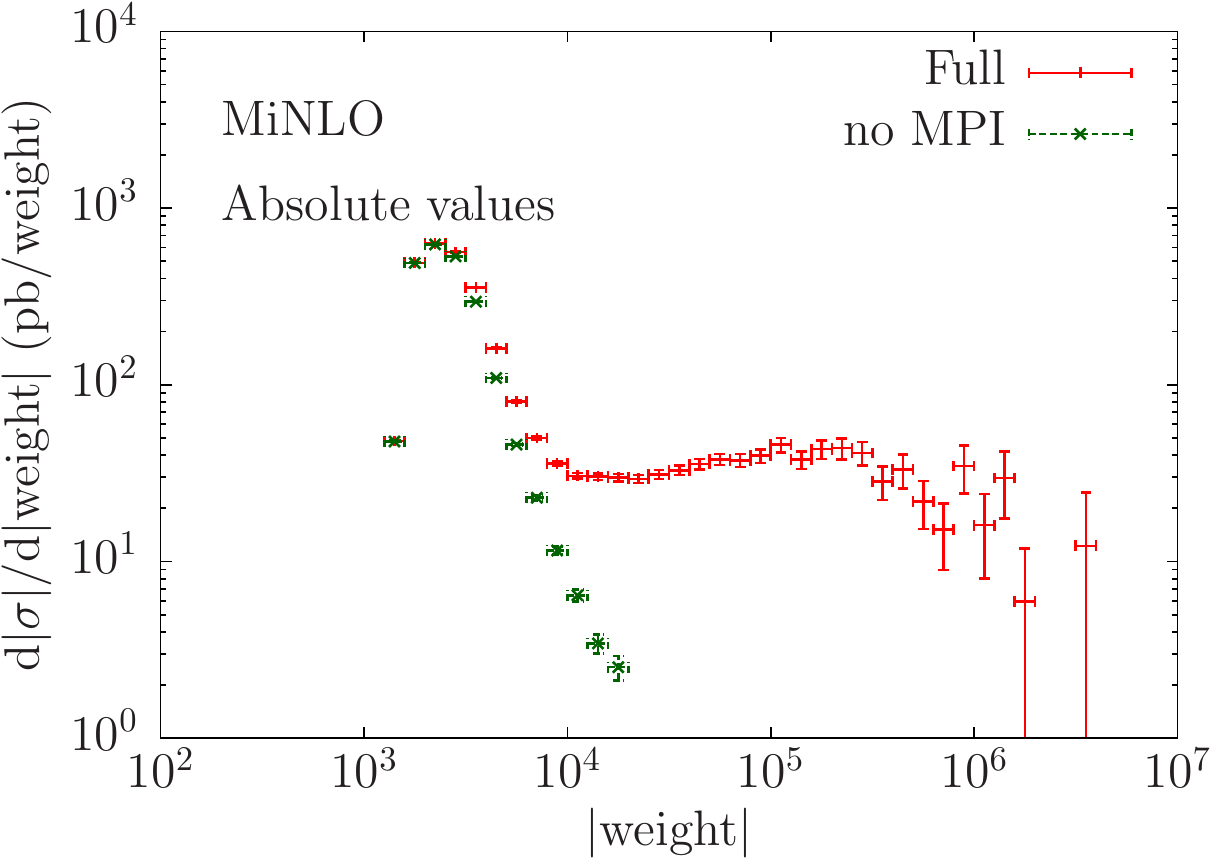,width=0.48\textwidth}
\epsfig{file=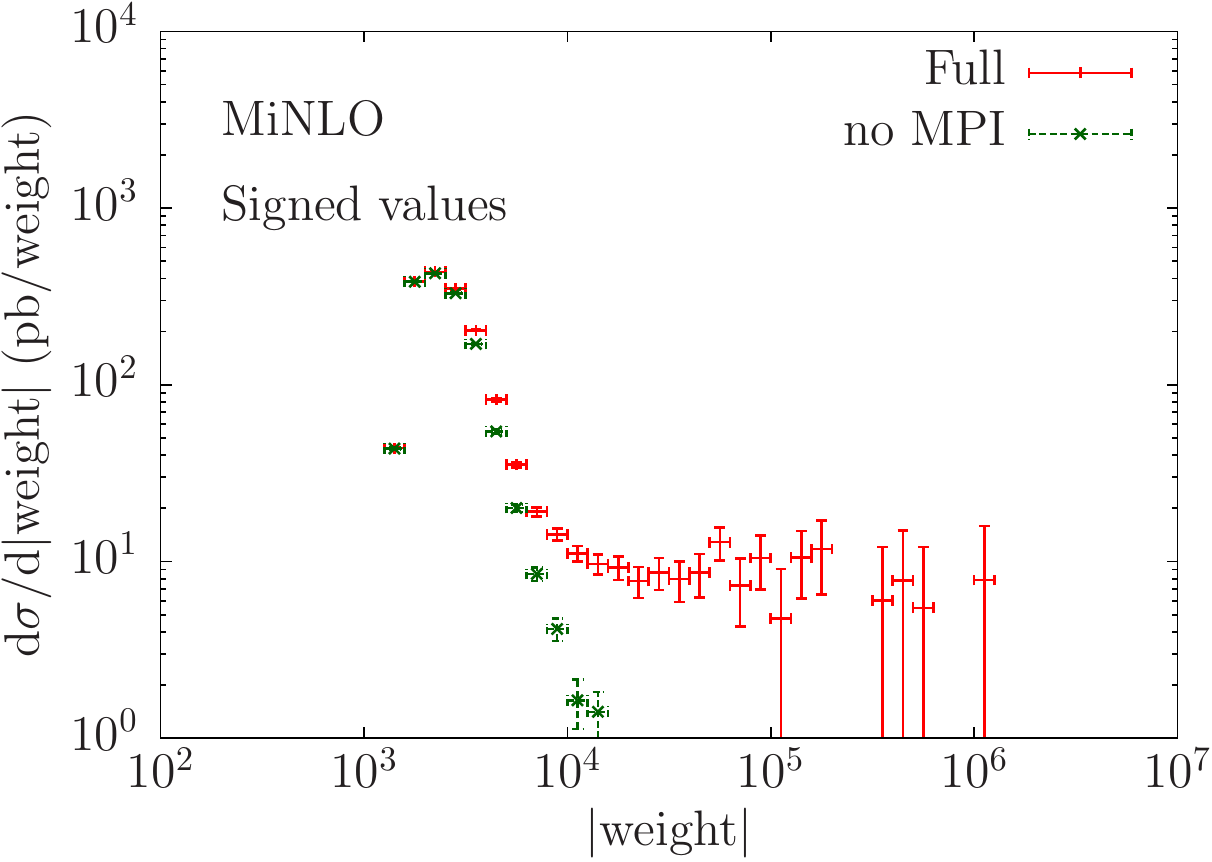,width=0.48\textwidth}
\end{center}
\caption{As in fig.~\protect\ref{fig:mpi-nompi-ht}, using the \MINLO{}
prescription.\label{fig:mpi-nompi-minlo} }
\end{figure}
is fully analogous to fig.~\ref{fig:mpi-nompi-ht}, except that now the
\MINLO{} prescription is used. We see that the rising contribution of
high-weight events disappear, and that the total contribution is
somewhat smaller. We expect this, since the \MINLO{} result has a
finite total integral, while the one with a standard scale choice
diverges at small transverse momenta. However, it is also clear that
also in this case we will get spikes in the final distributions.  The
reason for this is quite clear. We have suppressed with a Born
suppression factor the configurations of $W$ production accompanied by
low radiation activity. The MPI can promote these configurations to
hard ones, so that they eventually pass our jet cuts. However, they
are produced rarely and with a large weight. If we want to describe
reasonably well this process, the only way is not to suppress the low
$\pt$ region. In fact, it is not necessary to use a Born suppression
factor at all when using \MINLO{}, since the \MINLO{} Sudakov form
factors damp the cross section in the singular regions.  If one would
like to increase the number of events produced at relatively large
transverse momenta, one can use a suppression factor that does not
vanish in the regions with singular Born kinematics.

We finally stress that the true advantage of the \MINLO{} approach is
that the value of the inclusive cross section is sensible, and it is
quite close to the results that can be obtained with the
\W{}~\cite{Alioli:2008gx} and \WJ{}~\cite{Alioli:2010qp}
generators~\cite{Hamilton:2012np}.  This being the case, we can trust
that also the contribution coming from events with low radiation,
accompanied by hard secondary parton interactions, are represented in
a fair way. In section~\ref{sec:DATA} we will only use the \MINLO{}
improved generators to compare with real data.  Later in
section~\ref{sec:MPI} we will study some observables that are
sensitive to multi-parton interactions.

\section{Comparison of the \MINLO{} and \HTot{} results}\label{sec:htminlo}
Since the \MINLO{} method is relatively new, in this section 
we compare the results obtained using it with the results using the 
more common \HTot{} scale choice.  
For this purpose we only consider $W\to \mu \nu$ at 7
TeV, with the cuts of table~\ref{tab:wcuts}.  We don't expect
significant differences for the $Z$ production case.

We shower the events generated with \POWHEG{} using the \PYTHIASIX{}
Monte Carlo.  We have used the AMBT1 \PYTHIA{} tune (i.e. we have
performed the call {\tt CALL PYTUNE(340)}). We have switched off the
hadronization and the underlying event, by setting {\tt MSTP(111)=0},
{\tt MSTP(91)=0} for switching off hadronization, and {\tt
  MSTP(81)=20} for switching off the \MPI{}. We do not see significant
differences when hadronization is turned on, provided that \MPI{} are
switched off. On the other hand, as explained earlier, we were unable
to produce usable plots when using the scale \HTot{}, if \MPI{} are
turned on.

We have considered a very large set of distributions, comprising all
those that have been considered by the ATLAS
collaboration~\cite{Aad:2011qv}. Observables that do not require the
presence of at least two jets cannot be computed using the \WJJ{}
generator with the \HTot{} scale choice. For all observables that do
require at least two jets, we see very good agreement between the two
approaches. Here we show only, in figs~\ref{fig:mulHt}
and~\ref{fig:jetsHt},
\begin{figure}[htb]
\begin{center}
\epsfig{file=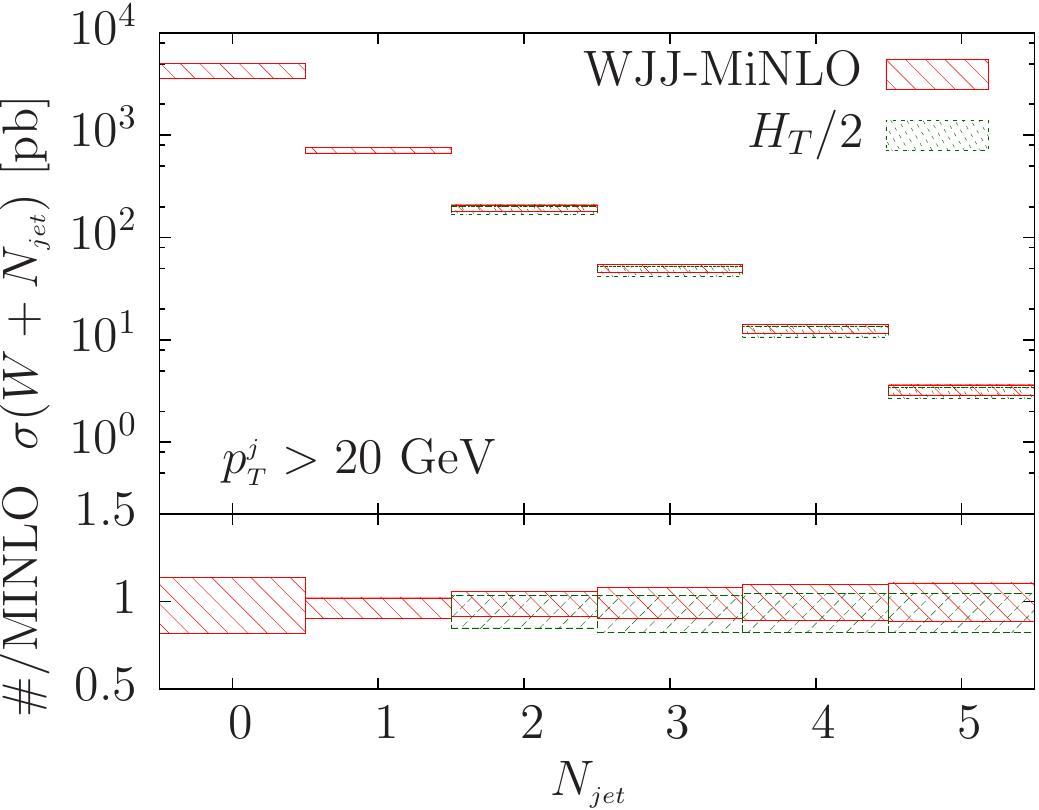,width=0.48\textwidth}
\epsfig{file=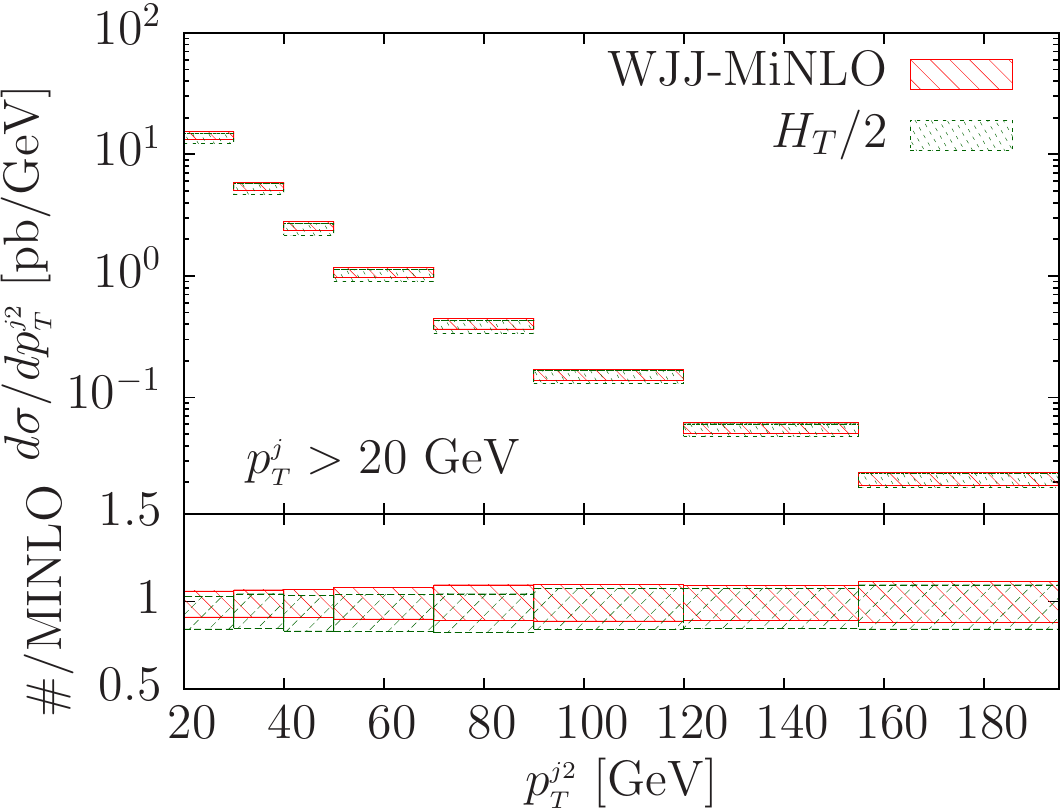,width=0.48\textwidth}
\end{center}
\caption{
Inclusive jet multiplicity (left) and transverse momentum distribution of the second
jet (right) in events with two or more jets.
\label{fig:mulHt} }
\end{figure}
\begin{figure}[htb]
\begin{center}
\epsfig{file=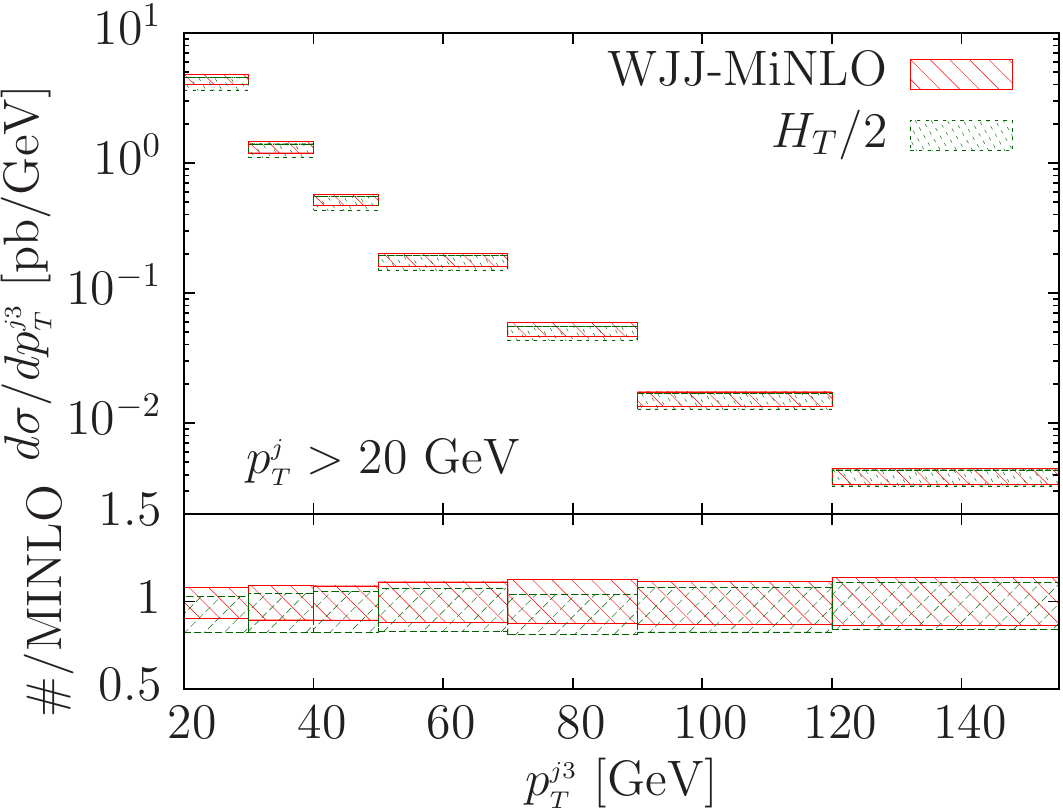,width=0.48\textwidth}
\epsfig{file=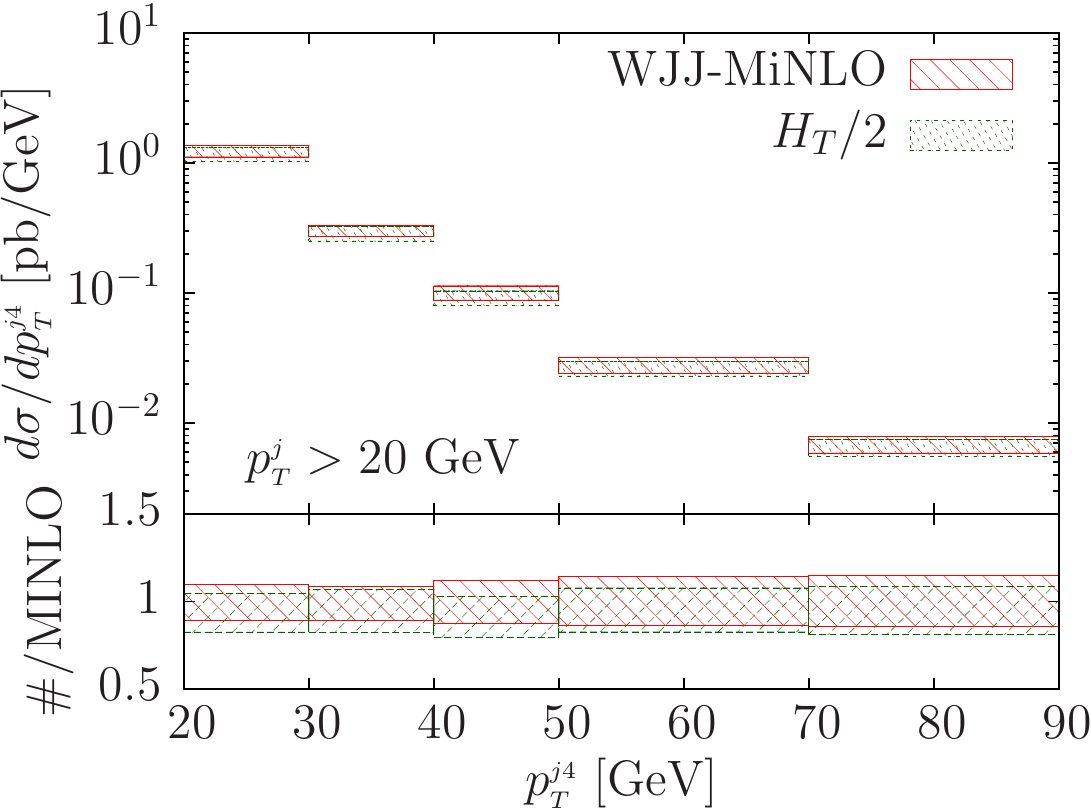,width=0.48\textwidth}
\end{center}
\caption{ Transverse momentum distribution of the third jet (left) in
  events with three or more jets and fourth jet in events with
  four or more jets (right).
\label{fig:jetsHt} }
\end{figure}
the inclusive jet multiplicities, and the transverse momentum of the
second, third and fourth jet in events with at least two, three or
four jets respectively.

The bands in the figure are the envelope of the 7-point scale
variation, by which the central renormalization and factorization
scales are multiplied by scale factors $K_r$ and $K_f$, in the
following combinations:
\begin{equation}
(K_r,K_f)=\;(1/2,1/2),\; (1/2,1),\; (1,1/2) ,\;(1,1),\; (2,1),\; (1,2), \;(2,2),
\end{equation}
(for details on how this is implemented in \MINLO{}, see
refs.~\cite{Hamilton:2012np,Hamilton:2012rf}).  Statistical errors
(not included) are negligible with respect to the scale variation
range.  Notice that in fig.~\ref{fig:mulHt}, left plot, no results are
given for the 0 and 1 inclusive jet multiplicity using the \HTot{}
scale choice, since, as stressed before, these cannot be computed with
the \WJJ{} generator.

We notice that not only the central values are in good agreement, but
also the scale variation bands are of the same order.

\section{Comparison with available data}\label{sec:DATA}

Results for $W$+jets production at the LHC, operating at $7$~TeV, have
been presented in ref.~\cite{Chatrchyan:2011ne} by the CMS
collaboration and in ref.~\cite{Aad:2012en} by the ATLAS collaboration. 
Similarly, CMS and
ATLAS have published results for $Z$+jets final states, at the same
operating energy, in ref.~\cite{Chatrchyan:2011ne} and
ref.~\cite{Aad:2011qv} respectively.
In this section we compare the results of the \WJJ{} and \ZJJ{} generators
with LHC data. We will limit ourselves to ATLAS data for reasons of space.

We will include in our predictions only the theoretical errors due to scale
uncertainties.  We believe that this is sufficient for the present purpose of
validating our generators.  We remind the reader, however, that other
sources of errors are present, especially when dealing with quantities
that are sensitive to the fourth jet or beyond. These jets are
generated by \PYTHIA{}, and are only accurate in the collinear
approximation. Changing or tuning the shower model may well have a
considerable impact on these quantities.
\subsection{$W$ production data}\label{sec:Wdata}
For the sake of illustration, in this section we present results for
$W$+jets production at $7$~TeV compared to the ATLAS data of
ref.~\cite{Aad:2011qv}. 
We adopt the set of cuts displayed in
table~\ref{tab:wcuts}, that are taken from ref.~\cite{Aad:2011qv}.
\begin{table}[htb]
\begin{center}
\begin{tabular}{|l|}
\hline Cuts for $W$ production \\
\hline
Jets defined using the anti-$\kt$ algorithm ($R=0.4$), with $\ptmin>20\,{\rm GeV}$, $|\eta|<4.4$;\\
One lepton required with $\ptl>20\,{\rm GeV}$, $|\etal|<2.5$; \\
Lepton isolation required: $\Delta R_{l j}$ for all jets (as defined above) $> 0.5$;\\
One neutrino (missing $\et$) required with $\ptnu>25\,{\rm GeV}$; \\
Transverse mass constraint required: $\sqrt{2 \ptl \ptnu (1-\cos \phi_{l,\nu})}>40\,{\rm GeV}$; \\
Events are classified according to the number of jets, as defined above.\\
\hline
\end{tabular}
\caption{Cuts for $W$ production in association
with jets.\label{tab:wcuts}}
\label{input}
\end{center}
\end{table}
The parameters used in our simulation are displayed in table~\ref{tab:input}.
\begin{table}
\begin{center}
\begin{tabular}{|l|l|}
\hline
$m_W= 80.419$~GeV & $\Gamma_W=2.06$~GeV \\
$m_Z=91.188$~GeV & $\Gamma_Z=2.49$~GeV \\
$G_F=1.116639\times 10^{-5}$ & $\sin^2 \theta_W = 1-m_W^2/m_Z^2$\\
\hline
\end{tabular}
\caption{Input parameters used for the phenomenological results.}
\label{tab:input}
\end{center}
\end{table}
The parton distributions used are the CTEQ6M set taken from
ref.~\cite{Pumplin:2002vw}.  
We use FastJet~\cite{Cacciari:2011ma} to cluster partons into jets via
the anti-kt algorithm~\cite{Cacciari:2008gp}.
As motivated earlier, we use the \MINLO{}
procedure for the scale choice, that gives reasonable predictions also
for quantities requiring only one or no jets.  The uncertainty bands
comprise only the scale uncertainties, estimated with the prescription
given in sec.~\ref{sec:htminlo}.

We have generated samples of $W^+\to e^+ \nu$ and $W^-\to e^-
\bar{\nu}$ production, and added them together.  We have switched off
electromagnetic radiation in \PYTHIASIX{} (i.e. we have set {\tt
  MSTJ(41)=11}).  In this way, the electron and muon decay modes do
not differ in any appreciable way. Our results can be considered an
average over muon and electron decay modes provided a dressed, rather
than a bare lepton, definition is used. We thus compare them with the
ATLAS results averaged over electrons and muons. We have used the
AMBT1 \PYTHIA{} tune (i.e. we have performed the call {\tt CALL
  PYTUNE(340)}).  We show here results obtained with the jet cut set
at $20$~GeV, while in Appendix~\ref{app:figuresW30} we show those
obtained with a $30$~GeV cut.

In fig.~\ref{fig:MINATLnjet}
\begin{figure}[htb]
\begin{center}
\epsfig{file=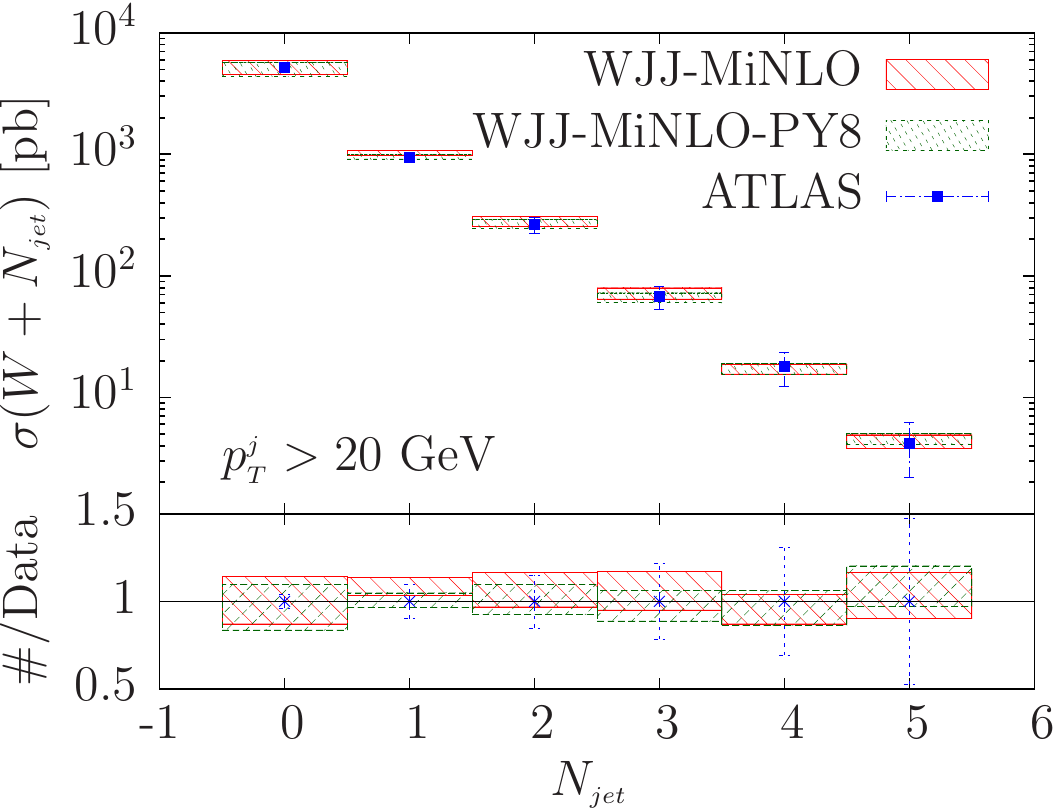,width=0.48\textwidth}
\end{center}
\caption{Inclusive jet multiplicity obtained with the \WJJ{} generator
using \MINLO{}, interfaced to \PYTHIASIX{} and \PYTHIAEIGHT{},
compared to ATLAS data.
\label{fig:MINATLnjet} }
\end{figure}
we show the inclusive jet multiplicity compared to ATLAS data.
We have also shown in this figure
the results obtained interfacing the \WJJ{} generator with
\PYTHIAEIGHT{}. The \PYTHIASIX{} and \PYTHIAEIGHT{} results use the same
Les Houches events sample. The difference in the results can thus be considered
an estimate of shower and matching uncertainties. In order not
to excessively crowd our figures we will not show more \PYTHIAEIGHT{} results
in the following.

We see that, within present uncertainties, the data and Monte Carlo
predictions agree quite well.  Notice that also for four and five jet
multiplicities, that are only described with collinear accuracy by the
generator, the agreement is quite good.  Notice also that the
quantities depending upon the one jet final state are predicted
adequately by the calculation.

The same consideration applies to all remaining observables measured
by the ATLAS collaboration, that are shown in the remaining part of
this section. We do however observe a tendency of our calculation to
overshoot the data at small transverse momentum. This trend is also
observed in ref.~{\cite{Aad:2012en}} when comparing data with NLO
calculations.
\begin{figure}[htb]
\begin{center}
\epsfig{file=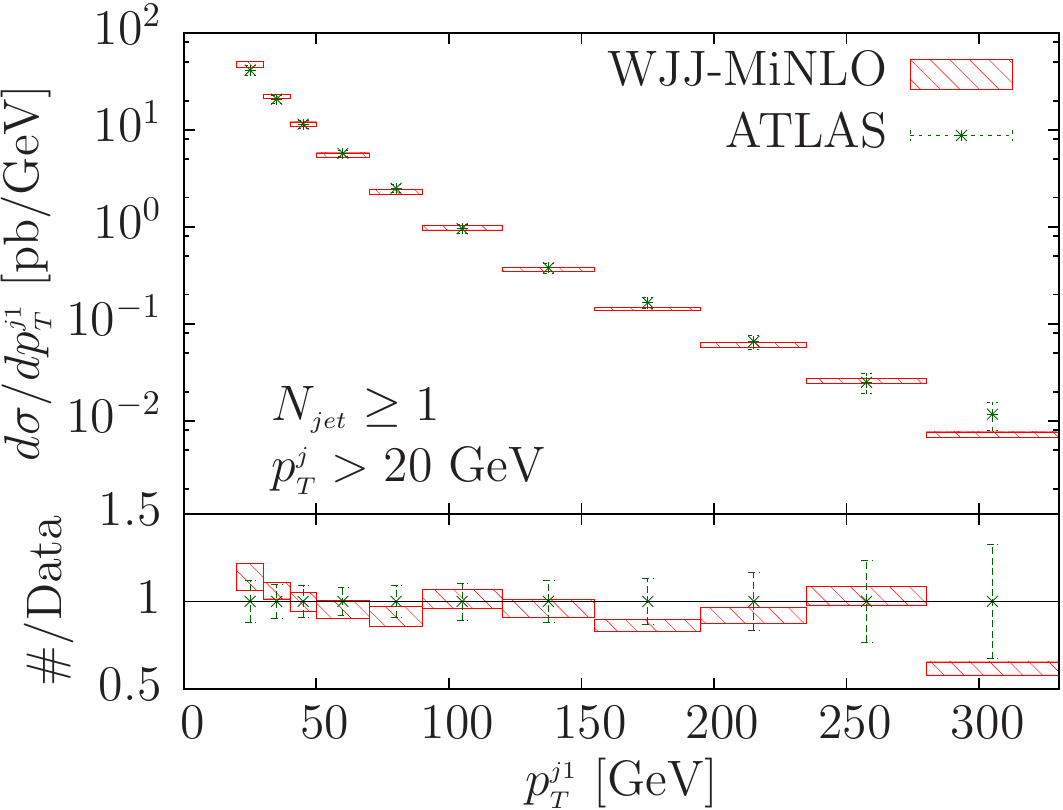,width=0.48\textwidth}
\epsfig{file=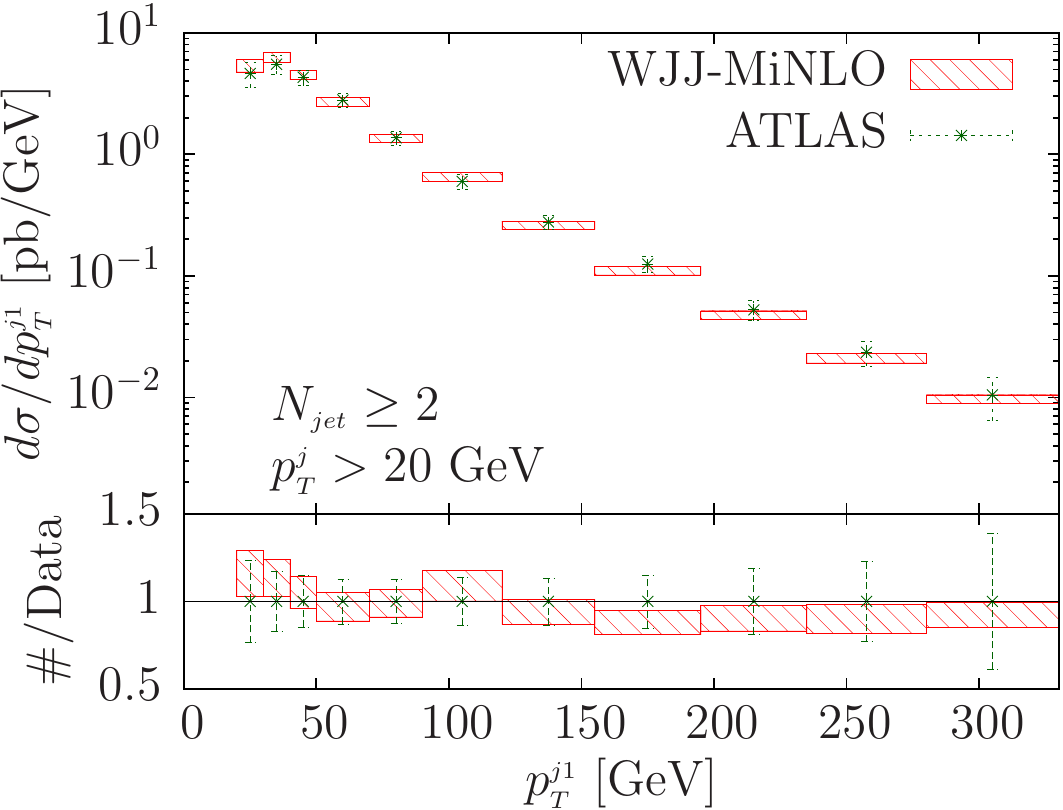,width=0.48\textwidth}
\end{center}
\caption{Transverse momentum of the leading jet in inclusive one-jet
  events (left) and two-jet events (right).
\label{fig:MINATLj1pt} }
\end{figure}

\begin{figure}[htb]
\begin{center}
\epsfig{file=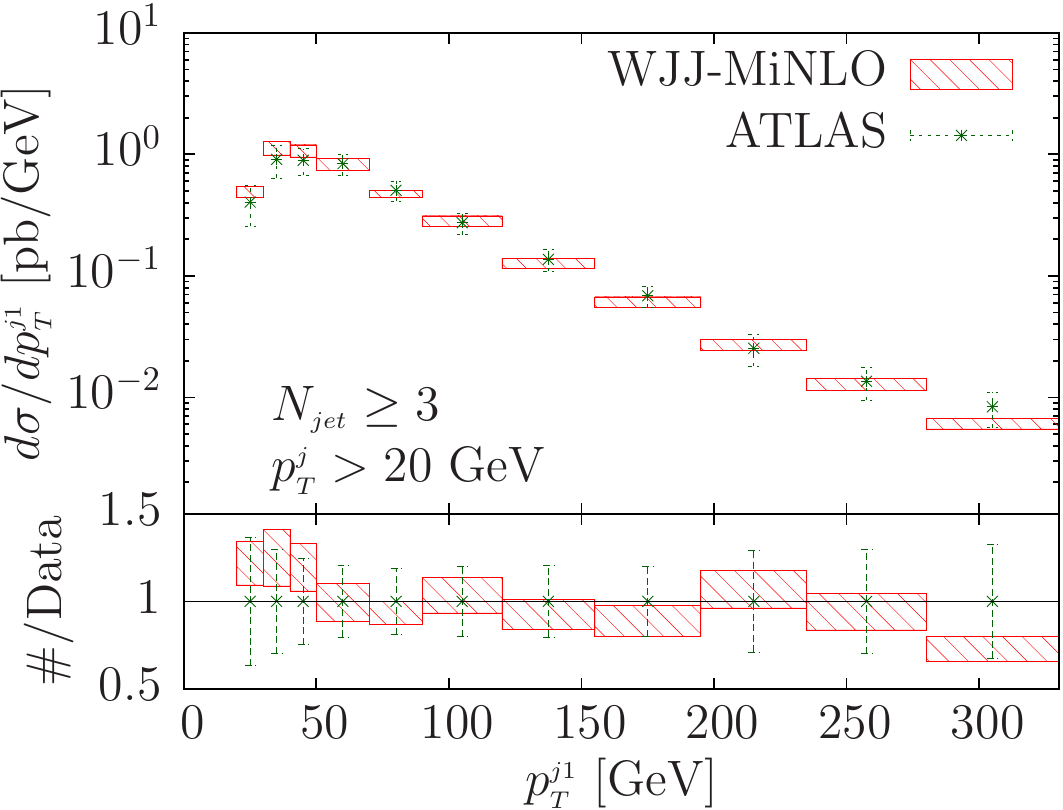,width=0.48\textwidth}
\epsfig{file=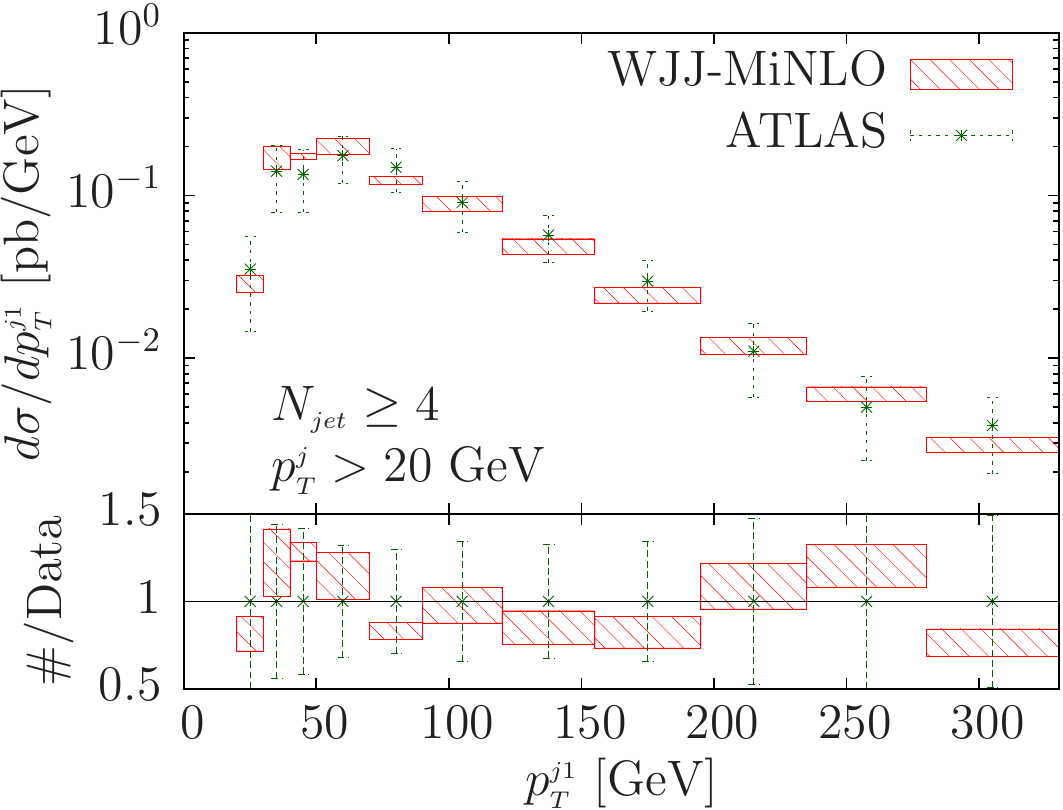,width=0.48\textwidth}
\end{center}
\caption{Transverse momentum of the leading jet in inclusive
  three-jet events (left) and four-jet events (right).
\label{fig:MINATLj1pt2} }
\end{figure}

\begin{figure}[htb]
\begin{center}
\epsfig{file=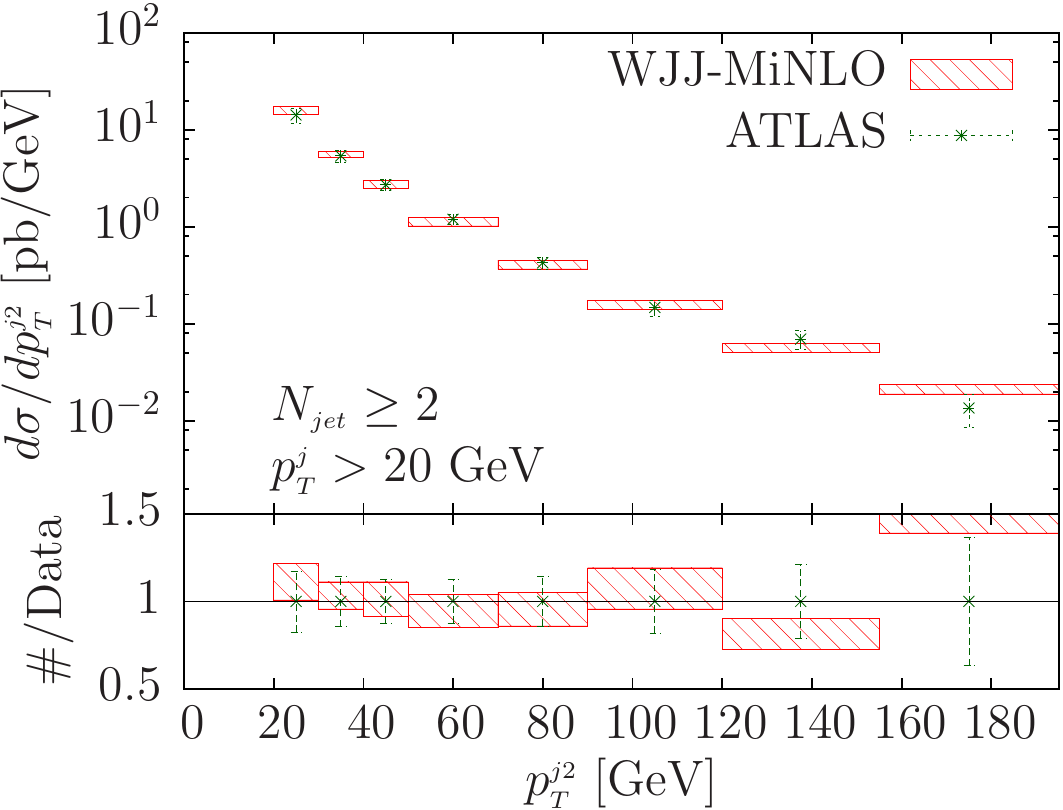,width=0.48\textwidth}
\end{center}
\caption{Transverse momentum of the second leading jet in inclusive two-jet events.  
\label{fig:MINATLj2pt} }
\end{figure}

\begin{figure}[htb]
\begin{center}
\epsfig{file=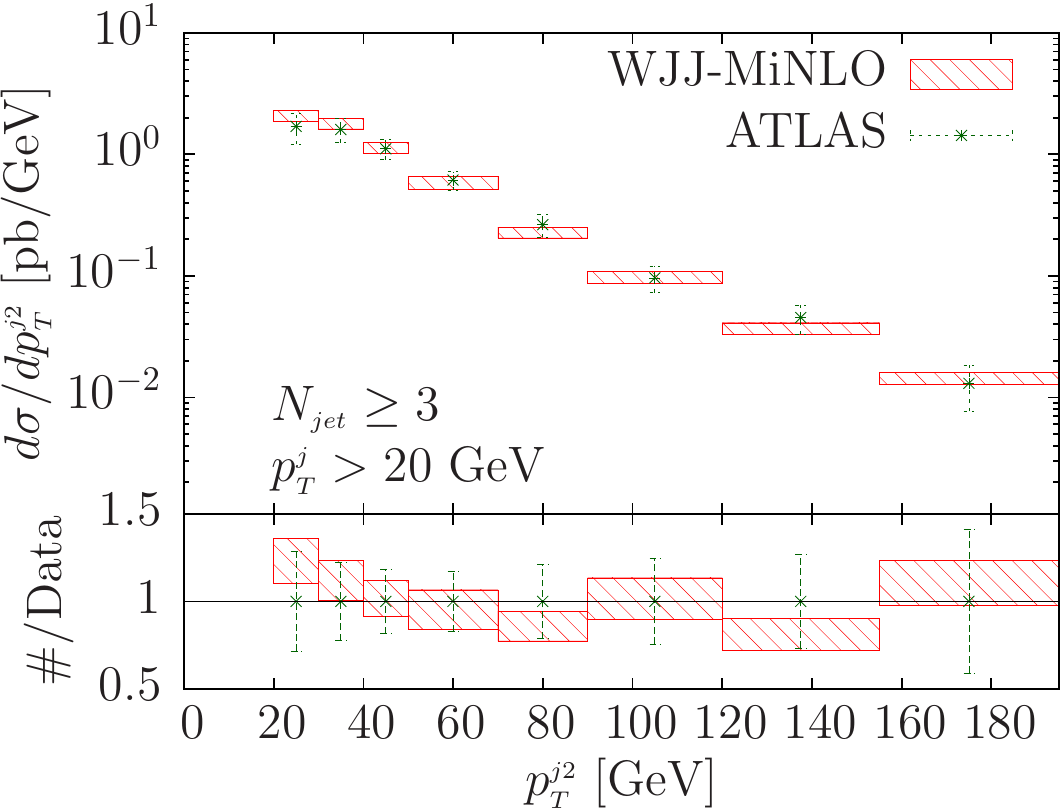,width=0.48\textwidth}
\epsfig{file=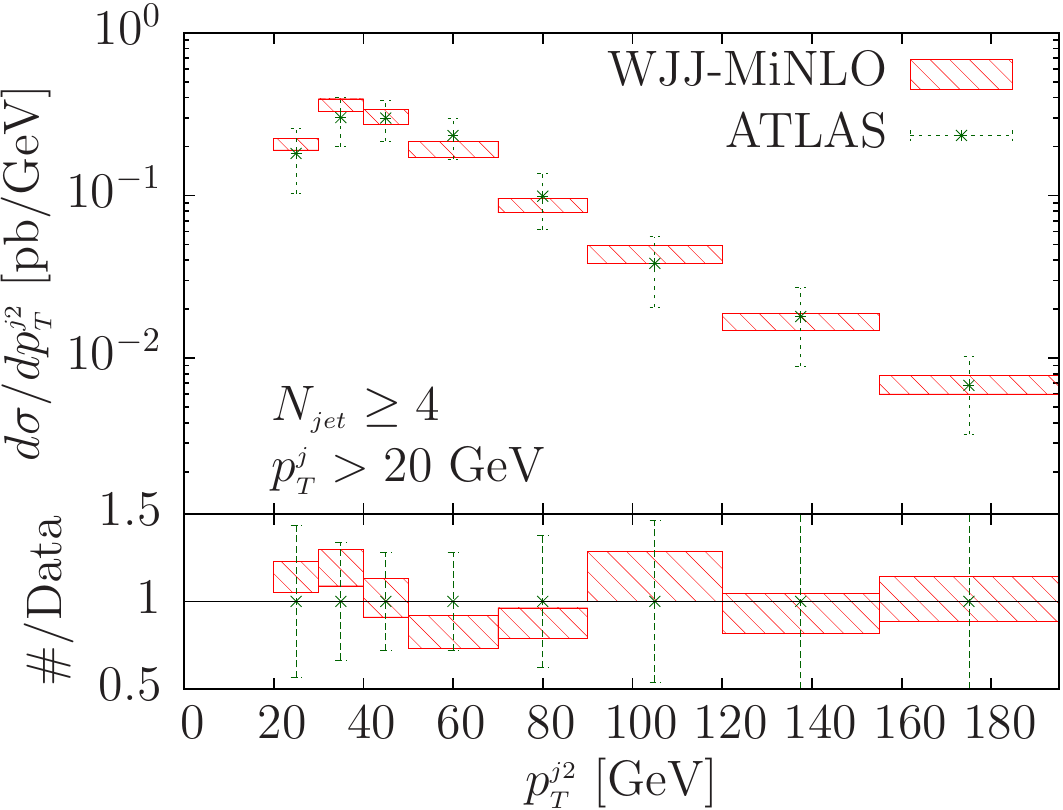,width=0.48\textwidth}
\end{center}
\caption{Transverse momentum of the second leading jet in inclusive
  three-jet events (left) and four-jet events (right).
\label{fig:MINATLj2ptbis} }
\end{figure}

\begin{figure}[htb]
\begin{center}
\epsfig{file=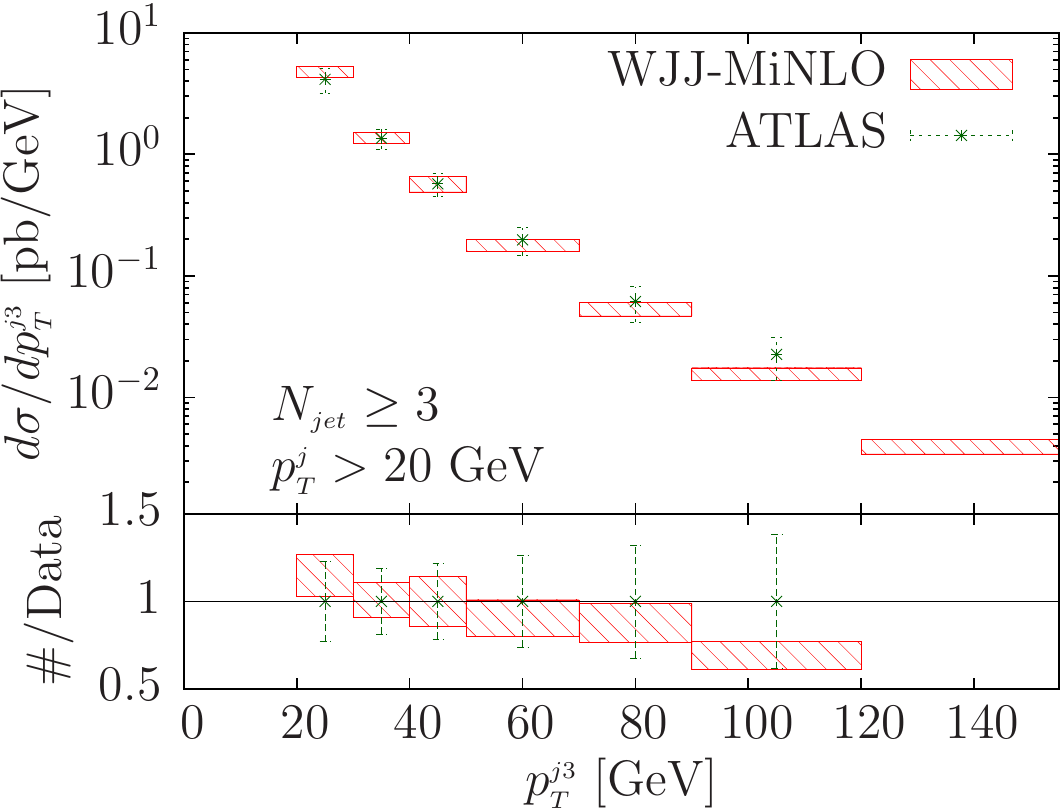,width=0.48\textwidth}
\epsfig{file=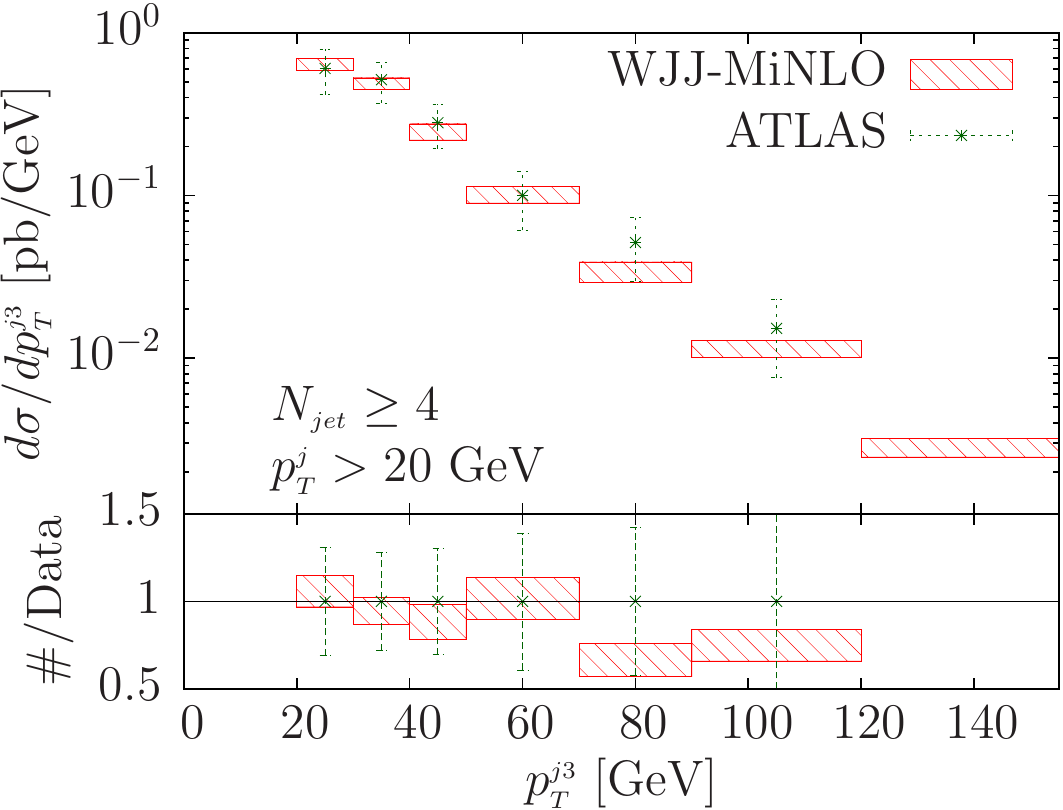,width=0.48\textwidth}
\end{center}
\caption{Transverse momentum of the third leading jet in inclusive
  three-jet events (left) and four-jet events (right).
\label{fig:MINATLj3pt} }
\end{figure}

\begin{figure}[htb]
\begin{center}
\epsfig{file=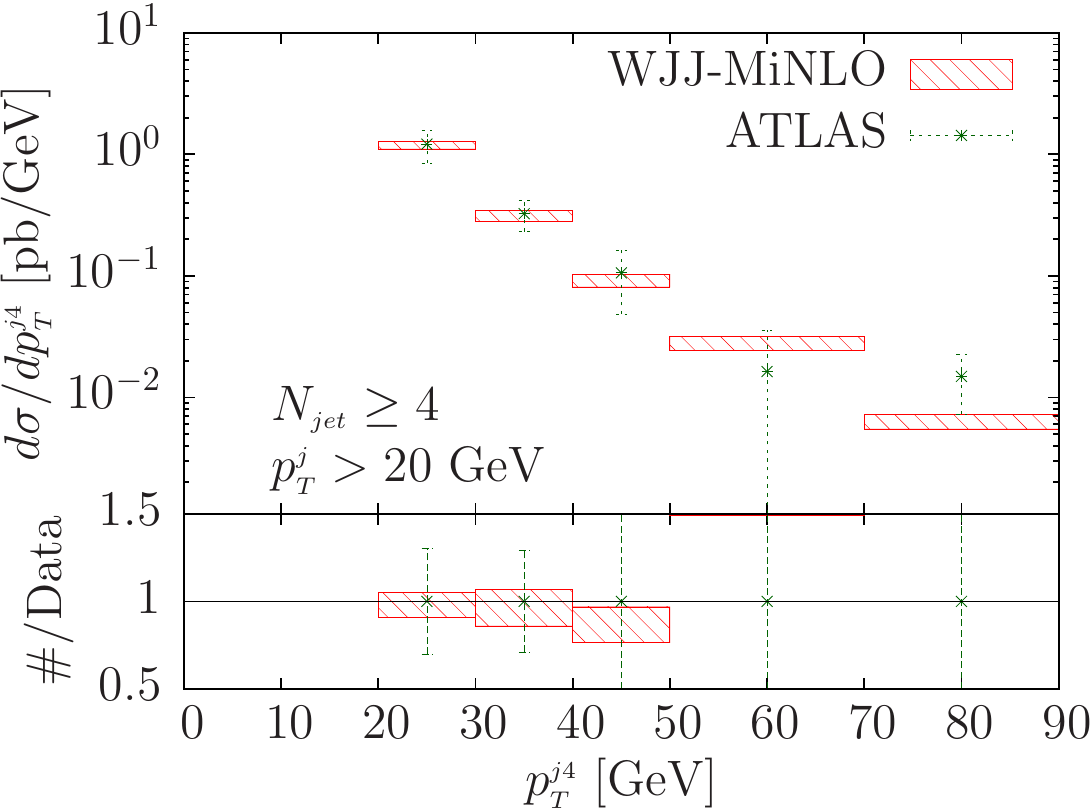,width=0.48\textwidth}
\end{center}
\caption{Transverse momentum of the fourth leading jet in inclusive
  four-jet events.
\label{fig:MINATLj4pt} }
\end{figure}

\begin{figure}[htb]
\begin{center}
\epsfig{file=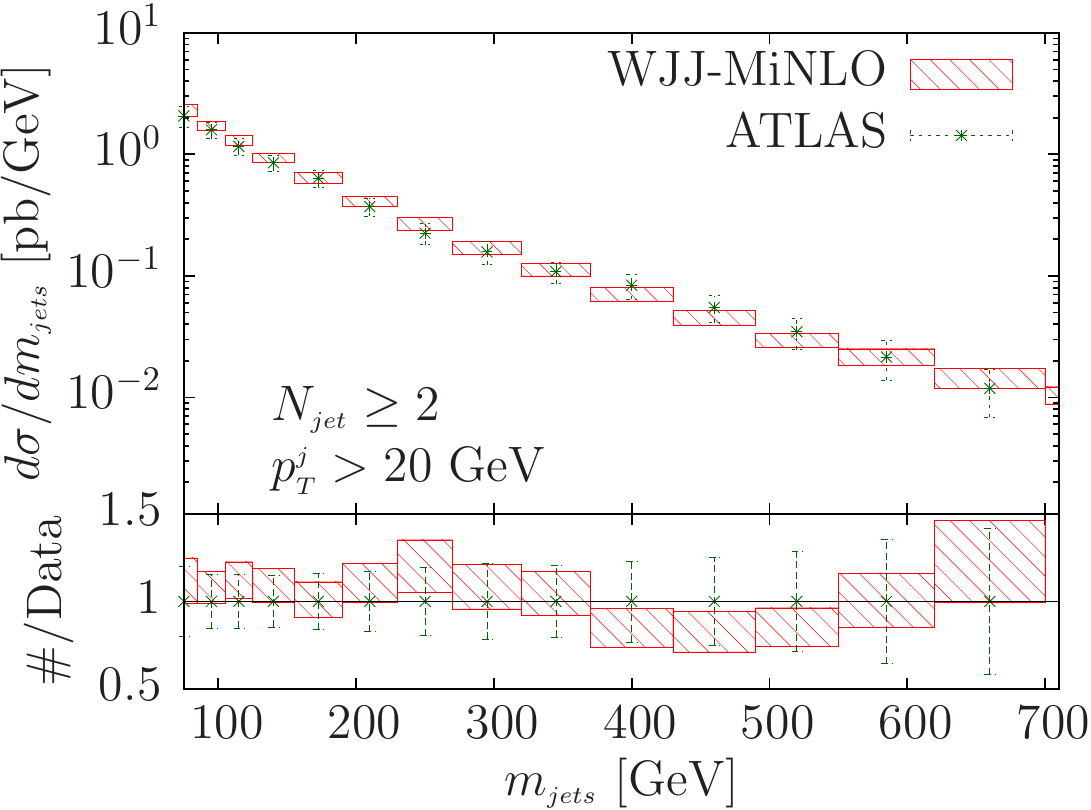,width=0.48\textwidth}
\end{center}
\caption{Invariant mass of the system composed of the first two
  leading jets in inclusive two-jet events.
\label{fig:MINATLmjets} }
\end{figure}

\begin{figure}[htb]
\begin{center}
\epsfig{file=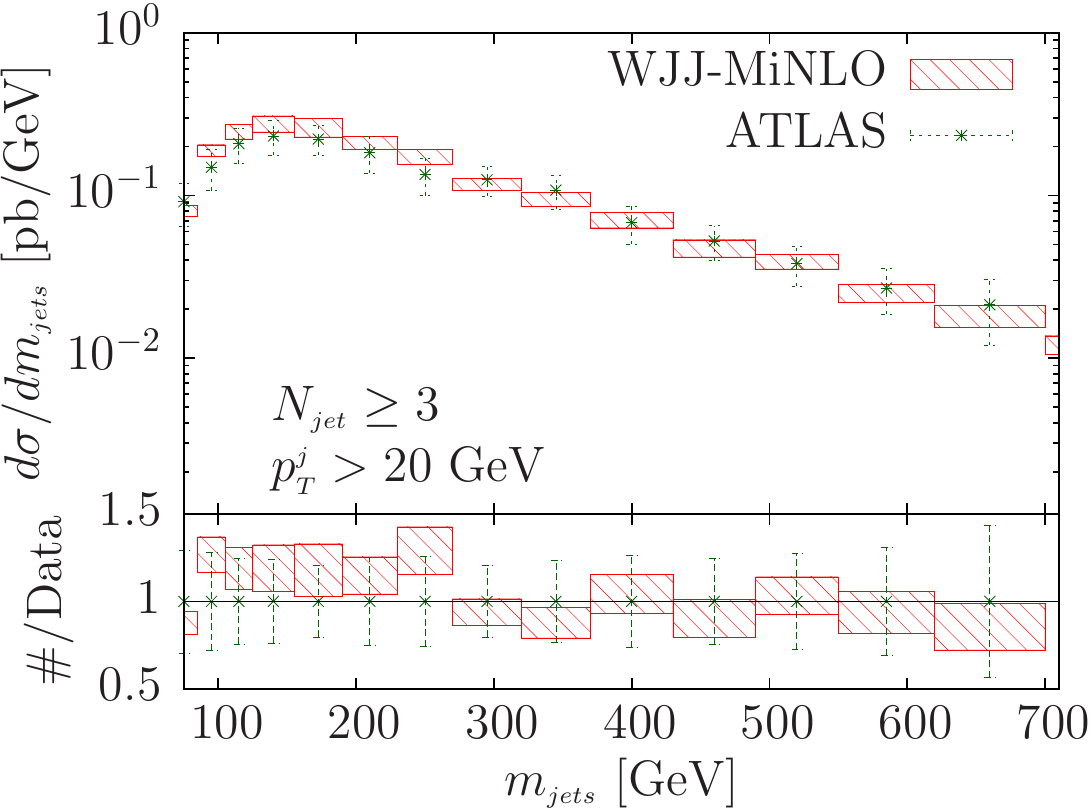,width=0.48\textwidth}
\epsfig{file=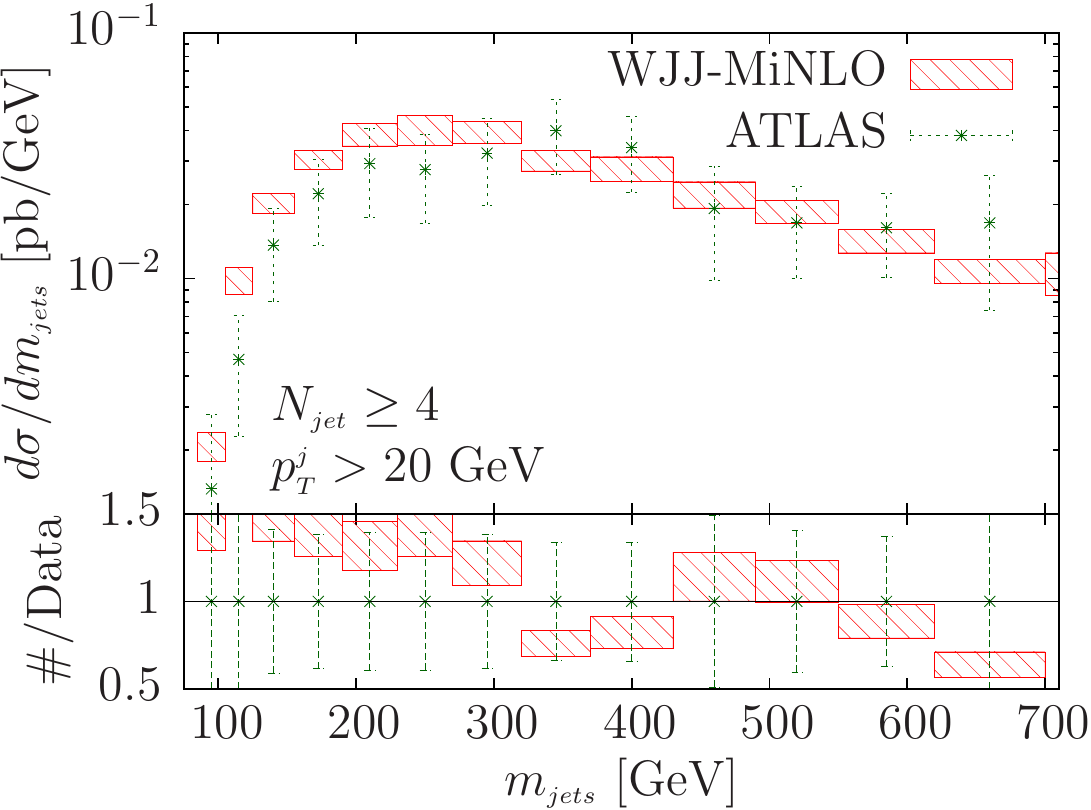,width=0.48\textwidth}
\end{center}
\caption{Invariant mass of the system composed of the first three
  leading jets in inclusive three-jet events (left) and by the first
  four leading jets in inclusive four-jet events (right).
\label{fig:MINATLmjets2} }
\end{figure}

\begin{figure}[htb]
\begin{center}
\epsfig{file=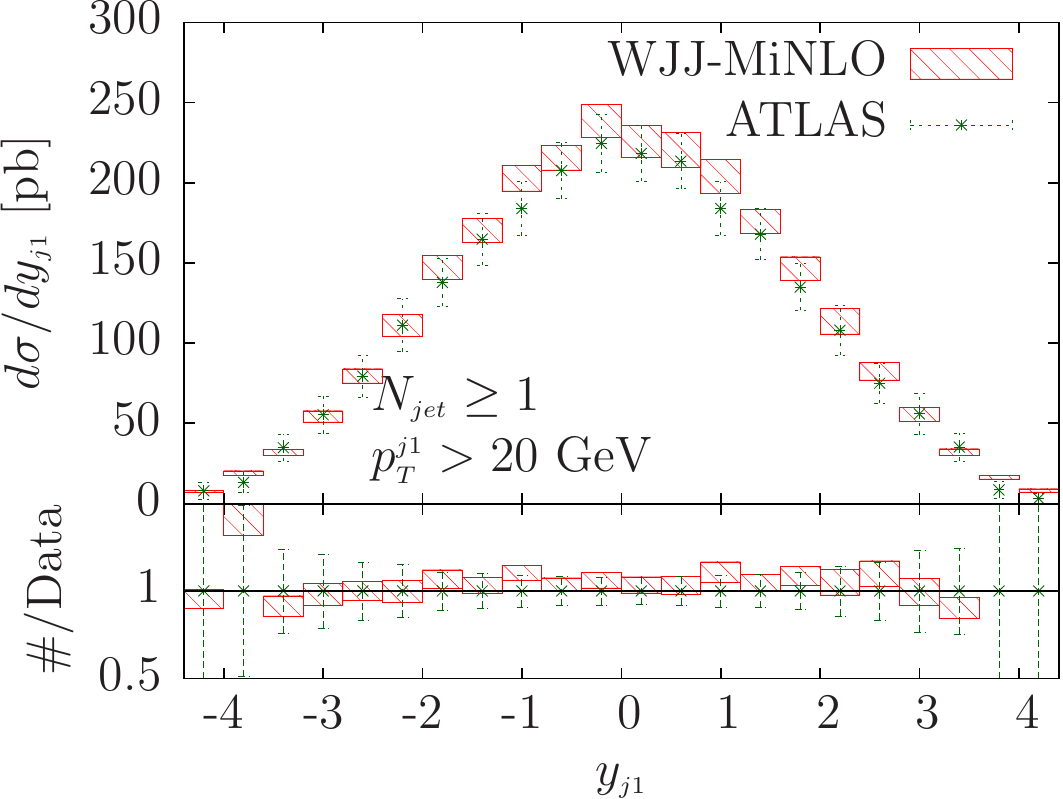,width=0.48\textwidth}
\end{center}
\caption{Rapidity of the leading jet in inclusive one-jet events.
\label{fig:MINATLmjets3} }
\end{figure}

\begin{figure}[htb]
\begin{center}
\epsfig{file=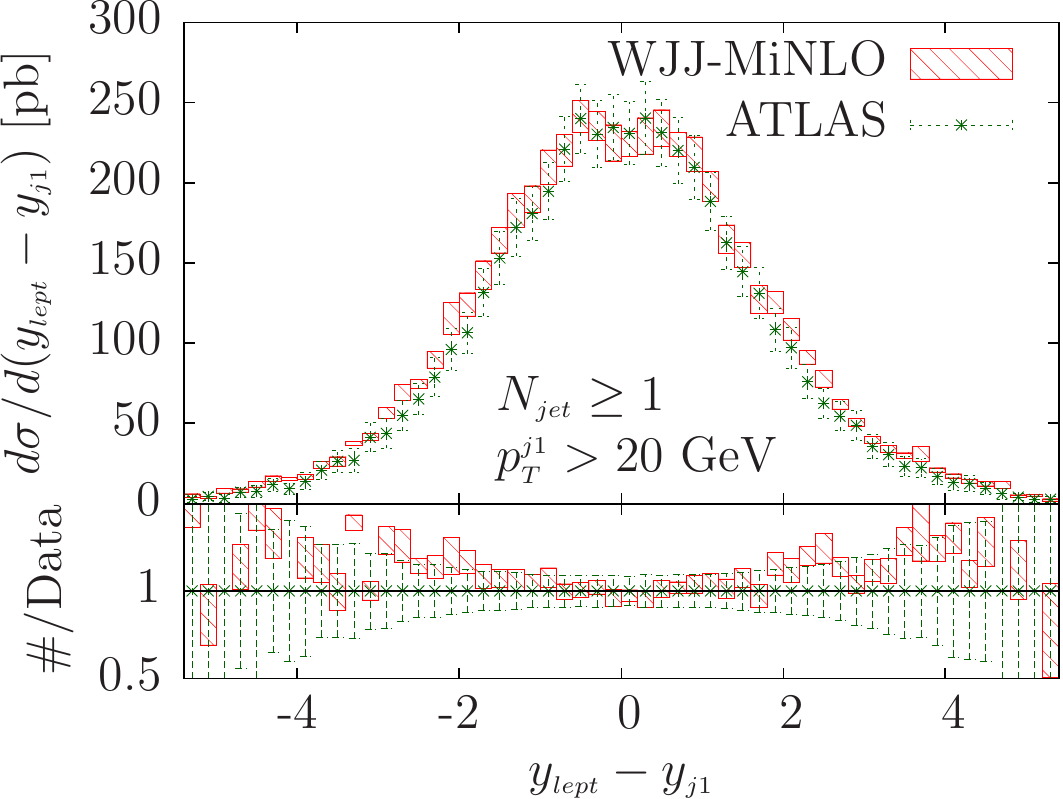,width=0.48\textwidth}
\epsfig{file=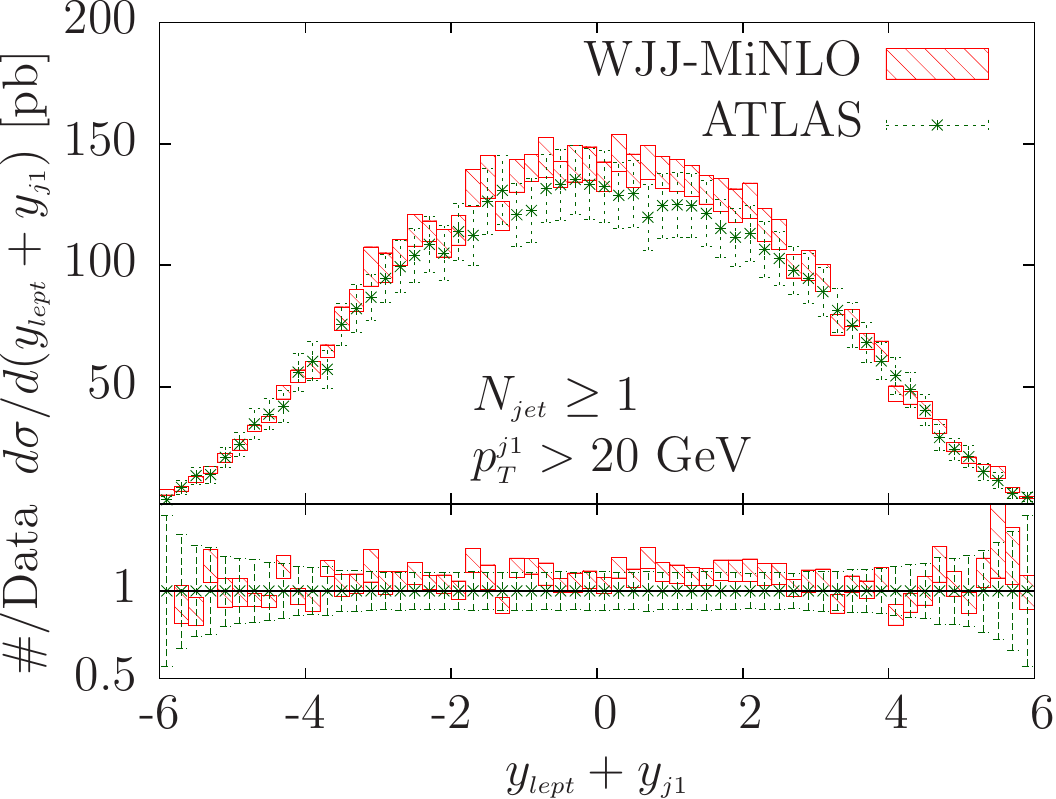,width=0.48\textwidth}
\end{center}
\caption{Difference (left) and sum (right) of the rapidity of the
  lepton and the leading jet in inclusive one-jet events.
\label{fig:MINATLlj} }
\end{figure}

\begin{figure}[htb]
\begin{center}
\epsfig{file=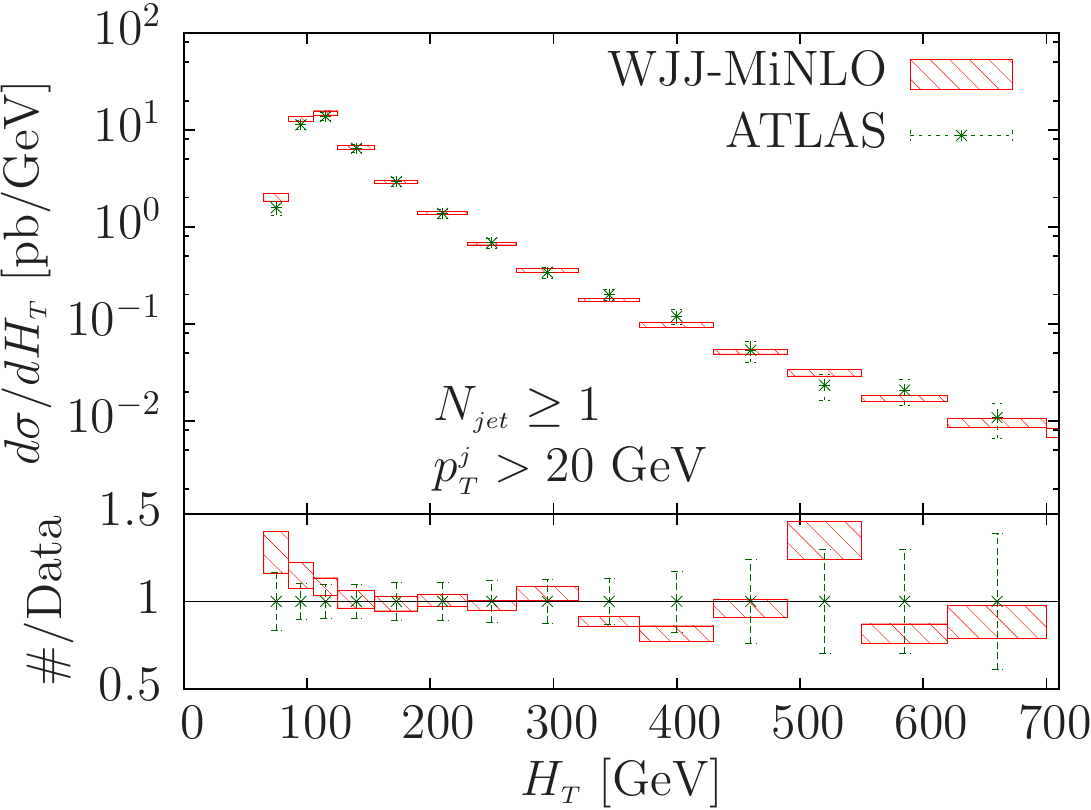,width=0.48\textwidth}
\epsfig{file=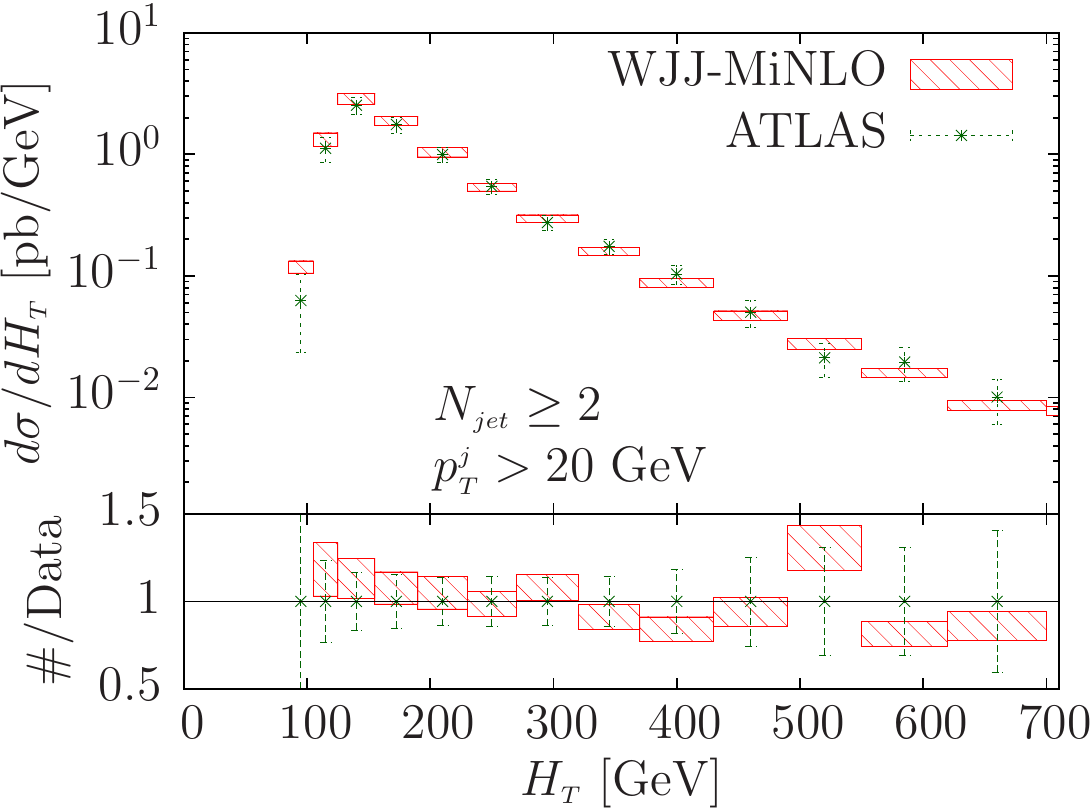,width=0.48\textwidth}
\end{center}
\caption{Scalar sum of transverse momenta of the event in inclusive
  one-jet events (left) and two-jet events (right).
\label{fig:MINATLht} }
\end{figure}

\begin{figure}[htb]
\begin{center}
\epsfig{file=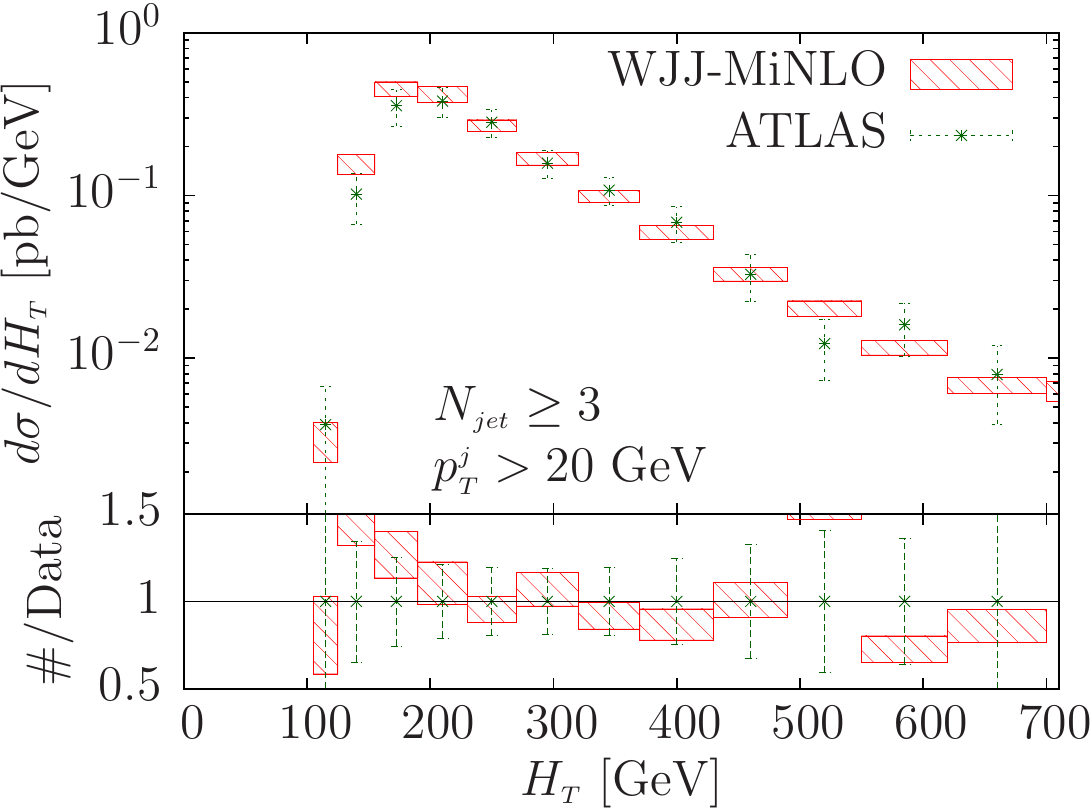,width=0.48\textwidth}
\epsfig{file=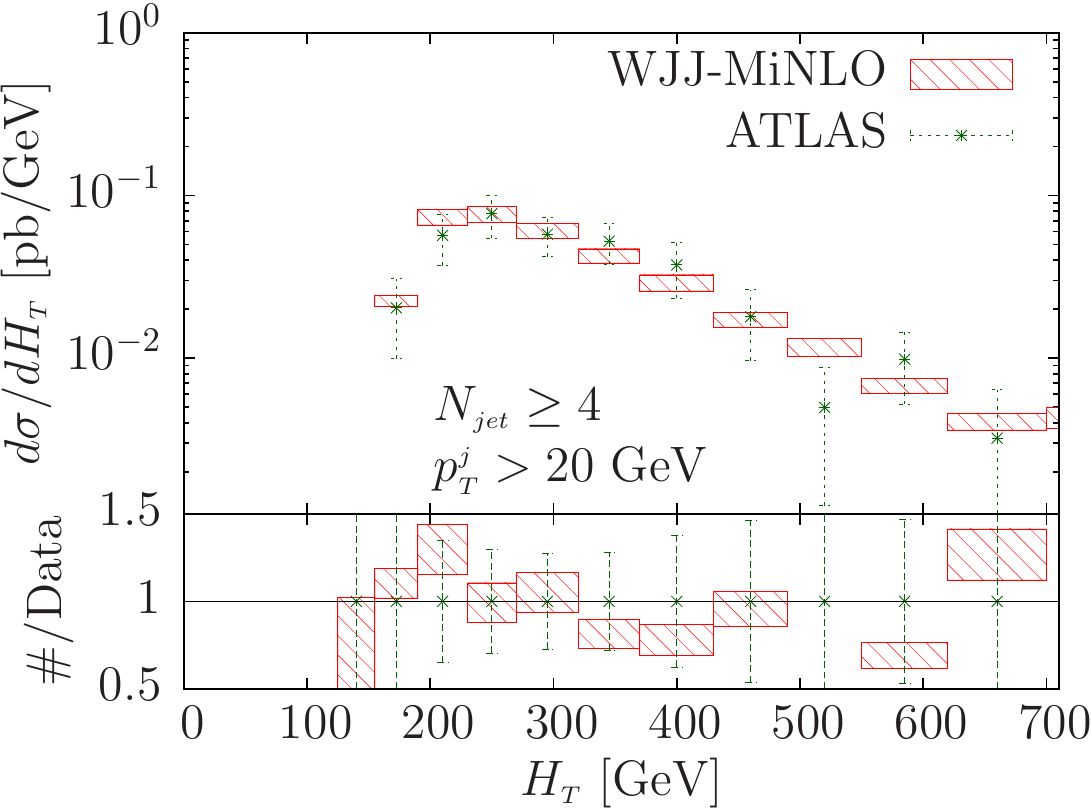,width=0.48\textwidth}
\end{center}
\caption{Scalar sum of transverse momenta of the event in inclusive
  three-jet events (left) and four-jet events (right).
\label{fig:MINATLht2} }
\end{figure}

\begin{figure}[htb]
\begin{center}
\epsfig{file=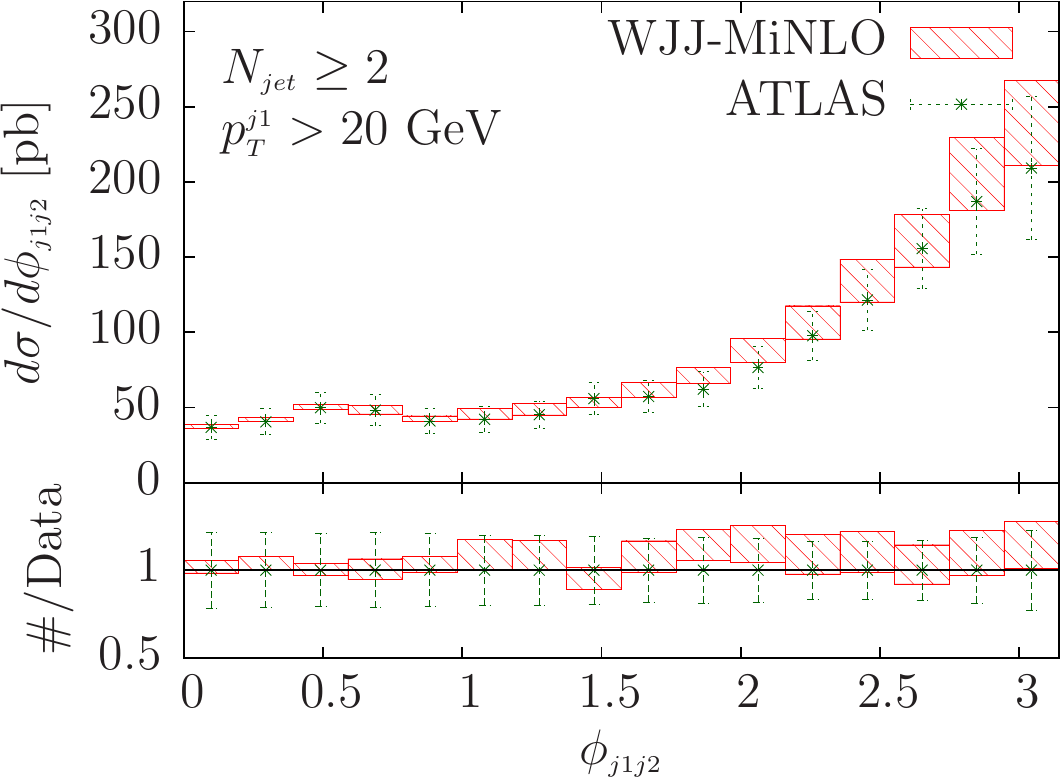,width=0.48\textwidth}
\epsfig{file=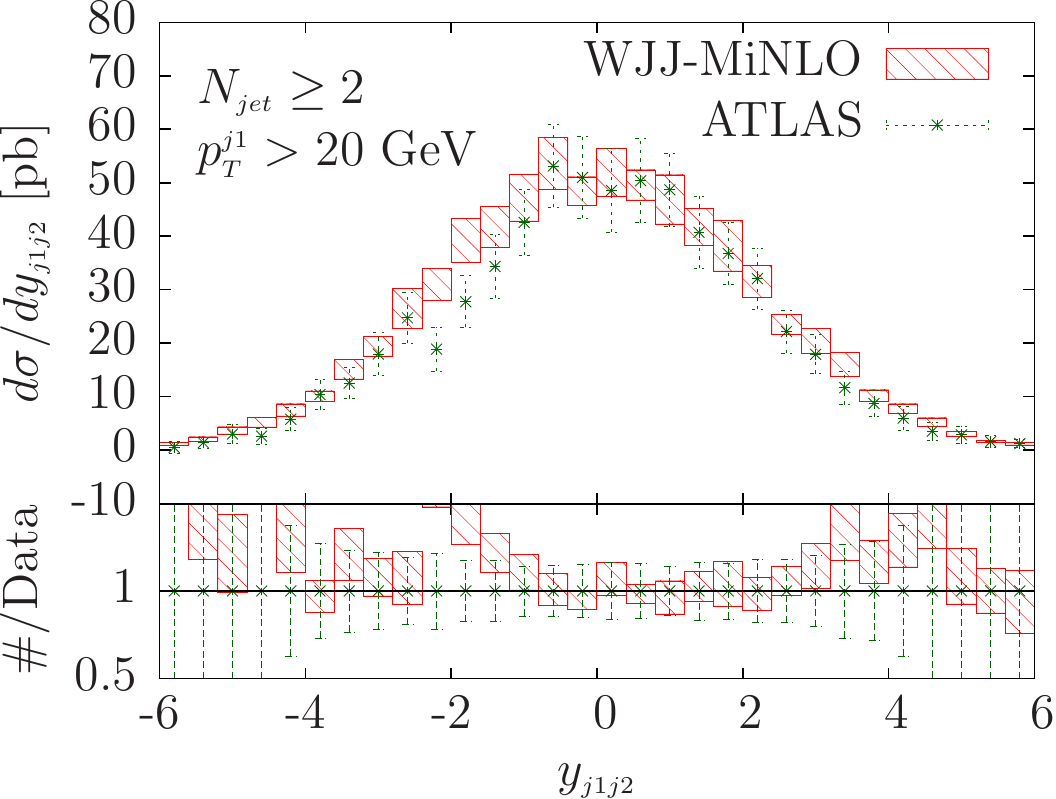,width=0.48\textwidth}
\end{center}
\caption{Azimuthal angle (left) and rapidity (right) difference
  between the two leading jets in inclusive two-jet events.
\label{fig:MINATLdphidy} }
\end{figure}

\begin{figure}[htb]
\begin{center}
\epsfig{file=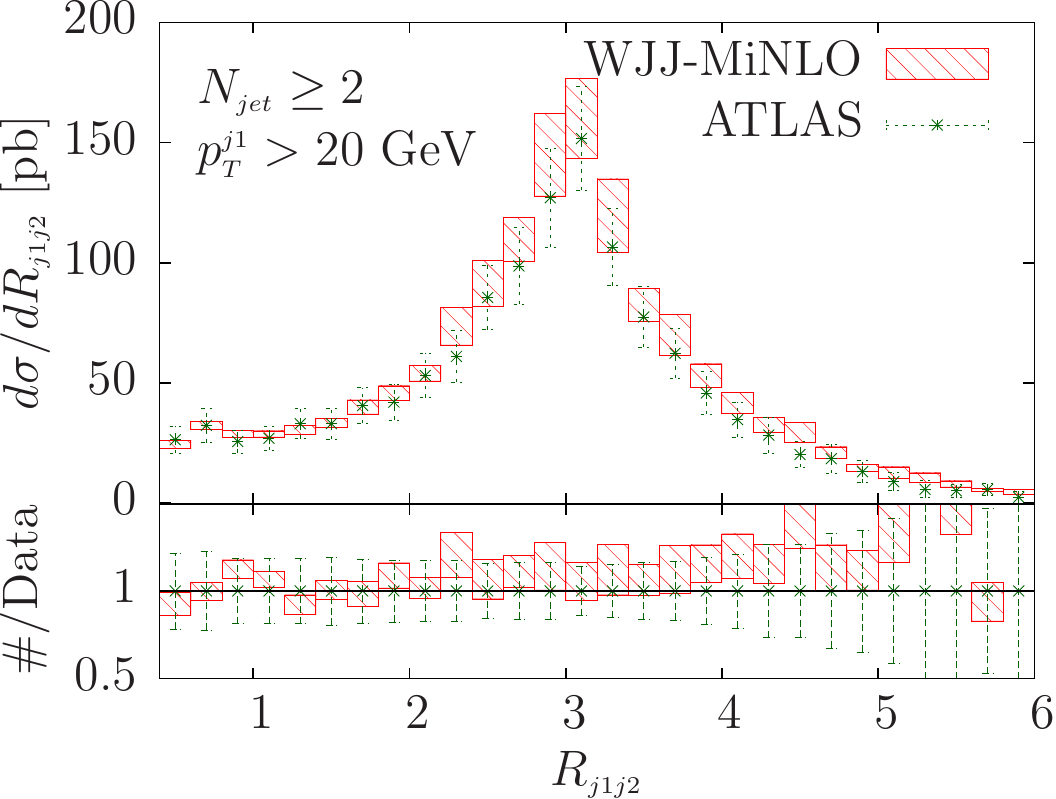,width=0.48\textwidth}
\end{center}
\caption{Distance $R_{j_1j_2} = (\Delta^2 y_{j_1j_2}+\Delta^2
  \phi_{j_1j_2})^{1/2}$ between the two leading jets in inclusive
  two-jet events.
\label{fig:MINATLdr} }
\end{figure}
\clearpage

\subsection{$Z$ production data}
In this section we compare the output of our \ZJJ{}+\MINLO{} generator
with the data of ref.~\cite{Aad:2011qv}.  The parameters used in the
simulation are the same as in sec.~\ref{sec:Wdata}.  The set of cuts
are displayed in table~\ref{tab:zcuts}.
\begin{table}[thb]
\begin{center}
\begin{tabular}{|l|}
\hline Cuts for $Z$ production \\
\hline
Jets defined using the anti-$\kt$ algorithm ($R=0.4$), with $\ptmin>30\,{\rm GeV}$, $|\eta|<4.4$;\\
Two opposite charge, same flavour leptons required with $\ptl>20\,{\rm GeV}$, $|\etal|<2.5$; \\
Lepton-jet isolation: $\Delta R_{l j}$ for all jets (as defined above) $> 0.5$;\\
Lepton-lepton isolation: $\Delta R_{ll}> 0.5$;\\
Constraint on dilepton invariant mass: $ 66 < m_{ll} < 116 \,{\rm GeV}$; \\
Events are classified according to the number of jets, as defined above.\\
\hline
\end{tabular}
\caption{Cuts for $Z$ production in association
with jets.\label{tab:zcuts}}
\end{center}
\end{table}
Again, we consider only $e^+ e^-$ decay but, since we switch off
electromagnetic radiation in \PYTHIA{}, our result can be considered
valid for any ``dressed'' lepton analysis. Therefore they will be shown in
comparison with the average data for electrons and muons.

We observe good agreement between the prediction of our generator and
ATLAS data.

\begin{figure}[H]
\begin{center}
\epsfig{file=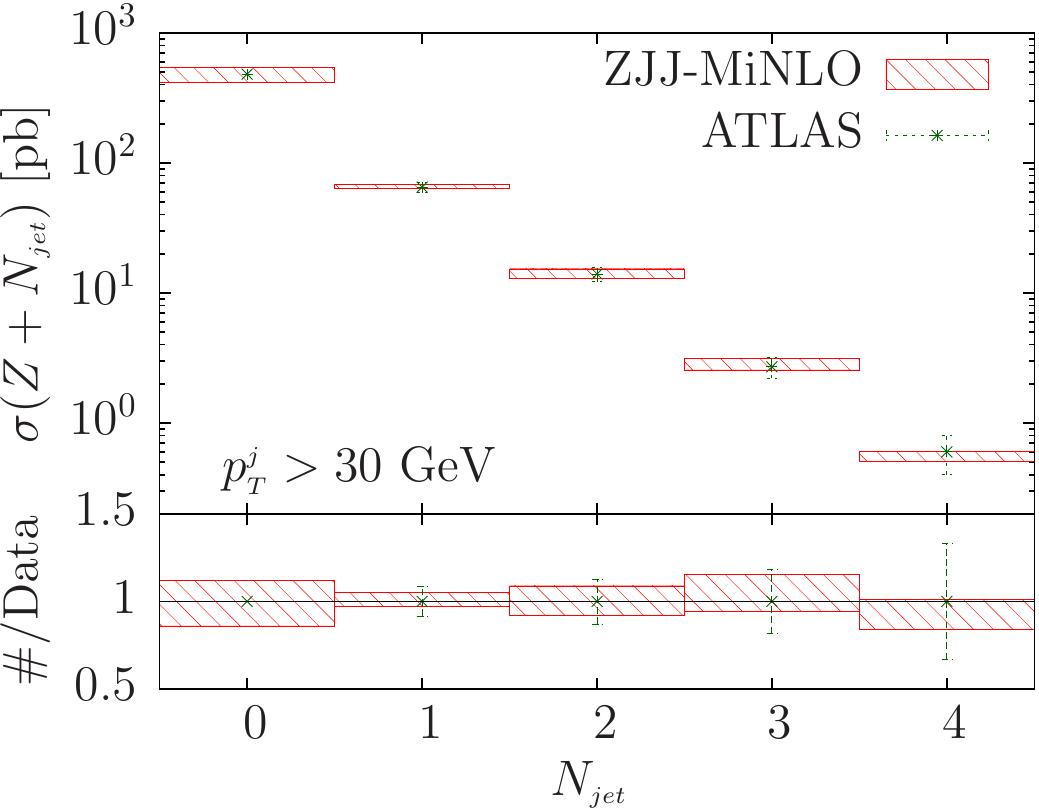,width=0.48\textwidth}
\end{center}
\caption{Inclusive jet multiplicity obtained with the \ZJJ{} generator
  using \MINLO{}, interfaced to \PYTHIASIX{}, compared to ATLAS data.
\label{fig:MINATLZnjet} }
\end{figure}
\begin{figure}[H]
\begin{center}
\epsfig{file=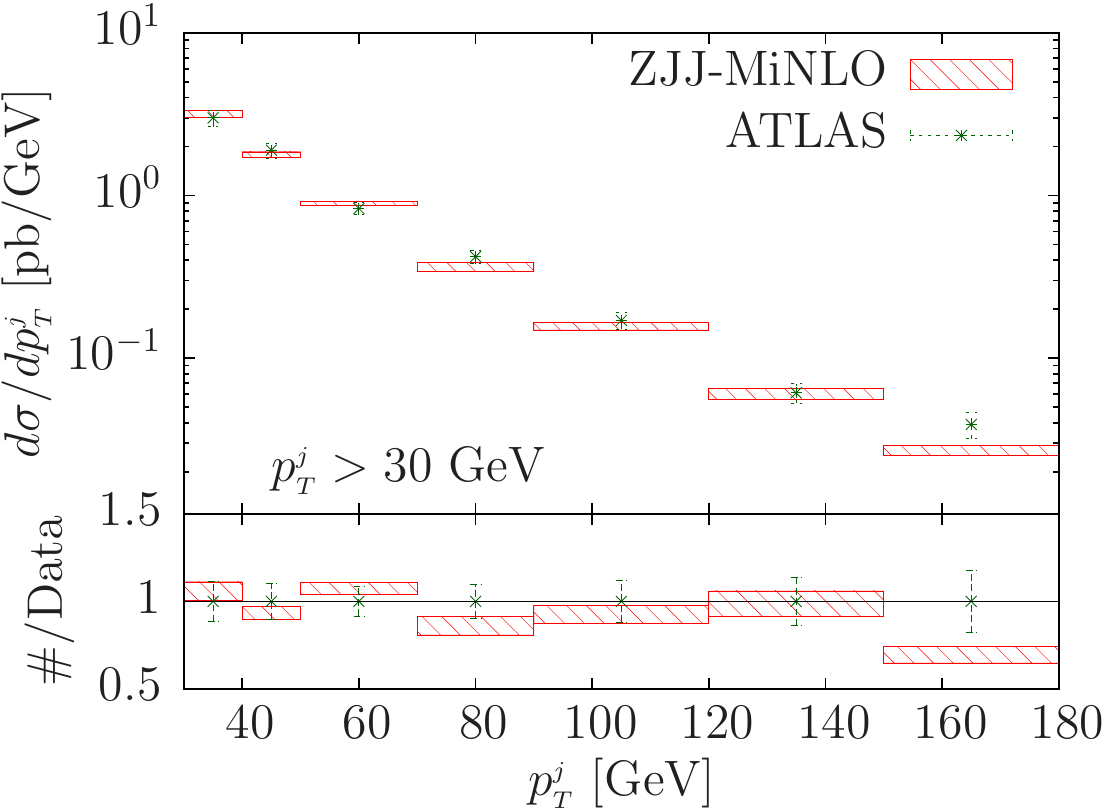,width=0.48\textwidth}
\epsfig{file=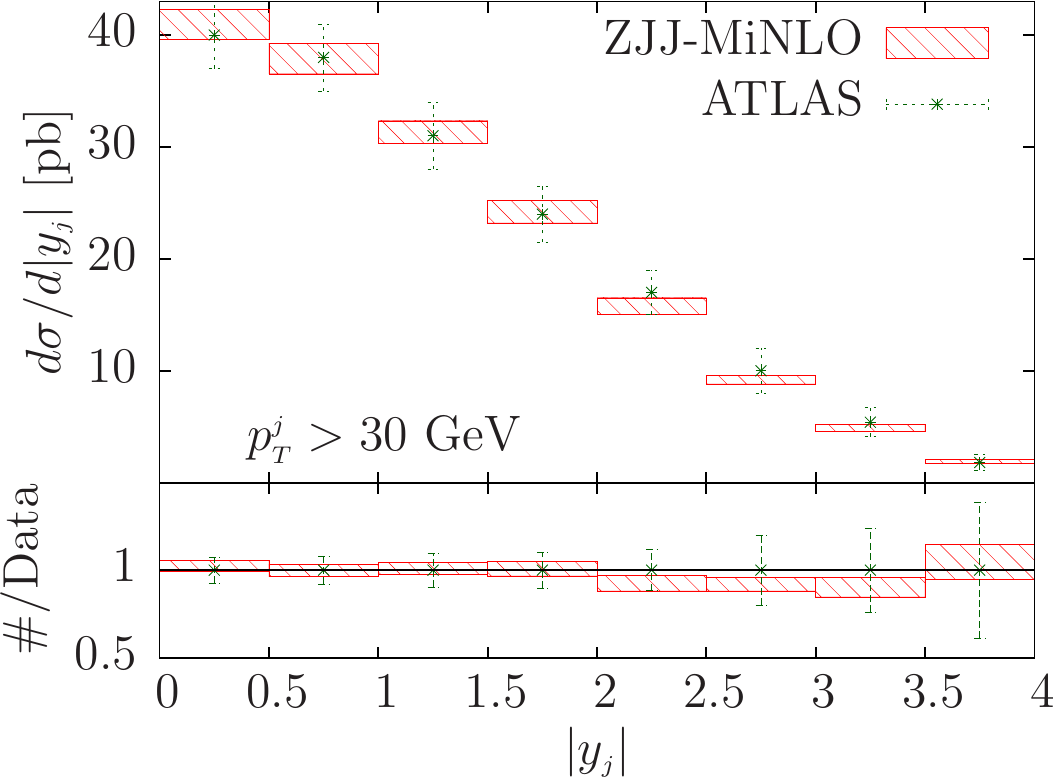,width=0.48\textwidth}
\end{center}
\caption{Inclusive transverse momentum (left) and rapidity
  distribution (right).
\label{fig:MINATLZinclj} }
\end{figure}
\begin{figure}[H]
\begin{center}
\epsfig{file=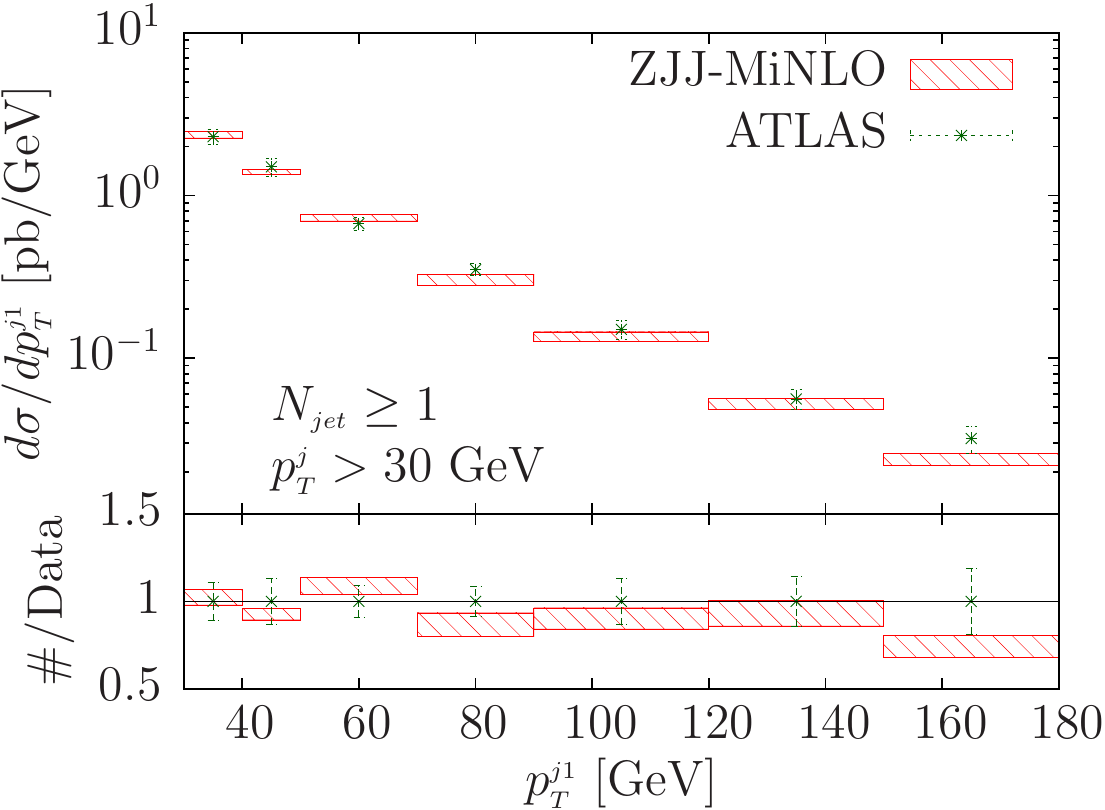,width=0.48\textwidth}
\epsfig{file=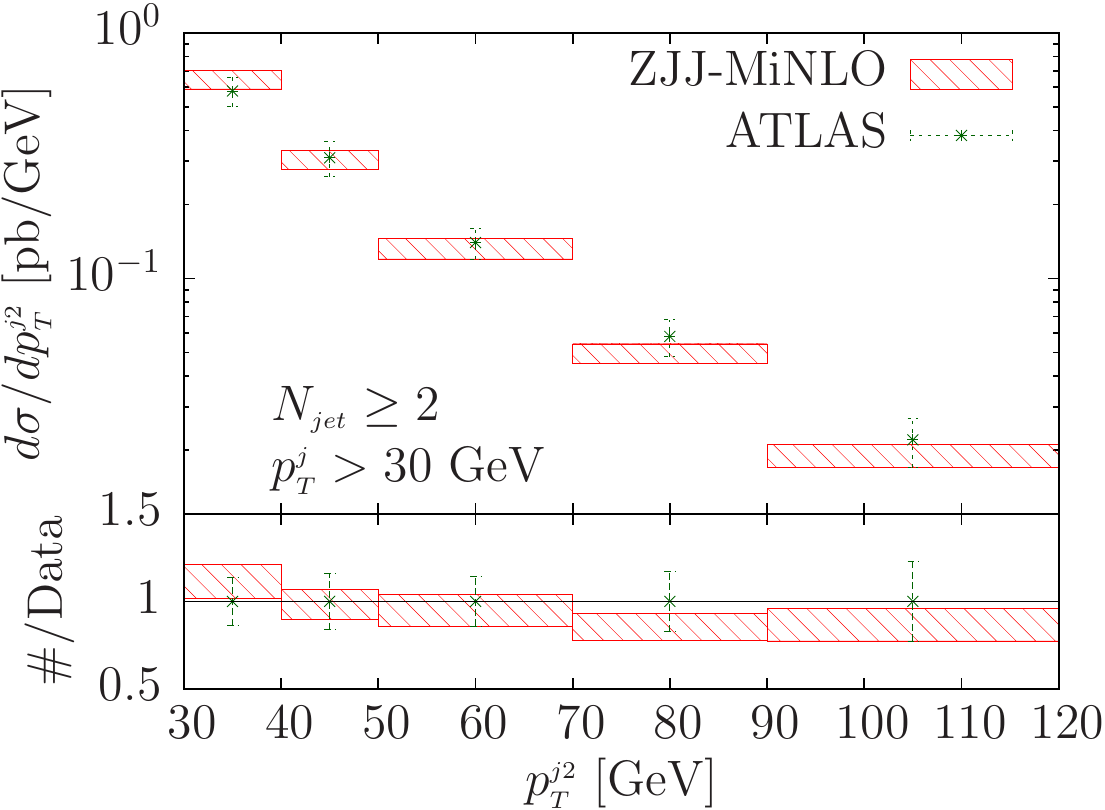,width=0.48\textwidth}
\end{center}
\caption{Transverse momentum of the leading jet in inclusive one-jet
  events (left) and of the second leading jet in inclusive two-jet
  events (right).
\label{fig:MINATLZj1j2pt} }
\end{figure}
\begin{figure}[H]
\begin{center}
\epsfig{file=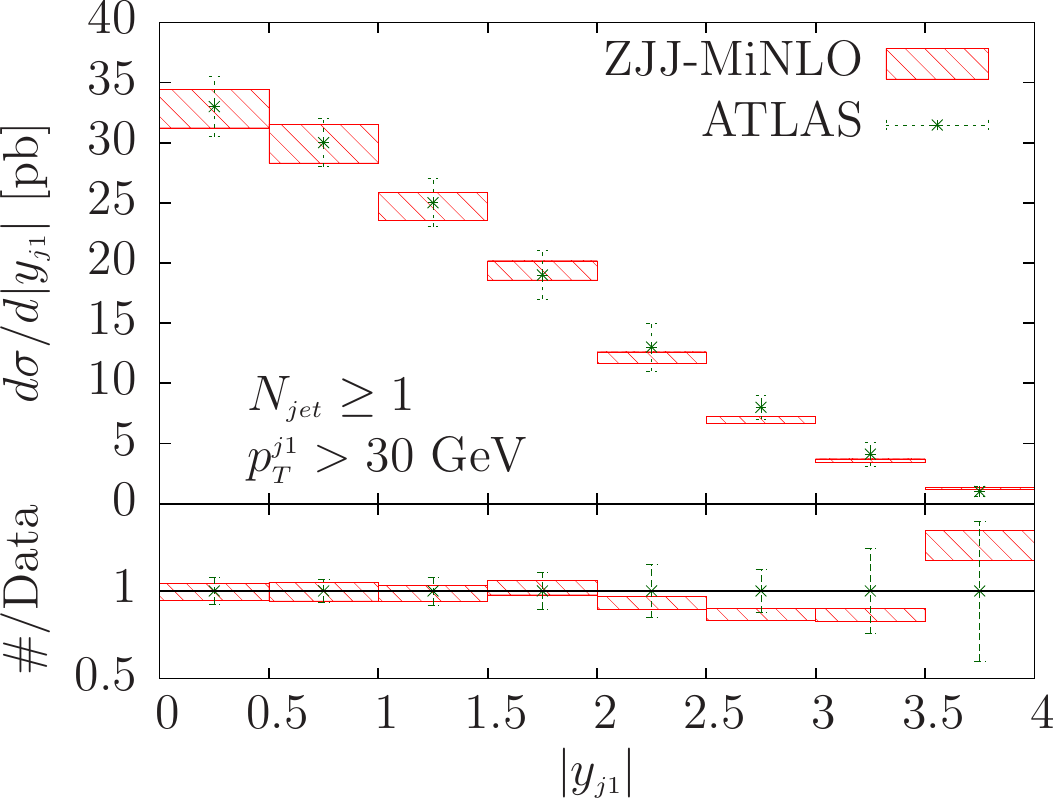,width=0.48\textwidth}
\epsfig{file=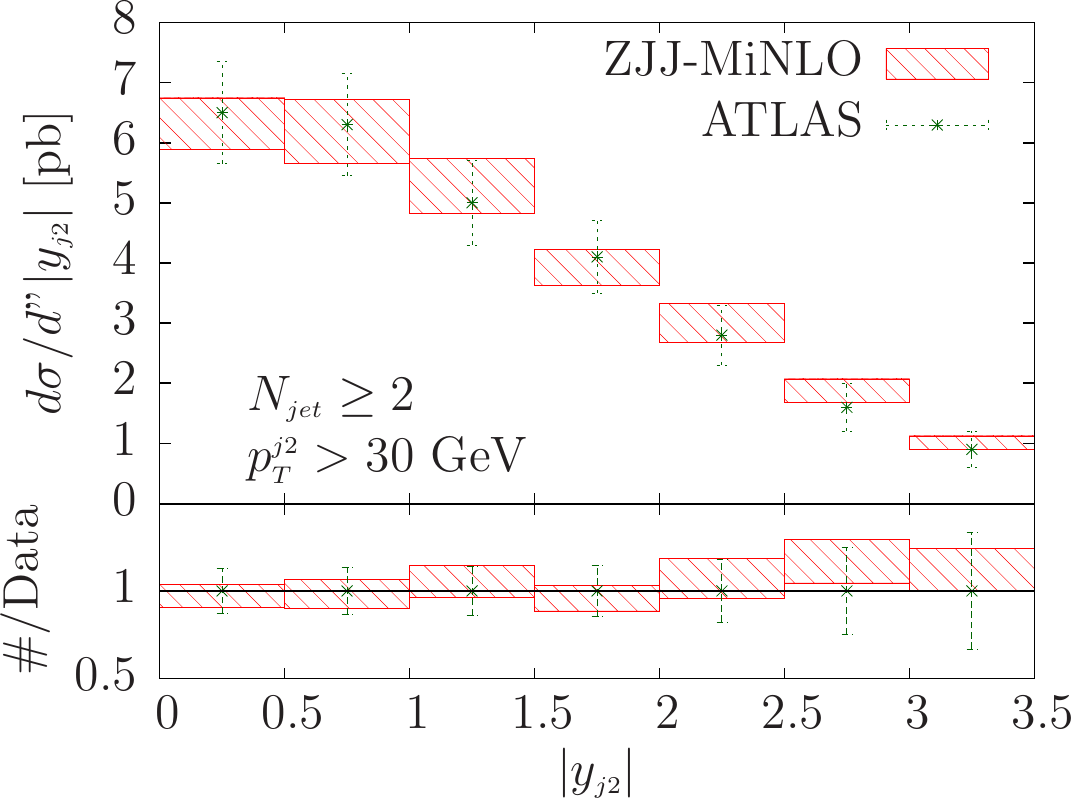,width=0.48\textwidth}
\end{center}
\caption{Rapidity distribution of the leading jet in inclusive one-jet
  events (left) and of the second leading jet in inclusive two-jet
  events (right).
\label{fig:MINATLZj1j2y} }
\end{figure}
\begin{figure}[H]
\begin{center}
\epsfig{file=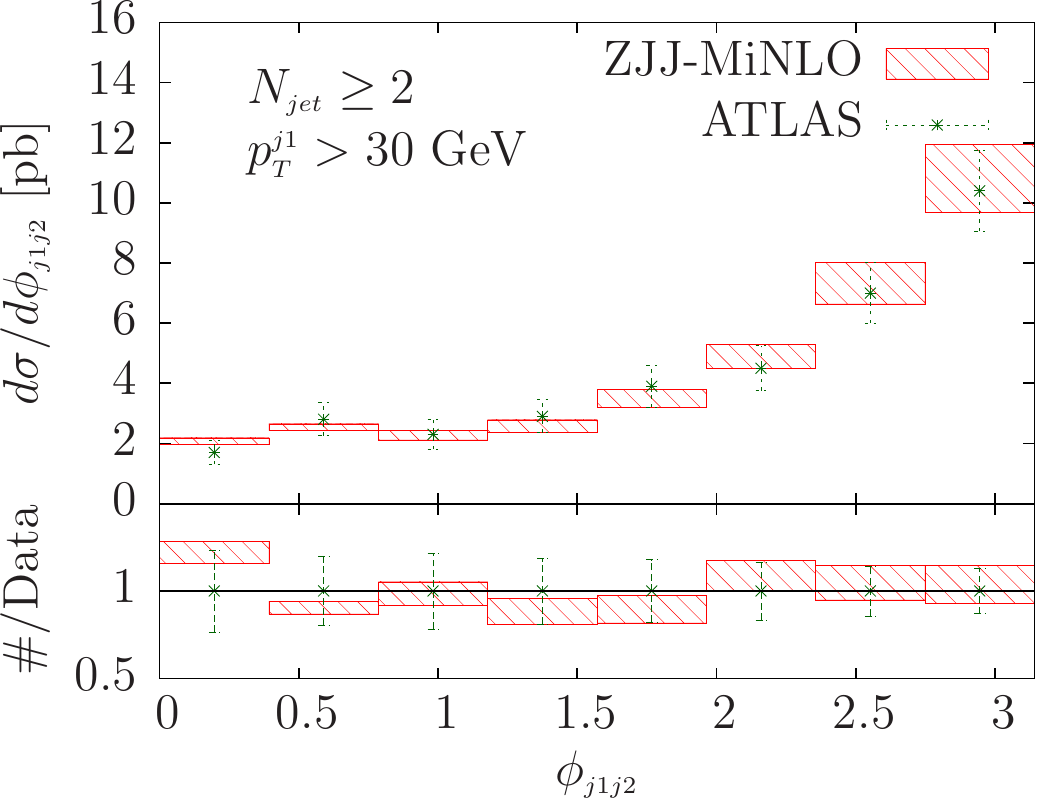,width=0.48\textwidth}
\epsfig{file=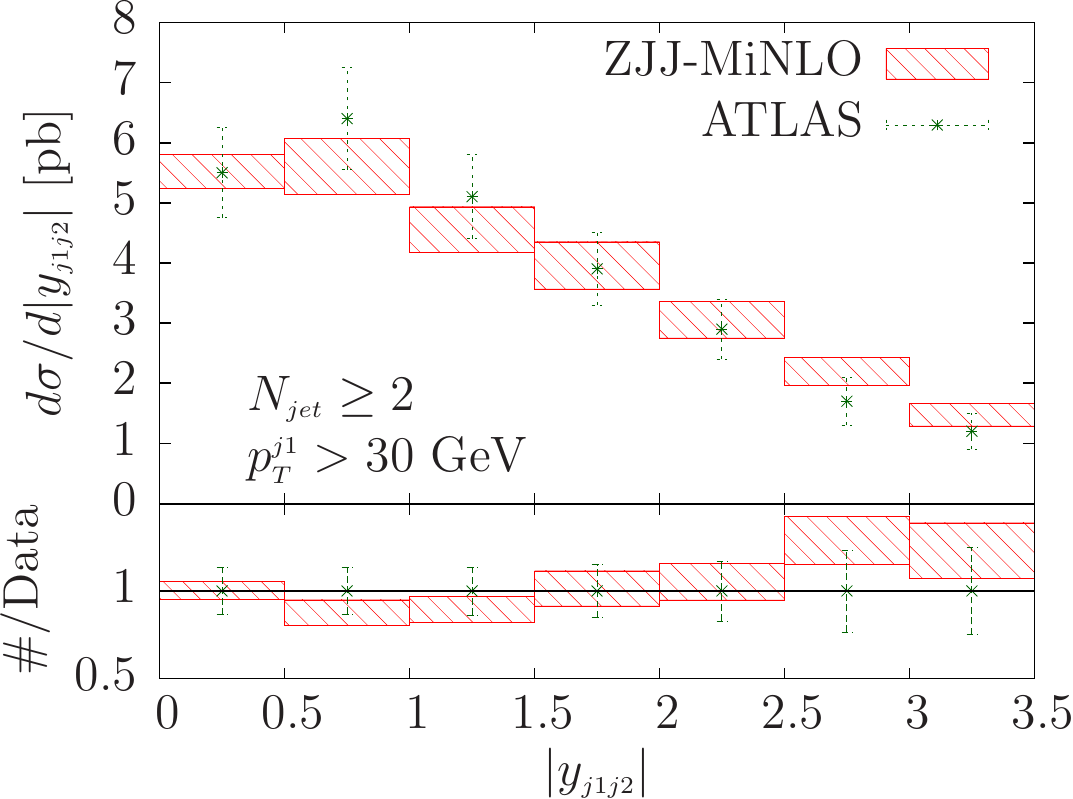,width=0.48\textwidth}
\end{center}
\caption{Azimuthal angle (left) and rapidity (right) difference
  between the two leading jets in inclusive two-jet events.
\label{fig:MINATLZj1j2ry} }
\end{figure}
\begin{figure}[H]
\begin{center}
\epsfig{file=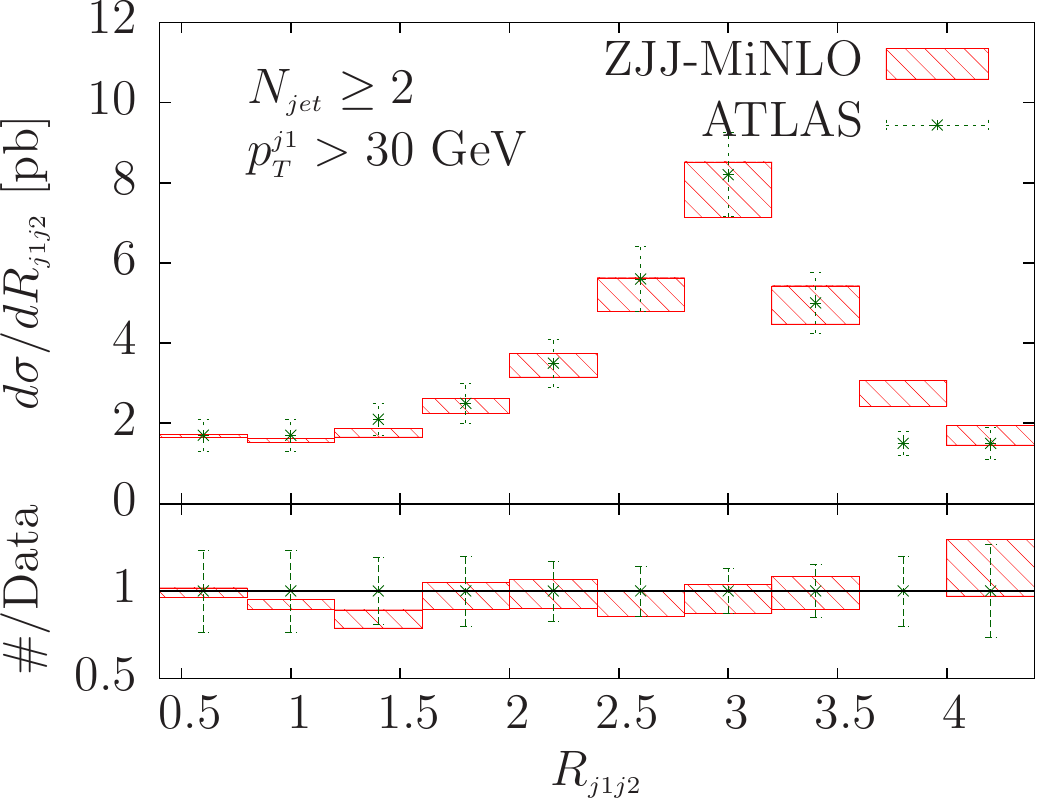,width=0.48\textwidth}
\epsfig{file=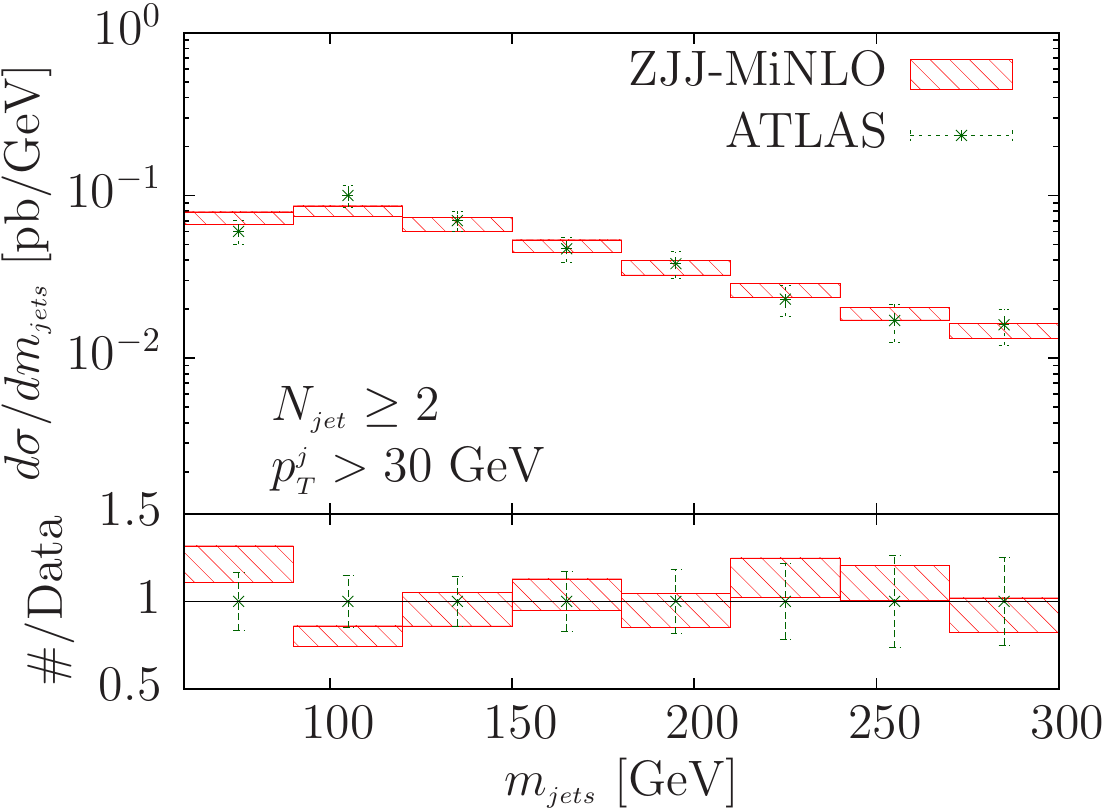,width=0.48\textwidth}
\end{center}
\caption{Distance $R_{j_1j_2} = (\Delta y_{j_1j_2}+\Delta
  \phi_{j_1j_2})^{1/2}$ between the two leading jets and invariant
  mass of the system of all jets (right) in inclusive two-jet events.
\label{fig:MINATLZj1j2phi-mas}} 
\end{figure}

\section{Observables sensitive to the MPI}\label{sec:MPI}
As discussed in section~\ref{sec:problems}, the contribution of the
\MPI{} to the cross section for vector boson production plus two jets
may not be negligible. On the other hand, the \VJJ{} generators
presented in this work, when augmented with the \MINLO{} procedure,
yield a good prediction also for the \MPI{} contributions, since they
describe reasonably well also inclusive distributions.

An analysis of double parton scattering in the framework of $W$
production has been presented by the ATLAS collaboration in
ref.~\cite{Aad:2013bjm}.  Here we examine two observables that should
be particularly sensitive to the \MPI{}.  The first and most natural
one is the transverse momentum of the vector boson in the two jet
inclusive sample. The second one is the vector sum of the transverse
momenta of the two jets, $p_T^{jj} = |\vec p_T^{j_1}+\vec p_T^{j_2}|$
in two-jet exclusive events.\footnote{In ref.~\cite{Aad:2013bjm} also
  the observable $p_T^{jj}/(p_T^{j_1}+p_T^{j_2})$ is considered, which
  is less sensitive to the jet-energy scale.  We here do not show it,
  since, being scale invariant, it does not show that the \MPI{}
  contributions vanish at large transverse momenta.}
In all cases one expects a non negligible contribution
from the \MPI{} in the region of small transverse momenta.

In fig.~\ref{fig:Wpt-ge2j}
\begin{figure}[htb]
\begin{center}
\epsfig{file=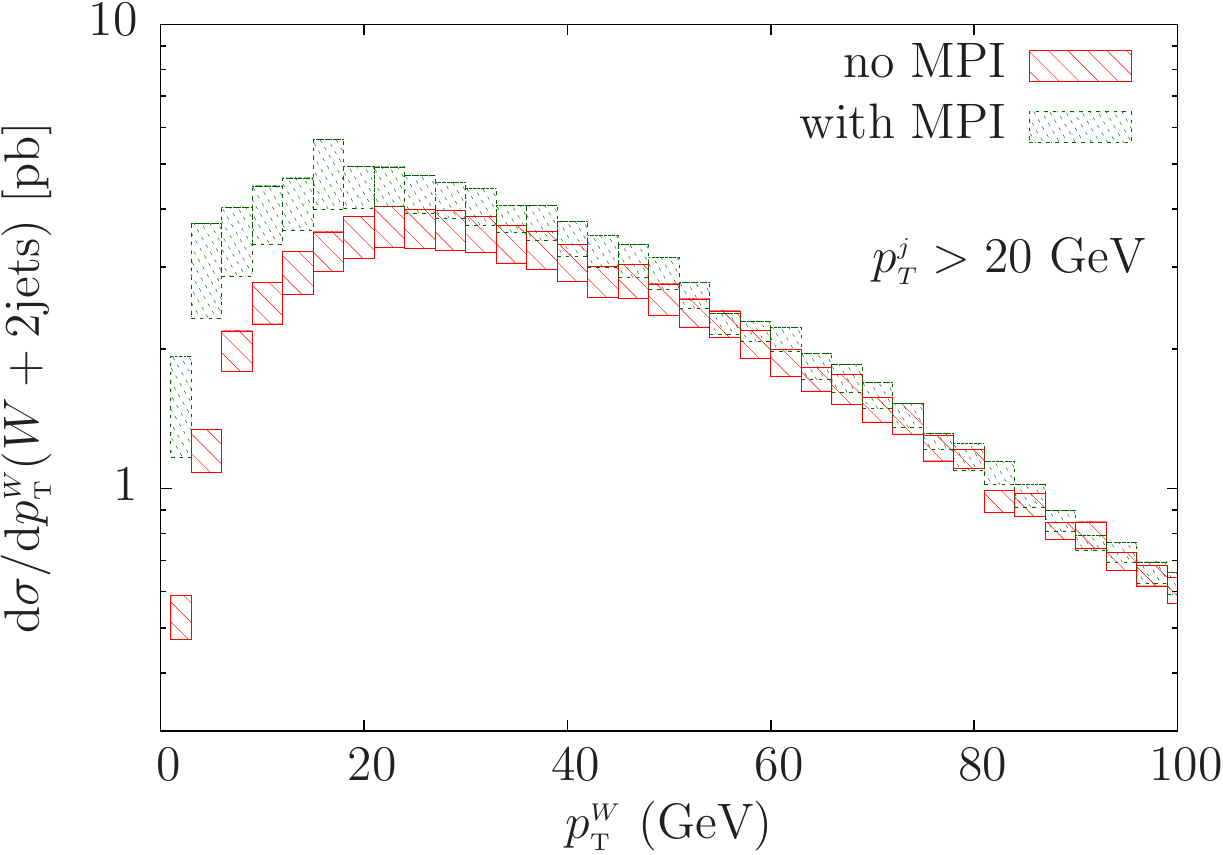,width=0.48\textwidth}
\epsfig{file=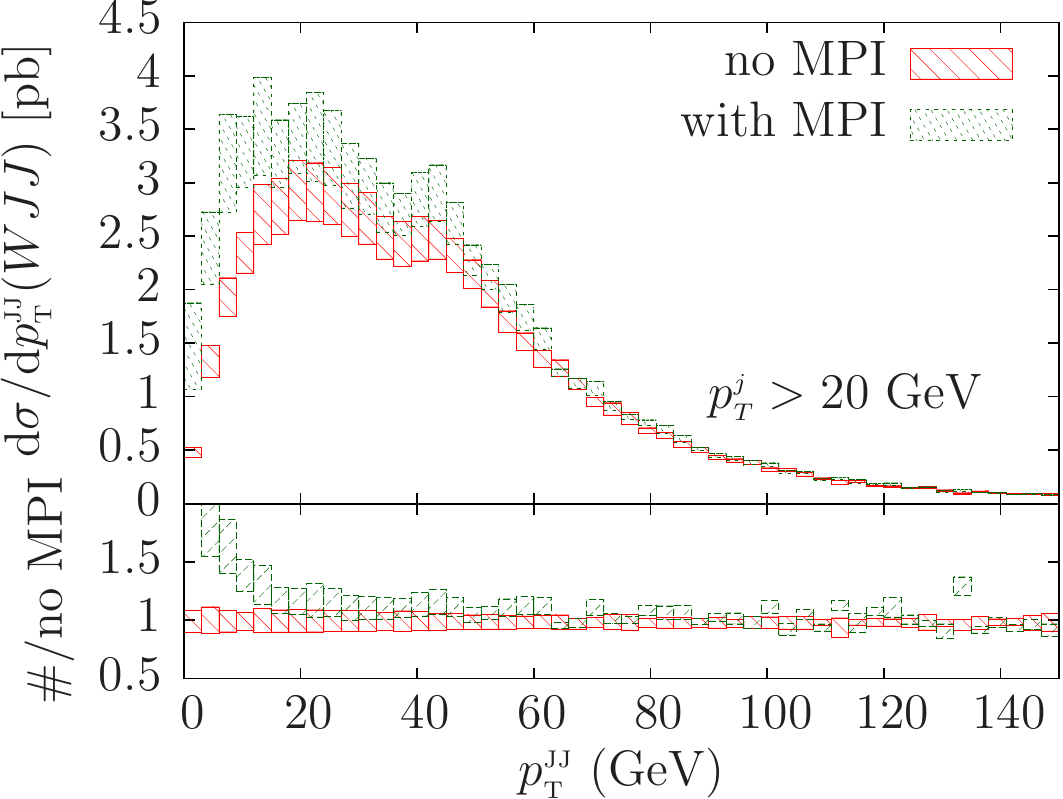,width=0.48\textwidth}
\end{center}
\caption{The transverse momentum distribution of the $W$ boson in the
  inclusive $W+2~{\rm jets}$ sample, and the total transverse momentum
  of the two leading jets in the exclusive $W+2~{\rm jets}$ sample,
  computed with \WJJ+\MINLO{} interfaced to \PYTHIASIX{}. The effect
  of switching on and off the MPI is shown. The bands are obtained by
  scale variation as described in sec.~\protect\ref{sec:htminlo}.
\label{fig:Wpt-ge2j}}
\end{figure}
we show the transverse momentum of the $W$ boson and the $p_T^{jj}$
distribution for $W$ production.  It is clear that the inclusion of
the MPI yields a noticeable change in the shapes. In particular, below
10 GeV, the contribution of the MPI increases, until it becomes
dominant at small transverse momenta.

In fig.~\ref{fig:Zpt-ge2j}
\begin{figure}[htb]
\begin{center}
\epsfig{file=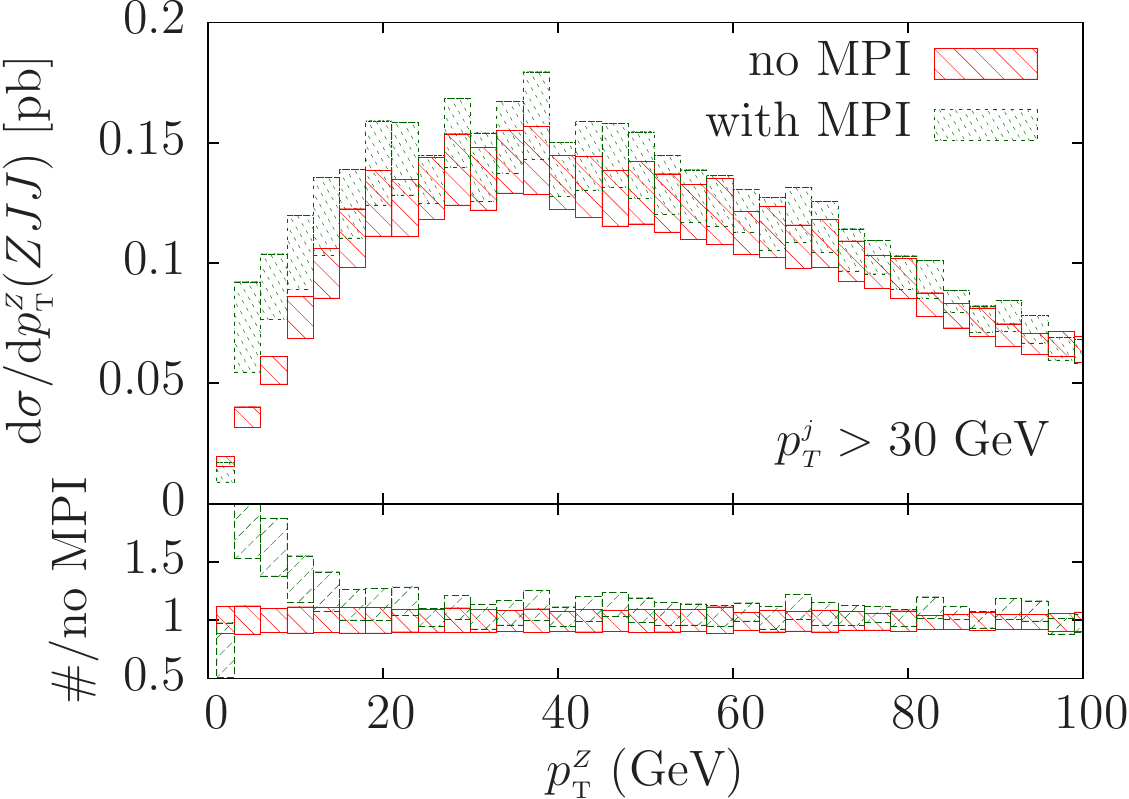,width=0.48\textwidth}
\epsfig{file=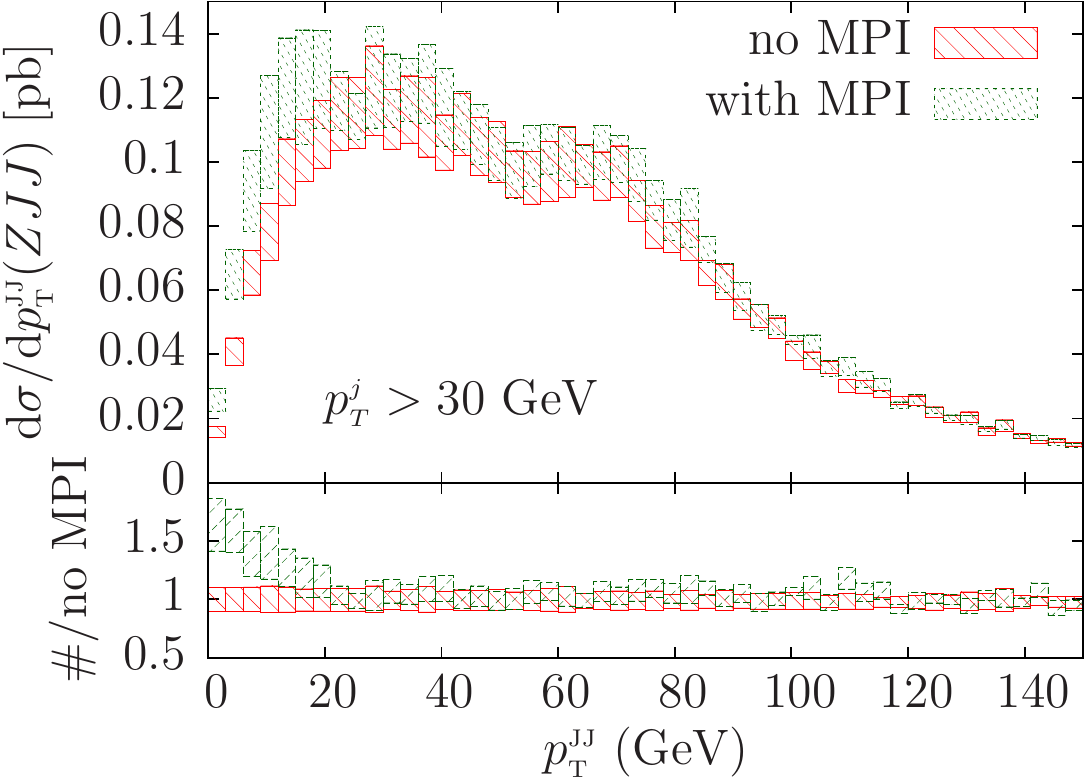,width=0.48\textwidth}
\end{center}
\caption{As in fig.~\protect\ref{fig:Wpt-ge2j} for $Z$ production.\label{fig:Zpt-ge2j}}
\end{figure}
we show the same quantities for $Z$ production. Since now the jet cut
is at 30 GeV rather than 20, the relative effect of MPI is reduced,
but still quite visible.  In conclusion, our study indicates that both
observables can be conveniently used to study and tune multi-parton
interactions.  The transverse momentum of the $Z$ bosons in the 2-jet
sample (inclusive or exclusive) is also a particularly promising
observable, provided the jet energy cut is low enough.

\section{Conclusions}
In the present work we have presented new \POWHEGBOX{} generators
for the production of a vector boson in association with two jets.
This is not the first time that generators for these
processes have appeared in the literature, and in the present
work we claim to have improved in terms of speed (more than a factor of three)
over previous \POWHEG{} implementations. However, this is not the only progress that we
have achieved. Although building NLO
generators matched with showers for processes with associated jet
production is today a straightforward task, several details about their
correct use are still subject to study. In our opinion, substantial
theoretical progress in this framework is desirable and indeed possible.
The problem of scale choice in associated jet production is, for
example, a detail with very profound
implications~\cite{Hamilton:2012np,Hamilton:2012rf}.

In the present work, we have addressed issues and implications related
to the use of a generation cut or of a Born suppression factor,
that are needed when associated jet production is considered.
It would be desirable that events that are soft at the underlying
Born level remain soft after the hardest radiation, after shower,
and after the inclusion of the underlying events and of multi-parton interactions.
If this is not the case, hard jet
distributions become sensitive to the generation cut, or, alternatively,
if a suppression factor is used instead, rare events with very large weight
contribute, making it difficult to interpret the results.

We have found that some simple modification of the kinematic functions
used for the separation of the singular regions in \POWHEG{} is very
beneficial in reducing the probability that a soft underlying Born
configuration may become hard after the \POWHEG{} hardest radiation.
The analogous problem, of an event that is soft at the underlying Born
level, but becomes hard after the parton shower, and before hadronization,
has also been discussed in the past in the framework of dijet production.

On the other hand, the case of events that are soft after shower, but
become hard after the hadronization stage, because of
multi-parton-interactions, cannot be dismissed or alleviated. This is
in fact a physical effect that should be properly simulated.
In order to do so, the generator should also be capable of describing
the inclusive cross section (i.e. without jet cuts) with reasonable
accuracy. By using the \MINLO{} procedure, this feature comes as a
bonus.

These considerations are not only relevant for the \ZJJ{} and \WJJ{} generators.
They are, of course, as significant for the \HJJ{} generator. But we have
found that they have also considerable relevance for processes involving
a single jet, like the \HJ{}, \WJ{} and \ZJ{} processes. We expect them
to be important for all processes involving jet production, irrespective
of the number of jets.

In this work we have also validated our full setup, involving both \MINLO{} and
the newly introduced kinematic functions for the separation of the
singular regions in \POWHEG{}, against available data. This is the first
extensive validation study of the \MINLO{} method. We have found a generally
good agreement with data, even for one jet inclusive and totally inclusive
observables, for which the method does not achieve NLO accuracy.

\section*{Acknowledgements}
We thank Emanuele Re for discussions.

\appendix 

\clearpage
\section{Comparison with data for $\mathbf{p_t^{(j)} > 30}$ GeV for $W +2$ jet production}
\label{app:figuresW30}
In this appendix we present plots comparing the data of the 
ATLAS collaboration for $W +2$ jet production using a jet cut of  
${p_t^{(j)} > 30}$~GeV with our predictions using the \MINLO{} method.

\begin{figure}[htb]
\begin{center}
\epsfig{file=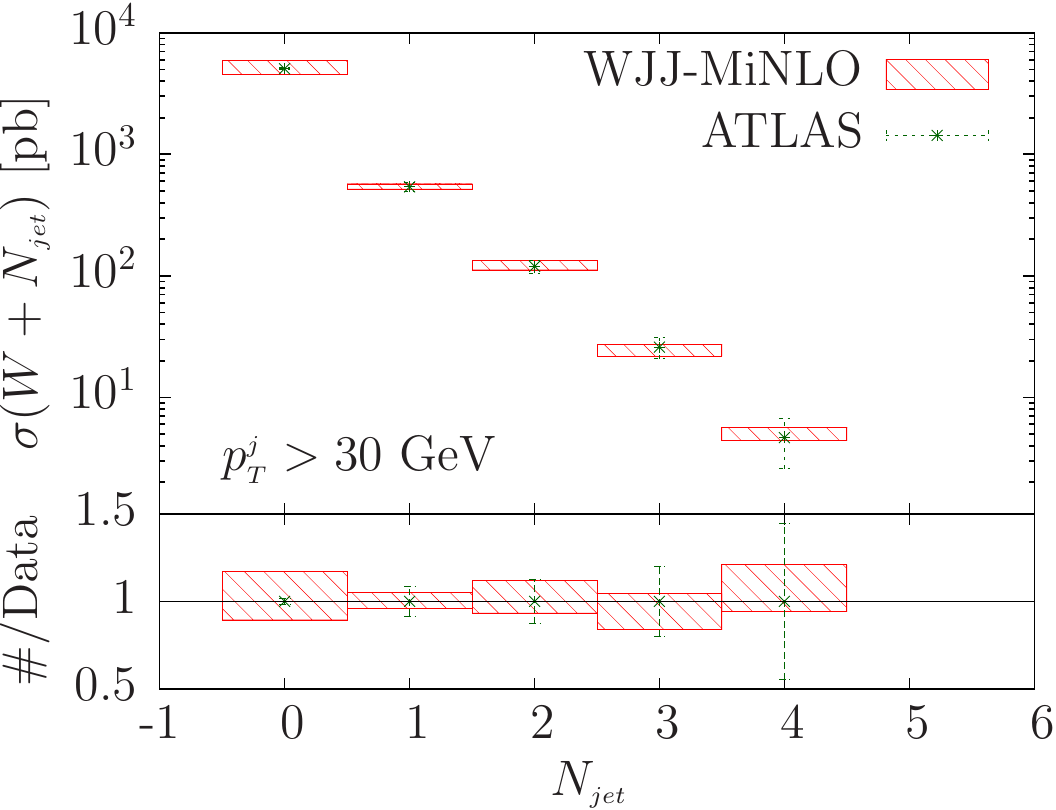,width=0.48\textwidth}
\end{center}
\caption{Inclusive jet multiplicity obtained with the \WJJ{} generator
  using \MINLO{}, interfaced to \PYTHIASIX{}, compared to ATLAS data.
\label{fig:ptj30MINATLnjet} }
\end{figure}

\begin{figure}[htb]
\begin{center}
\epsfig{file=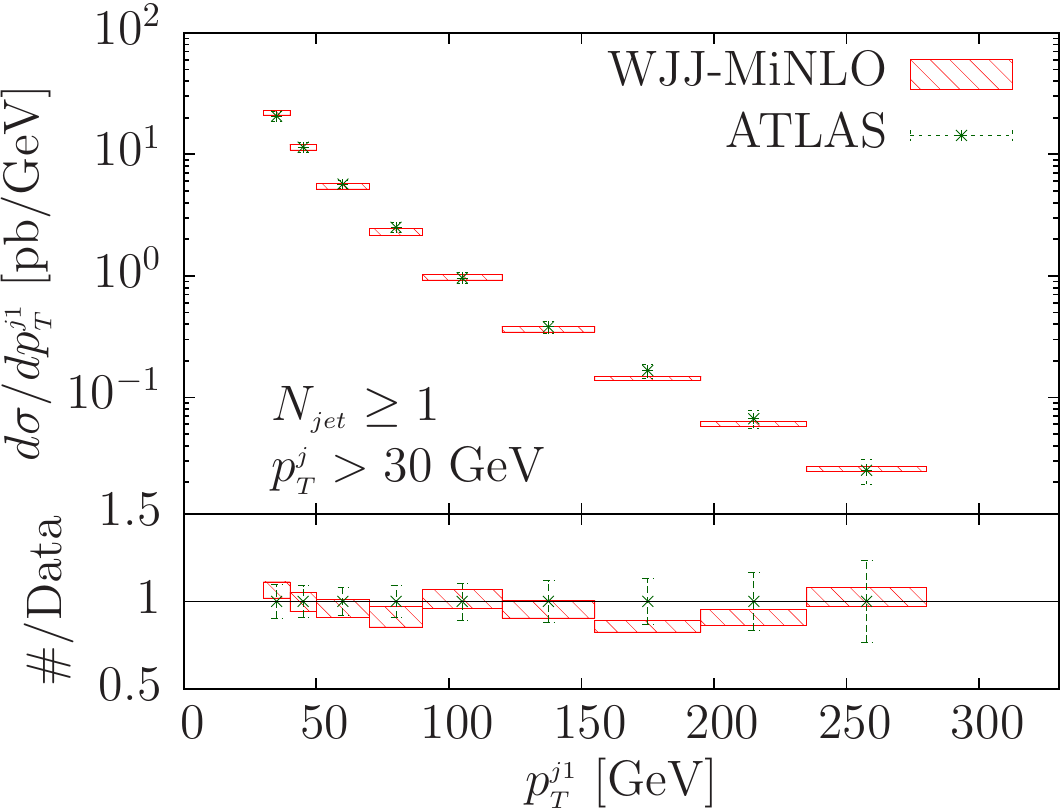,width=0.48\textwidth}
\epsfig{file=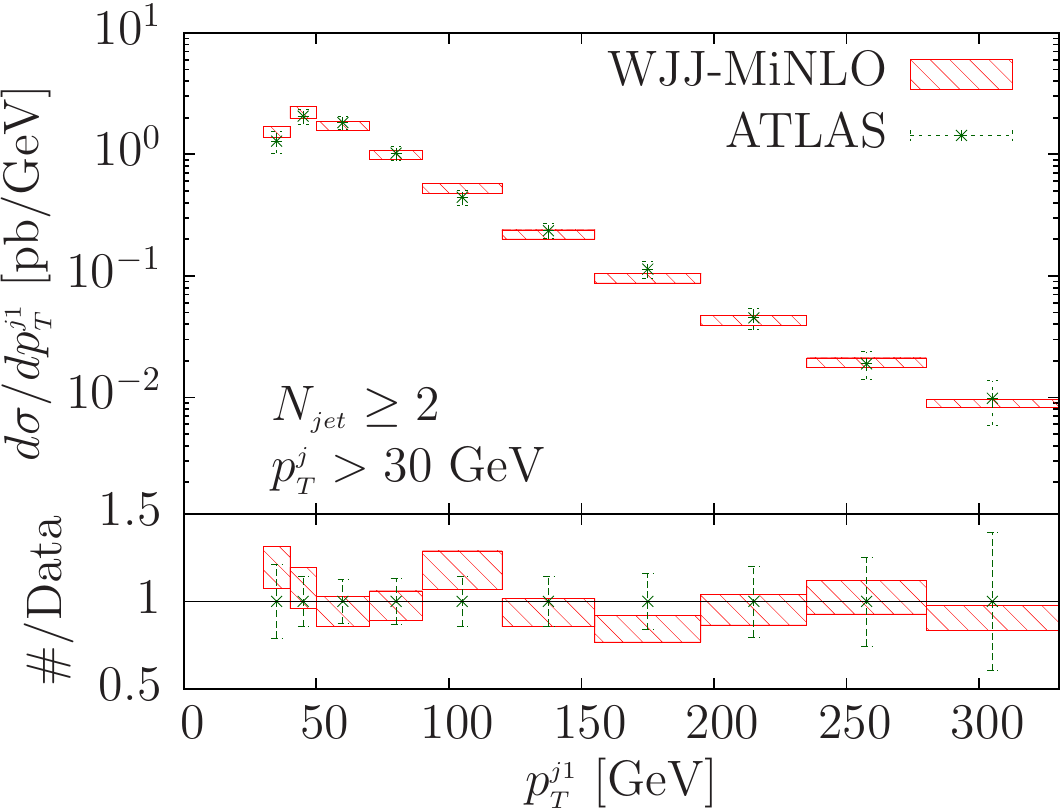,width=0.48\textwidth}
\end{center}
\caption{Transverse momentum of the leading jet in inclusive one-jet
  events (left) and two-jet events (right).
\label{fig:ptj30MINATLj1pt} }
\end{figure}

\begin{figure}[htb]
\begin{center}
\epsfig{file=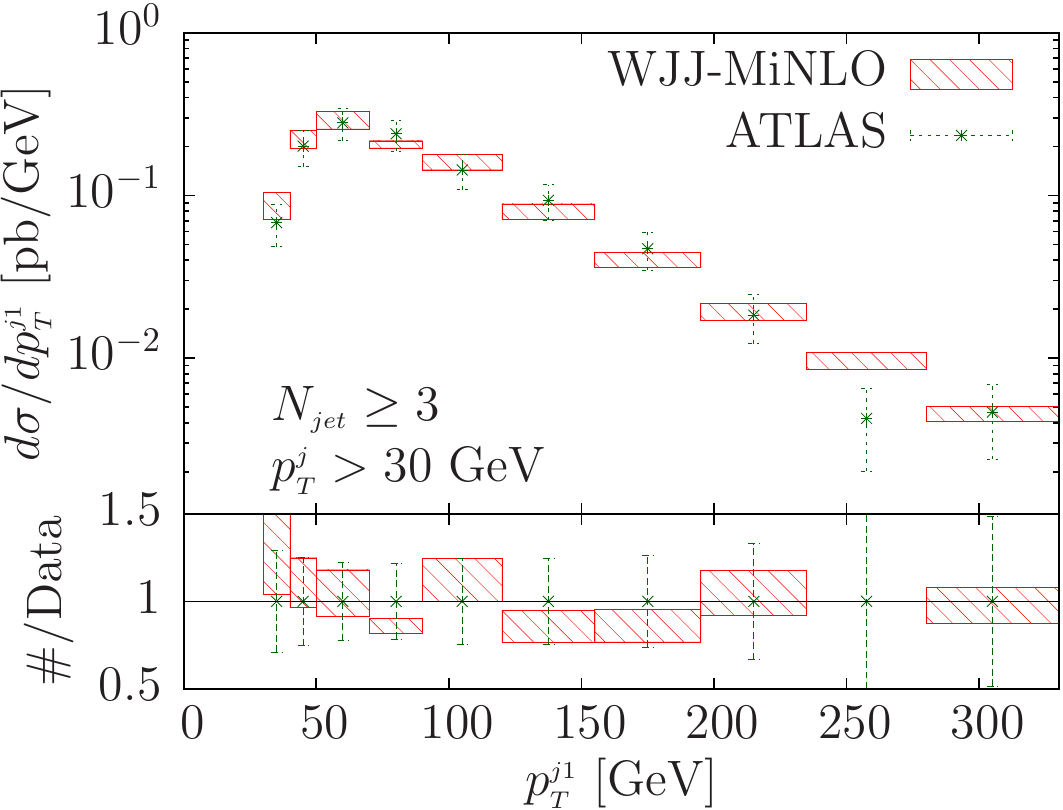,width=0.48\textwidth}
\epsfig{file=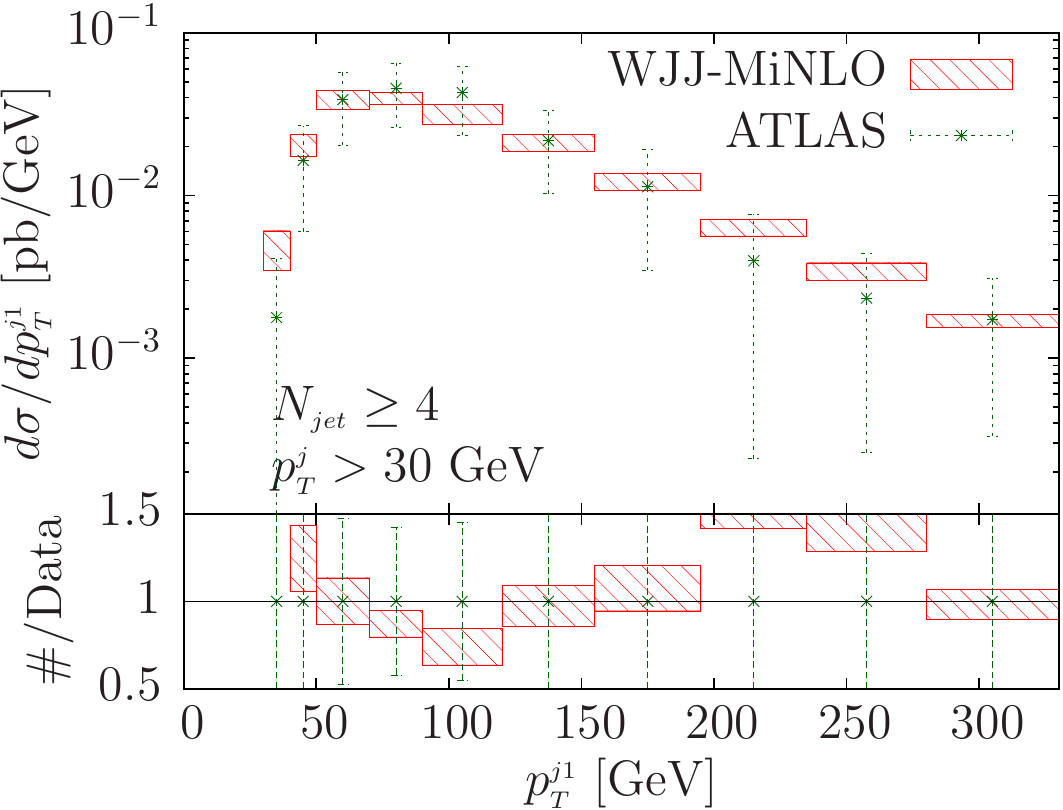,width=0.48\textwidth}
\end{center}
\caption{Transverse momentum of the leading jet in inclusive three-jet
  events (left) and four-jet events (right).
\label{fig:ptj30MINATLj1pt2} }
\end{figure}

\begin{figure}[htb]
\begin{center}
\epsfig{file=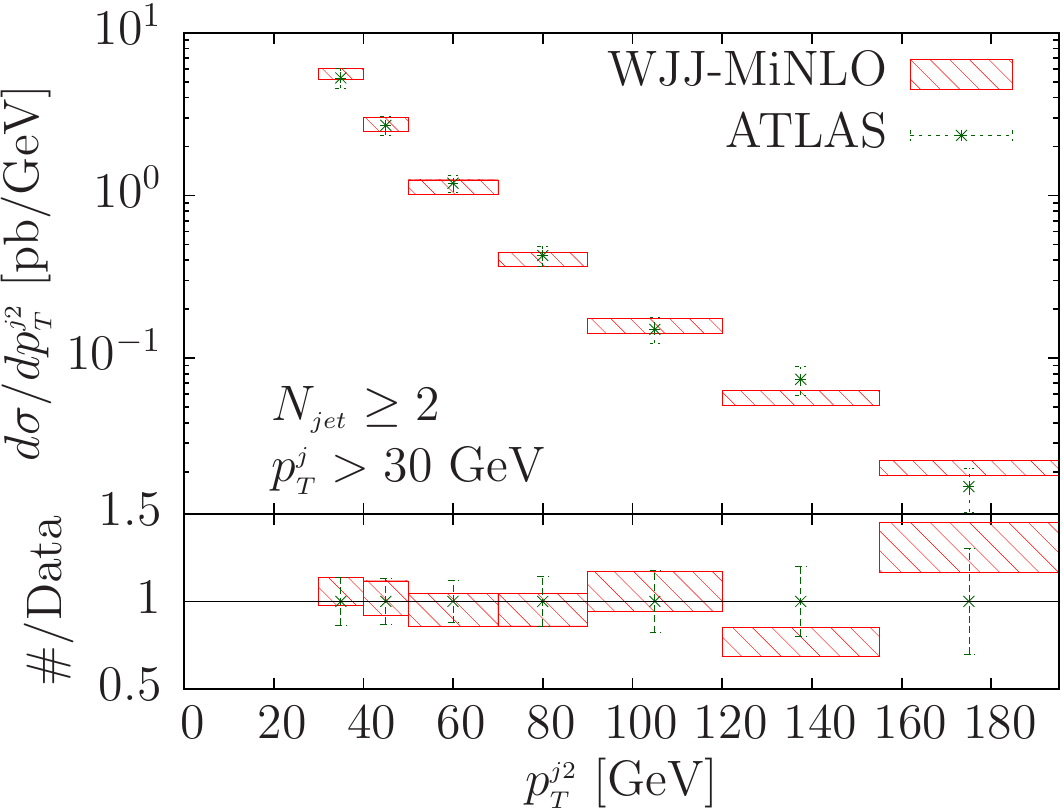,width=0.48\textwidth}
\end{center}
\caption{Transverse momentum of the second leading jet in inclusive
  two-jet events.
\label{fig:ptj30MINATLj2pt} }
\end{figure}

\begin{figure}[htb]
\begin{center}
\epsfig{file=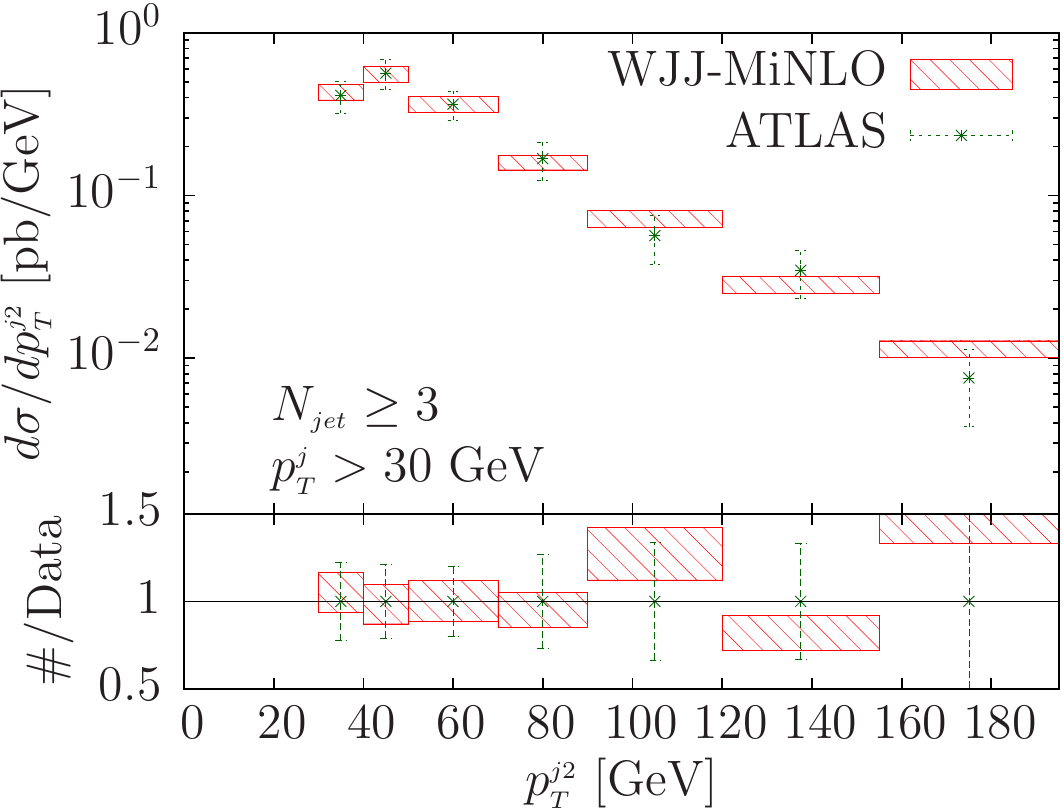,width=0.48\textwidth}
\epsfig{file=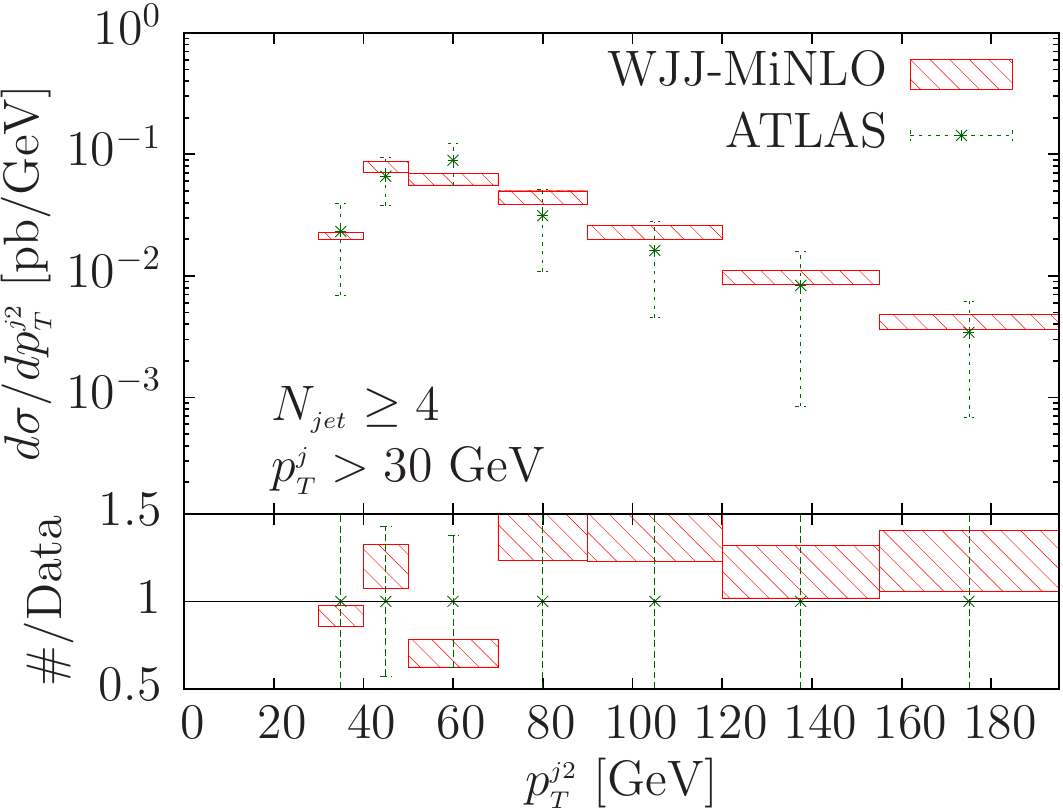,width=0.48\textwidth}
\end{center}
\caption{Transverse momentum of the second leading jet in inclusive
  three-jet events (left) and four-jet events (right).
\label{fig:ptj30MINATLj2ptbis} }
\end{figure}

\begin{figure}[htb]
\begin{center}
\epsfig{file=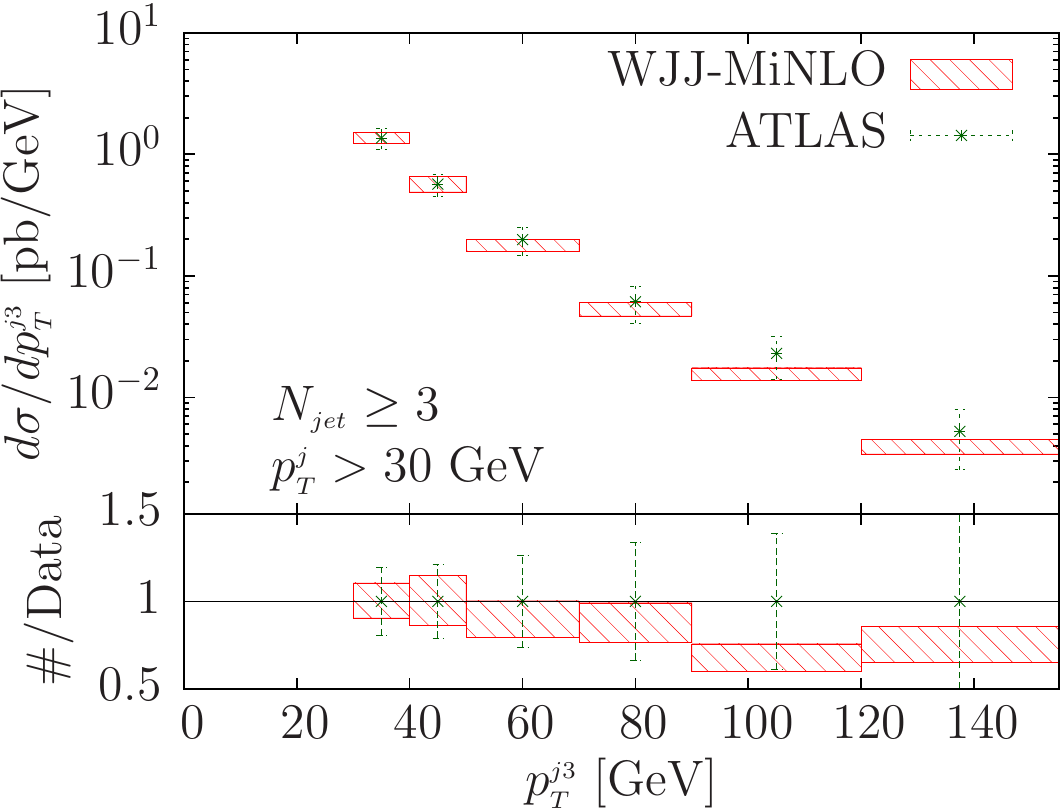,width=0.48\textwidth}
\epsfig{file=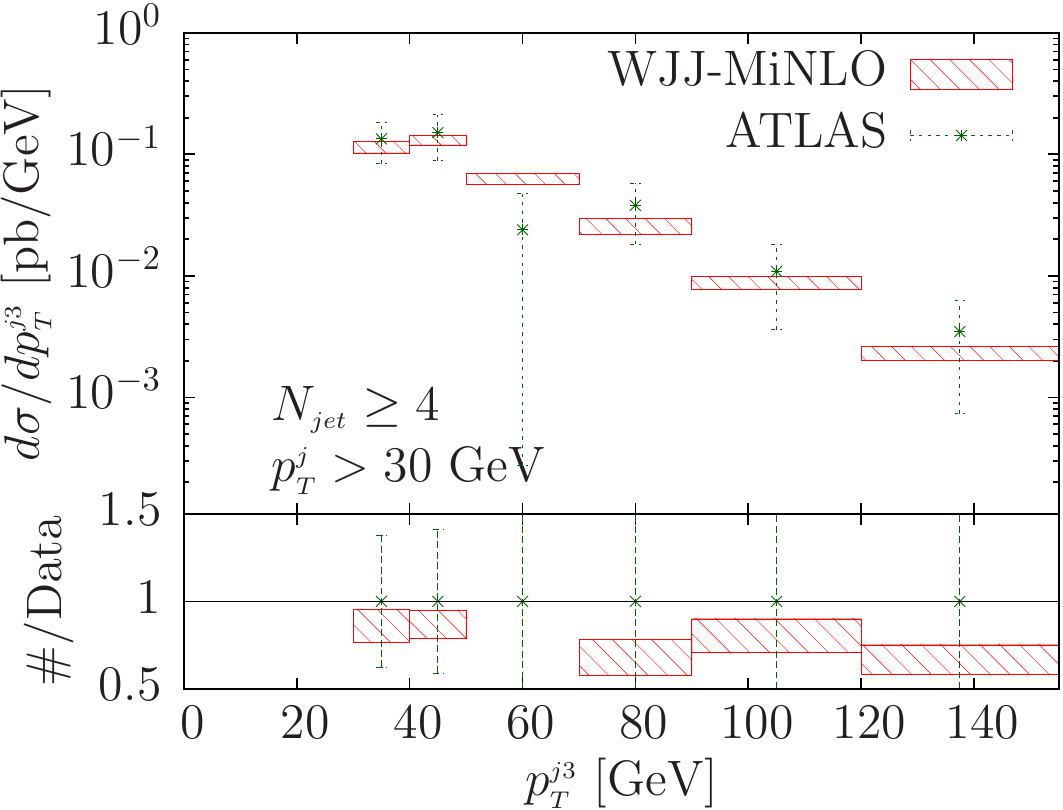,width=0.48\textwidth}
\end{center}
\caption{Transverse momentum of the third leading jet in inclusive
  three-jet events (left) and four-jet events (right).
\label{fig:ptj30MINATLj3pt} }
\end{figure}

\begin{figure}[htb]
\begin{center}
\epsfig{file=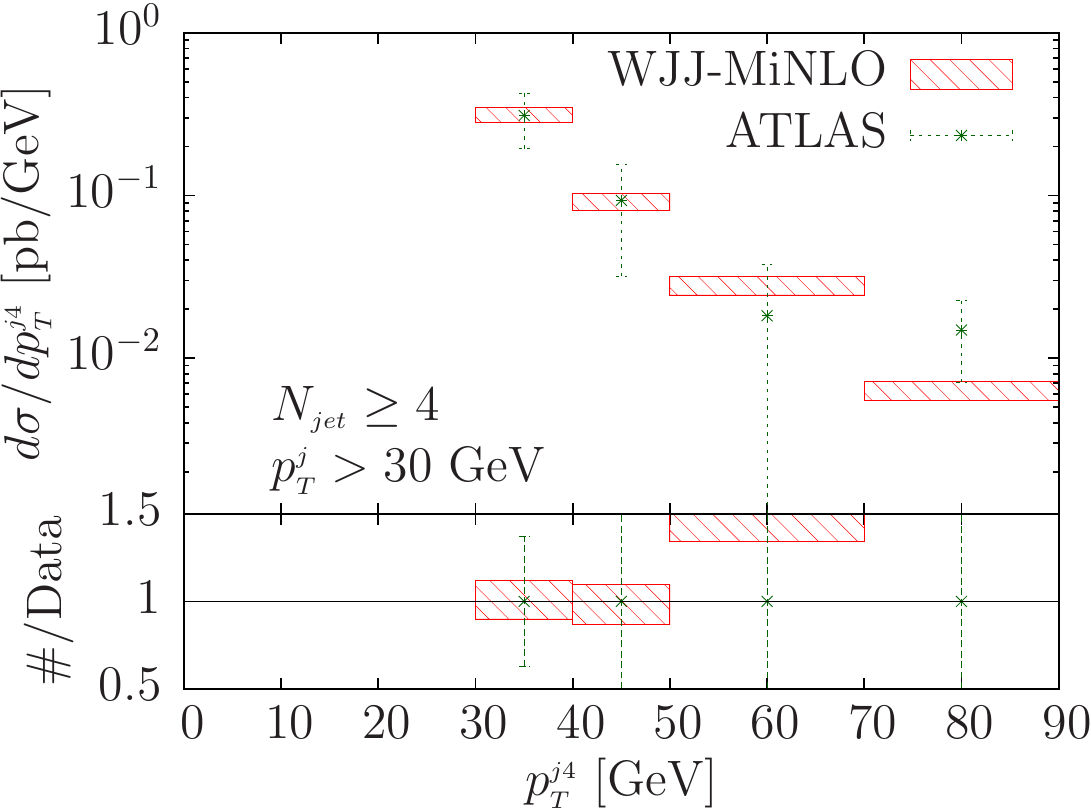,width=0.48\textwidth}
\end{center}
\caption{Transverse momentum of the fourth leading jet in inclusive
  four-jet events.
\label{fig:ptj30MINATLj4pt} }
\end{figure}

\begin{figure}[htb]
\begin{center}
\epsfig{file=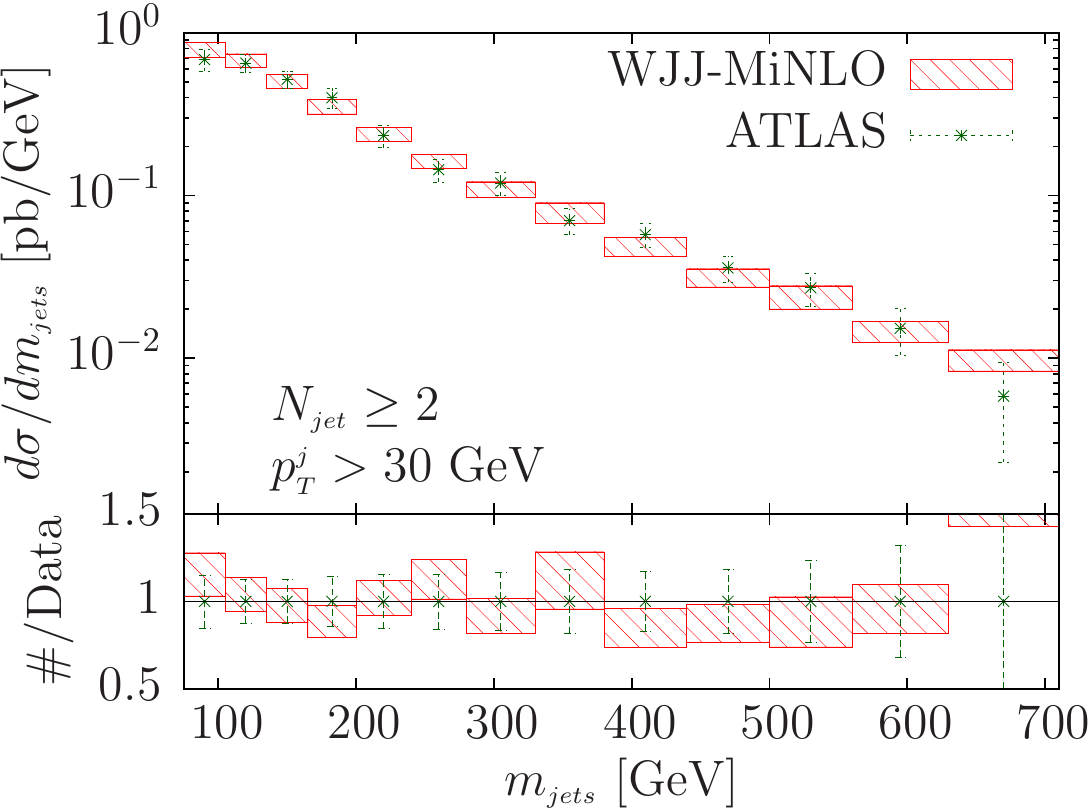,width=0.48\textwidth}
\end{center}
\caption{Invariant mass of the system composed of the first two
  leading jets in inclusive two-jet
  events. \label{fig:ptj30MINATLmjets} }
\end{figure}

\begin{figure}[htb]
\begin{center}
\epsfig{file=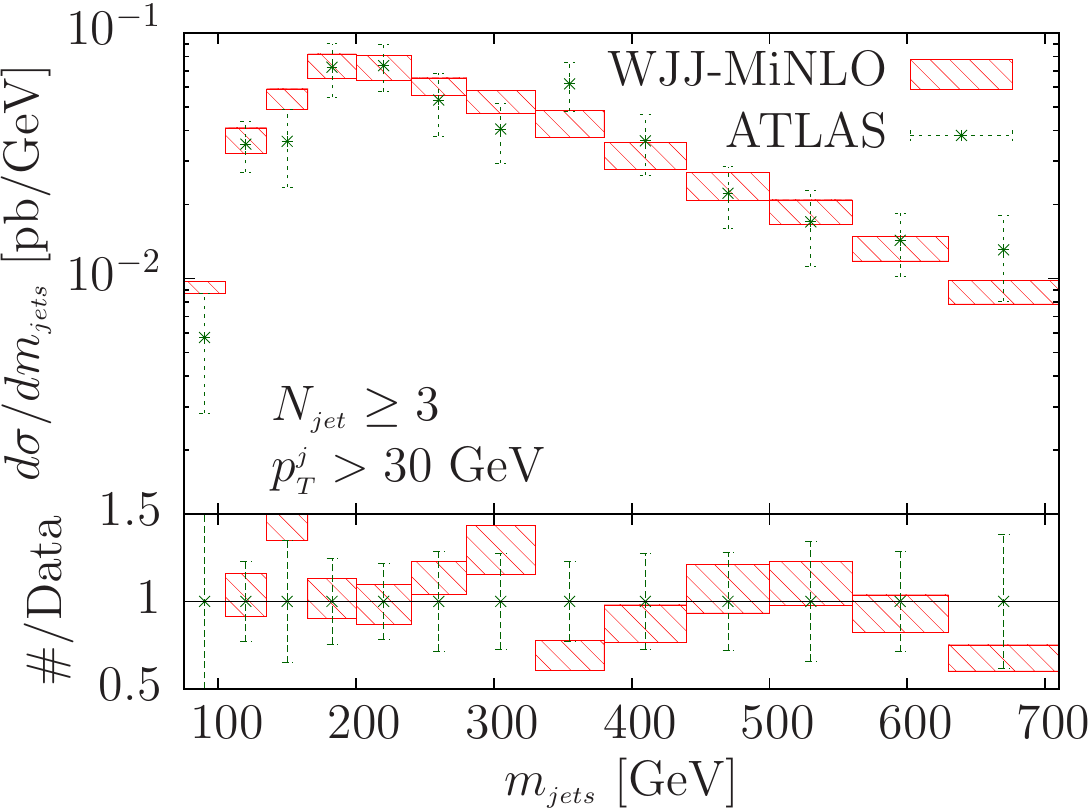,width=0.48\textwidth}
\epsfig{file=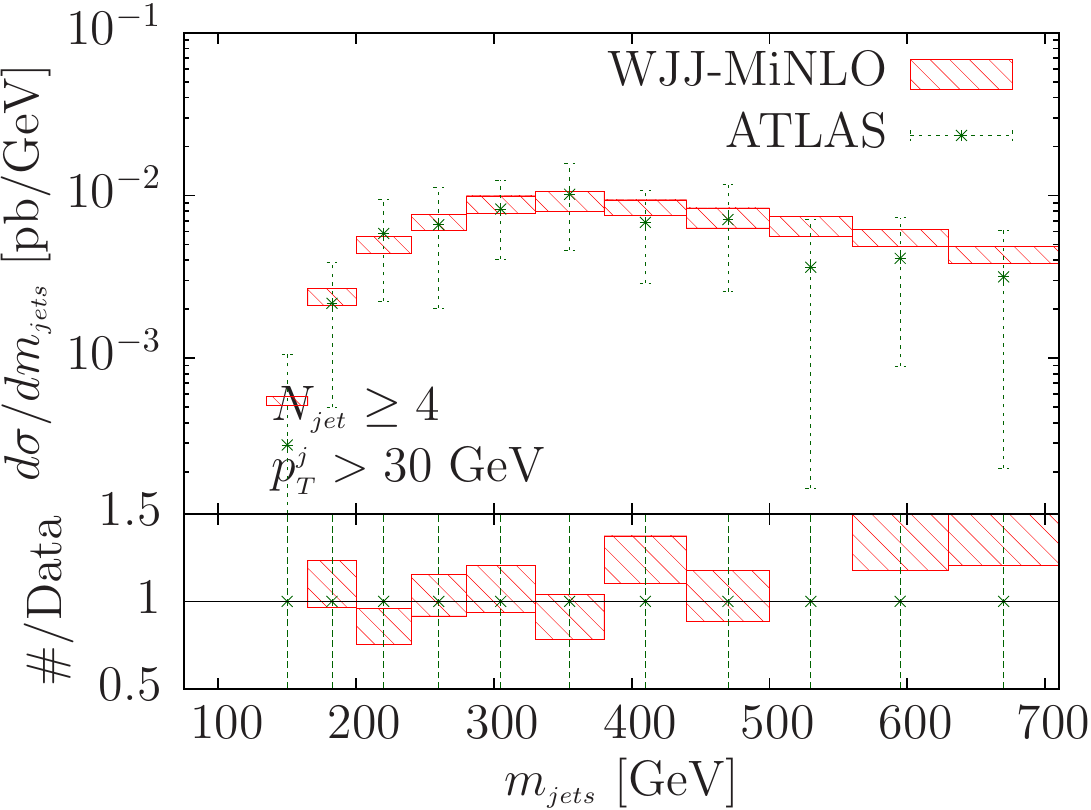,width=0.48\textwidth}
\end{center}
\caption{Invariant mass of the system composed of the first three
  leading jets in inclusive three-jet events (left) and by the first
  four leading jets in inclusive four-jet events (right).
\label{fig:ptj30MINATLmjets2} }
\end{figure}

\begin{figure}[htb]
\begin{center}
\epsfig{file=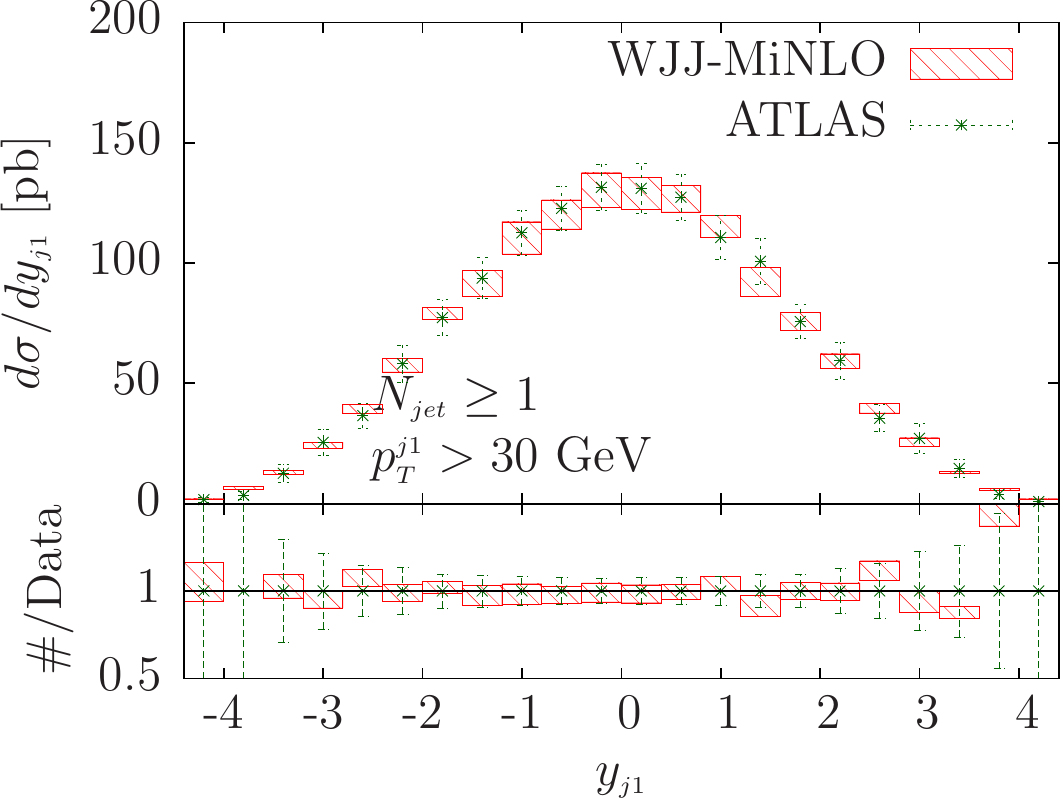,width=0.48\textwidth}
\end{center}
\caption{Rapidity of the leading jet in inclusive one-jet events.
\label{fig:ptj30MINATLmjets3} }
\end{figure}

\begin{figure}[htb]
\begin{center}
\epsfig{file=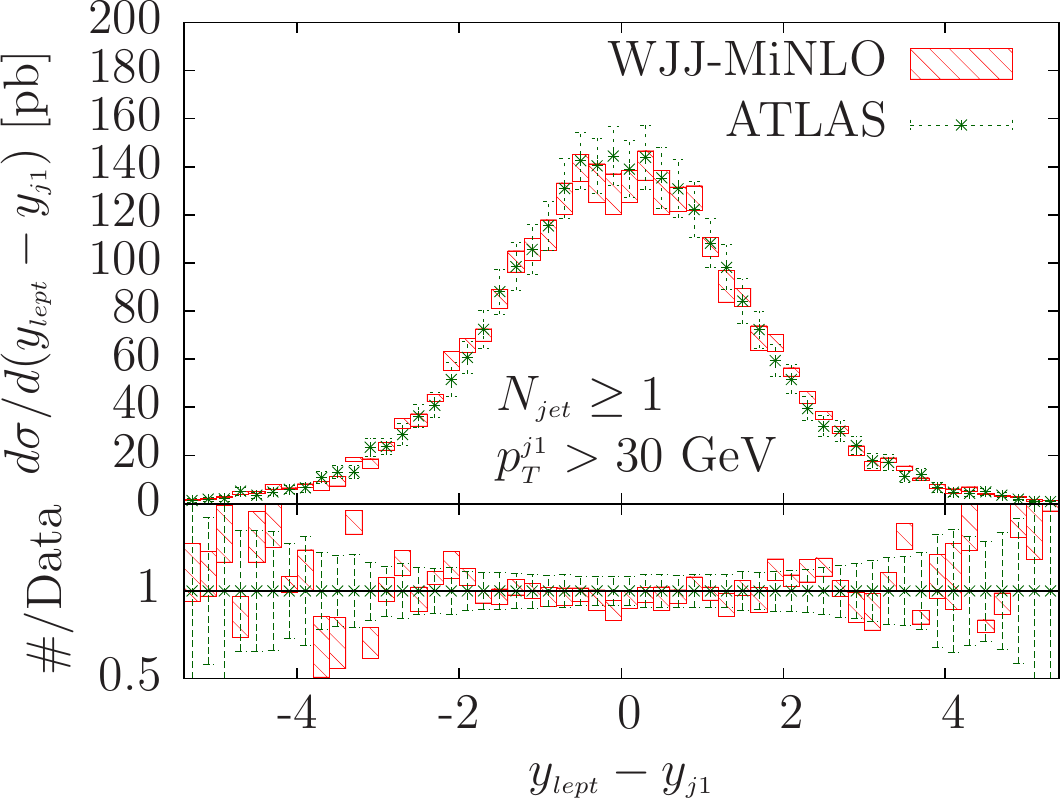,width=0.48\textwidth}
\epsfig{file=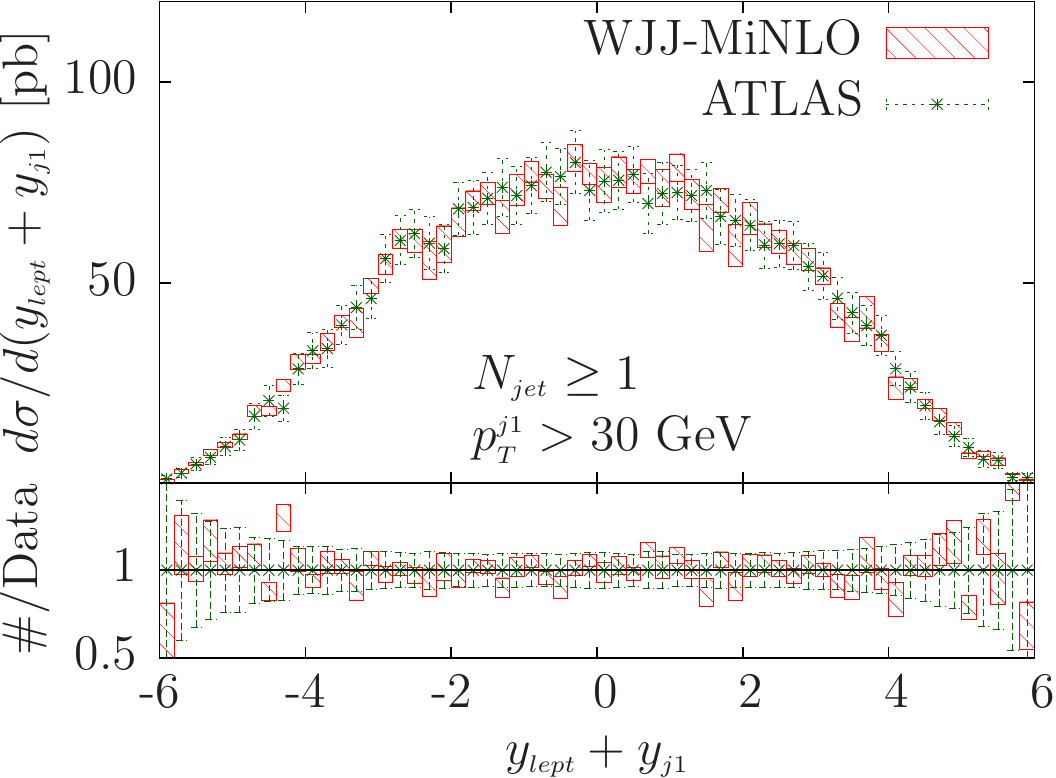,width=0.48\textwidth}
\end{center}
\caption{Difference (left) and sum (right) of the rapidity of the
  lepton and the leading jet in inclusive one-jet events.
\label{fig:ptj30MINATLlj} }
\end{figure}

\begin{figure}[htb]
\begin{center}
\epsfig{file=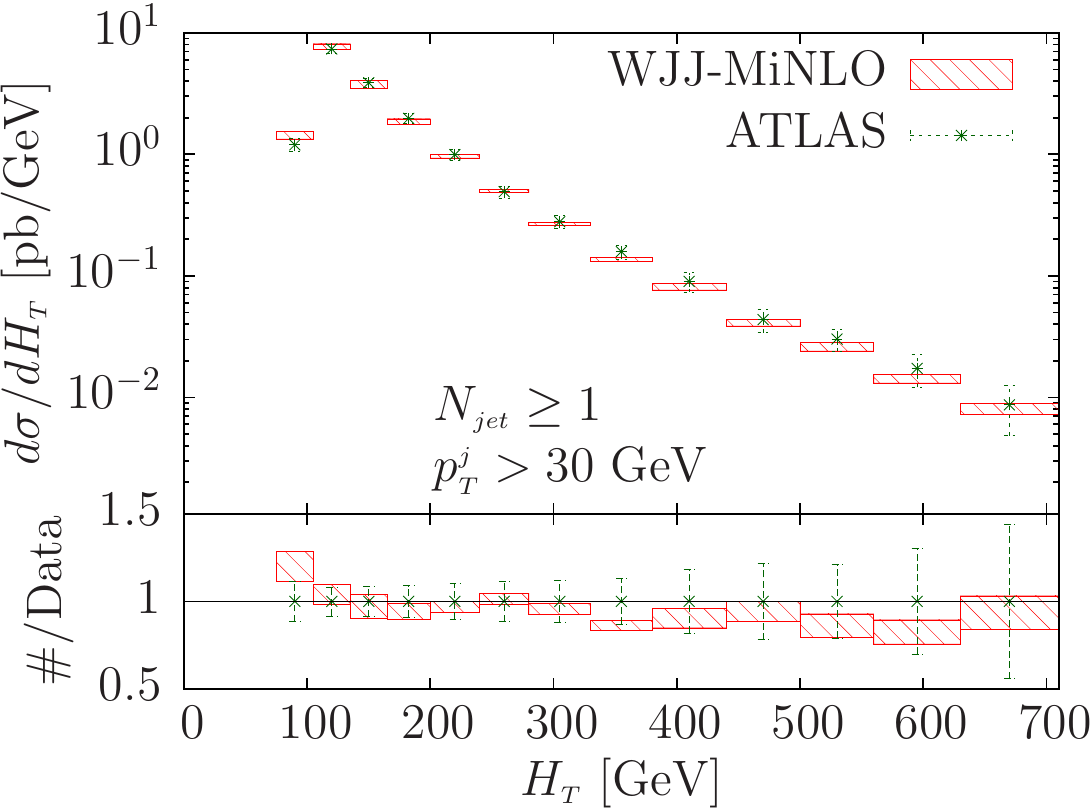,width=0.48\textwidth}
\epsfig{file=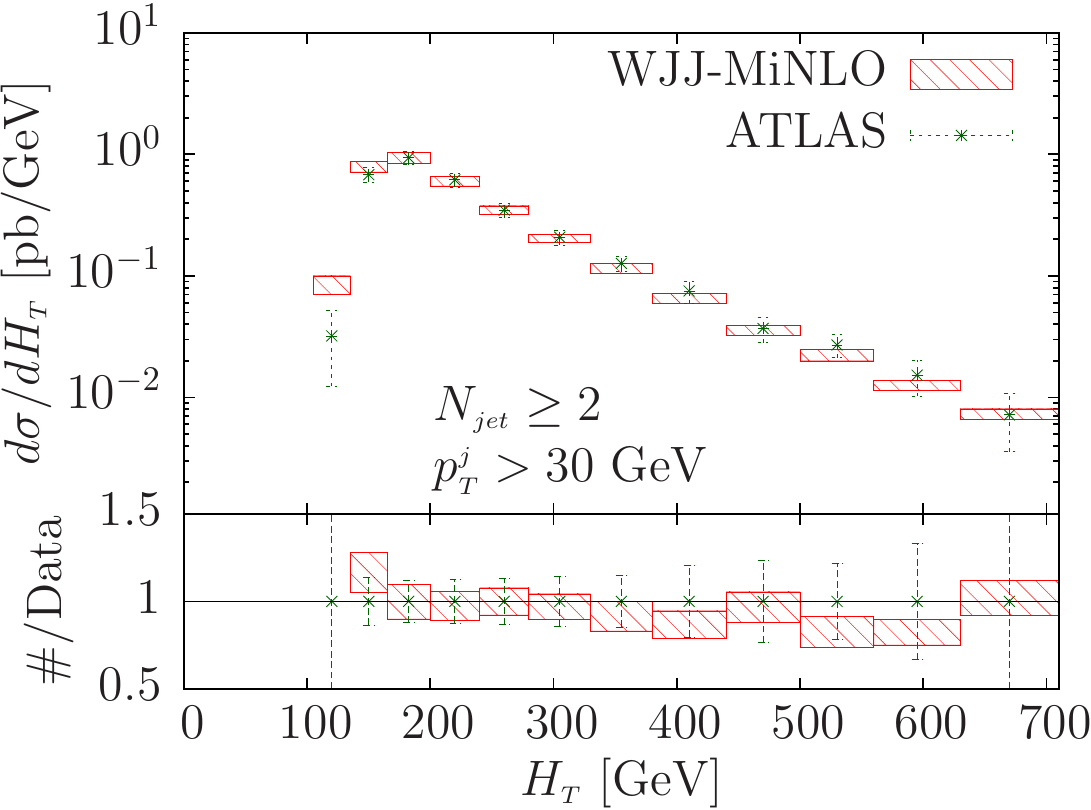,width=0.48\textwidth}
\end{center}
\caption{Scalar sum of transverse momenta of the event in inclusive
  one-jet events (left) and two-jet events (right).
\label{fig:ptj30MINATLht} }
\end{figure}

\begin{figure}[htb]
\begin{center}
\epsfig{file=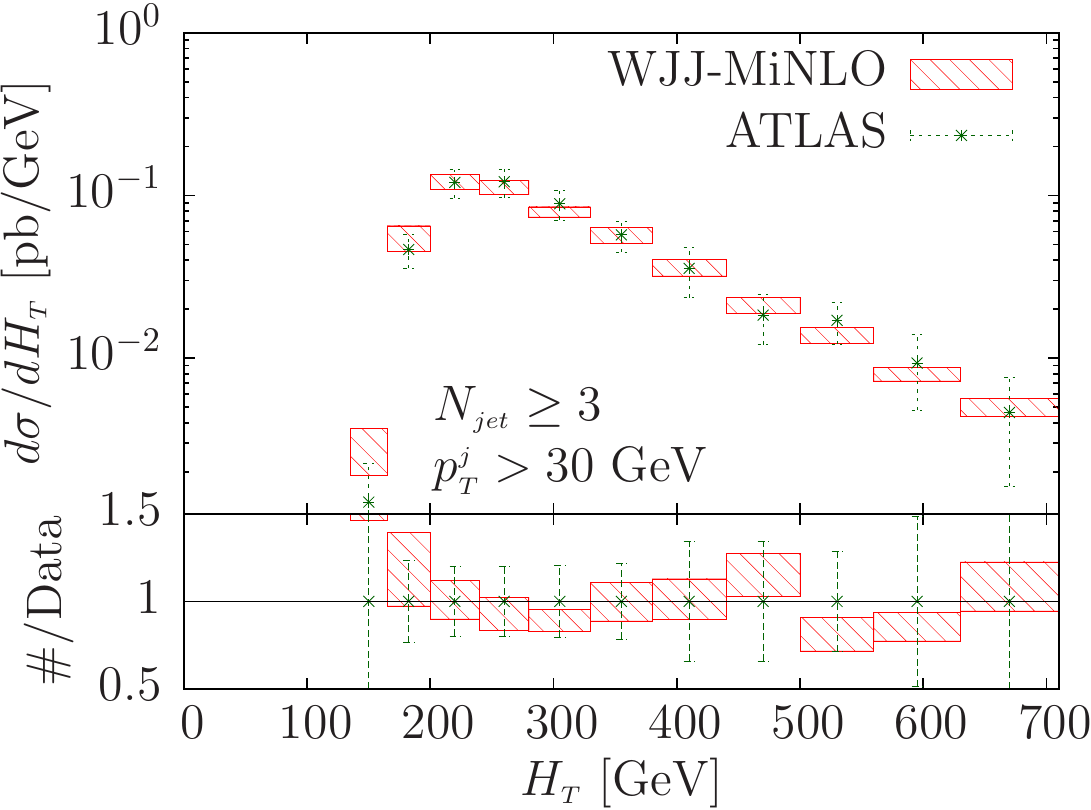,width=0.48\textwidth}
\epsfig{file=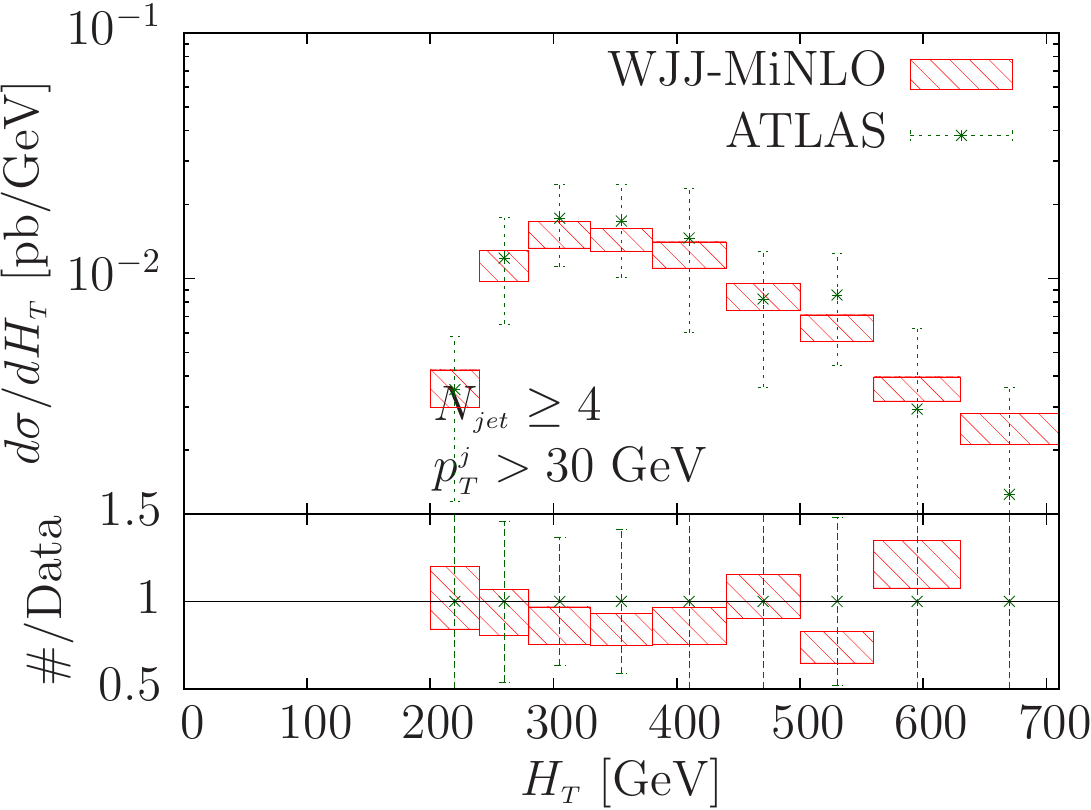,width=0.48\textwidth}
\end{center}
\caption{Scalar sum of transverse momenta of the event in inclusive
  three-jet events (left) and four-jet events (right).
\label{fig:ptj30MINATLht2} }
\end{figure}

\begin{figure}[htb]
\begin{center}
\epsfig{file=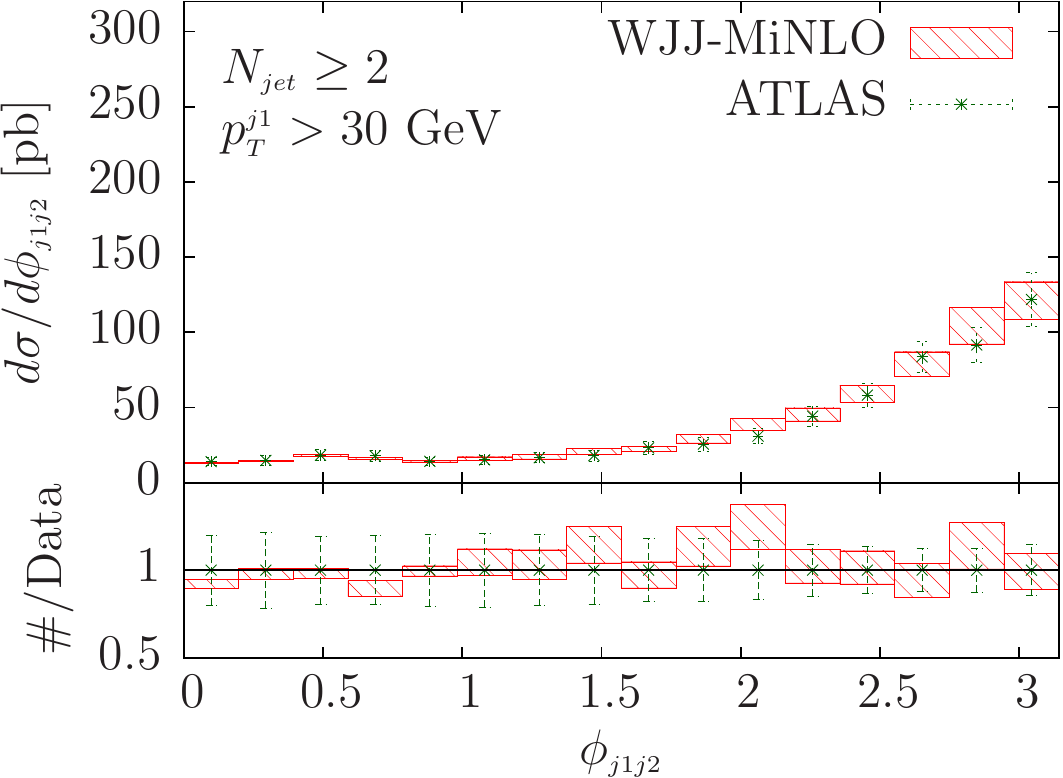,width=0.48\textwidth}
\epsfig{file=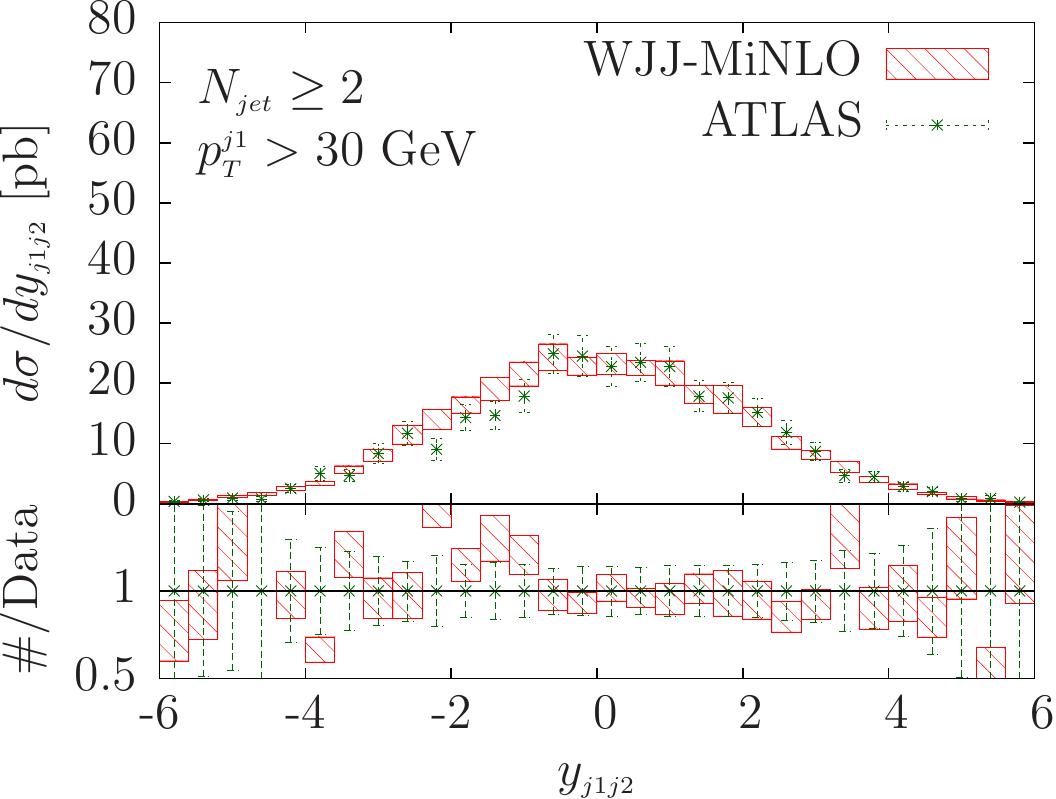,width=0.48\textwidth}
\end{center}
\caption{Azimuthal angle (left) and rapidity (right) difference
  between the two leading jets in inclusive two-jet events.
\label{fig:ptj30MINATLdphidy} }
\end{figure}

\begin{figure}[htb]
\begin{center}
\epsfig{file=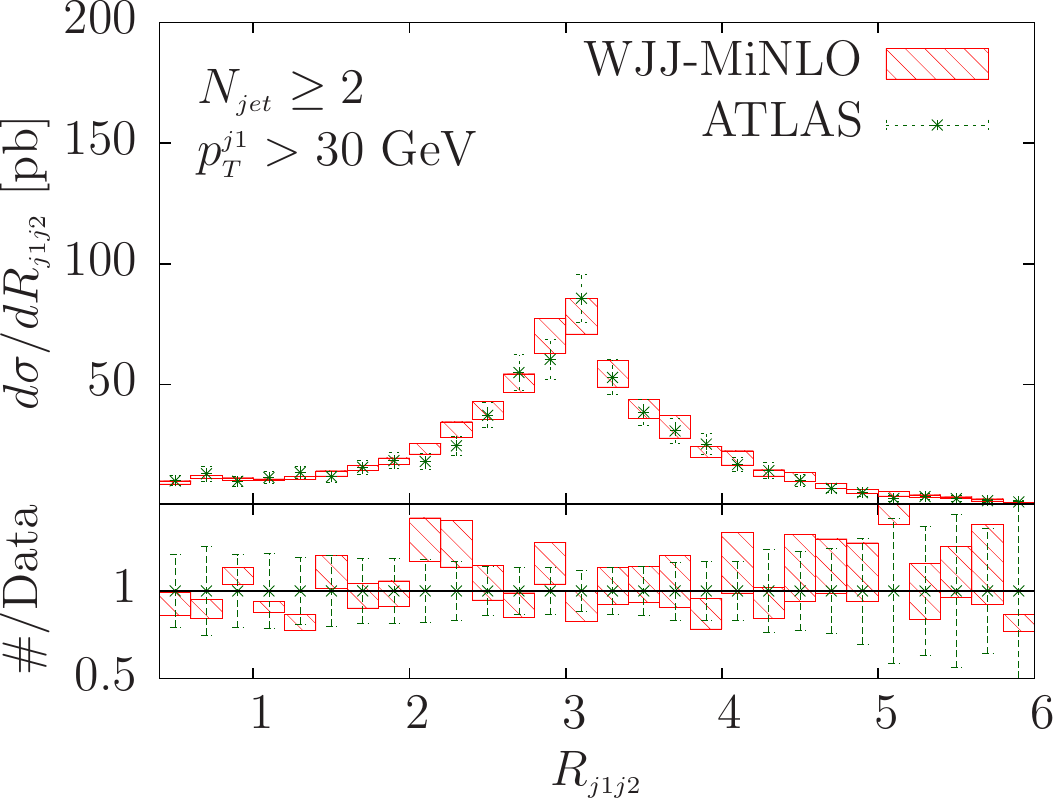,width=0.48\textwidth}
\end{center}
\caption{Distance $R_{j_1j_2} = (\Delta^2 y_{j_1j_2}+\Delta^2
  \phi_{j_1j_2})^{1/2}$ between the two leading jets in inclusive
  two-jet events.
\label{fig:ptj30MINATLdr} }
\end{figure}

\clearpage

\bibliography{WZjj}
\bibliographystyle{JHEP}

\end{document}